\documentclass[11pt,urlcolor=blue, linkcolor=blue]{article} 

\usepackage[utf8]{inputenc}
\usepackage[usenames, dvipsnames]{color}
\usepackage[svgnames]{xcolor}
\usepackage[colorlinks,citecolor=RoyalBlue, urlcolor=RoyalBlue, linkcolor=RoyalBlue ]{hyperref}

\usepackage{comment}
\usepackage{cite}
\usepackage{hhline}
\usepackage{mathrsfs,amsfonts,amsmath, amsthm, amssymb,slashed,dsfont}
\usepackage{slashed}
\usepackage[makeroom]{cancel}
\usepackage[normalem]{ulem}
\usepackage{soul}
\usepackage{booktabs}

\newcommand{\stkout}[1]{\ifmmode\text{\sout{\ensuremath{#1}}}\else\sout{#1}\fi}
\allowdisplaybreaks[1]

\usepackage{ifpdf}
\ifpdf
\usepackage[pdftex]{graphicx}
\usepackage{epstopdf}
\else
\usepackage[dvips]{graphicx}
\fi

\parskip=\baselineskip
\newcommand{\ccblue}[1]{\textcolor{blue}{#1}}
\newcommand{\ccred}[1]{\textcolor{red}{#1}}

\newcommand{\ket}[1]{|#1\rangle}

\newcommand{\innerproduct}[2]{\langle #1| #2\rangle}
\newcommand{\outerproduct}[2]{|#1\rangle\langle #2|}
\newcommand{\PGS}{\mathbf{P}_{\rm GS}}

\newcommand{\invertibleG}{\mathscr{I}_G}
\newcommand{\invertible}{\mathscr{I}}
\newcommand{\SPT}{\mathcal{S}_G}
\newcommand{\scr}[1]{\mathscr{#1}}

\newcommand{\cT}{\mathcal{T}}
\newcommand{\cH}{\mathcal{H}}

\newcommand{\cK}{\mathcal{K}}

\newcommand{\bZ}{\mathbb{Z}}
\newcommand{\bD}{\mathbb{D}}
\newcommand{\bM}{\mathbb{M}}

\newcommand{\ztwo}{\mathbb{Z}_2}

\newcommand{\ii}{\hspace{1pt}\mathrm{i}\hspace{1pt}}
\newcommand{\Z}{\mathbb{Z}}

\newcommand{\SO}{{\rm SO}}
\newcommand{\Spin}{{\rm Spin}}
\newcommand{\U}{{\rm U}}
\newcommand{\SU}{{\rm SU}}

\renewcommand{\O}{{\rm O}}

\newcommand{\rE}{{\rm E}}

\newcommand{\bea}{\begin{eqnarray}}
\newcommand{\eea}{\end{eqnarray}}
\def\be{\begin{equation}}
\def\ee{\end{equation}}
\def \Pin{\mathrm{Pin}}

\def \DPin{\mathrm{DPin}}
\def \EPin{\mathrm{EPin}}

\def \TP{\mathrm{TP}}
\def\B{\mathrm{B}}
\def\e{\mathrm{e}}

\def\GSD{\mathrm{GSD}}

\def \H{\operatorname{H}}

\newcommand{\Refe}[1]{Ref.~\cite{#1}}
\newcommand{\Eq}[1]{eq~(\ref{#1})} 
\newcommand{\eq}[1]{eq~(\ref{#1})} 
\newcommand{\eqn}[1]{eq~(\ref{#1})} 
\newcommand{\Eqn}[1]{eq~(\ref{#1})} 

\newcommand{\Fig}[1]{Fig.~\ref{#1}} 
 
\newcommand{\Table}[1]{Table \ref{#1}} 
\newcommand{\ft}[1]{Footnote \ref{#1}}

\def\SWAP{\mathrm{SWAP}}
\def\PD{\mathrm{PD}}
\def\ABK{\mathrm{ABK}}
\def\Arf{\mathrm{Arf}}
\newcommand{\Sec}[1]{Sec.~\ref{#1}} 

\def\GL{\mathrm{GL}}
\def\SL{\mathrm{SL}}
\newcommand{\nn}{\nonumber}

\def\cN{{\cal N}}

\def\nn{{\nonumber}}

\numberwithin{equation}{section}

\usepackage{upgreek}
\usepackage{multirow}
\usepackage{mathtools}

\usepackage{enumitem,cleveref}

\usepackage[all,cmtip]{xy}
\usepackage{tikz-cd}
\usepackage{tikz}
\usetikzlibrary{matrix}

\newcommand{\al}{\alpha} 
\newcommand{\bt}{\beta}

\title{\bf{\LARGE{Unwinding 
Fermionic SPT Phases:\\[8mm] 
 Supersymmetry Extension}}\\[8mm]}
 
\author{
	Abhishodh~Prakash$^a$\thanks{abhishodh.prakash@icts.res.in}~~ and
	Juven Wang$^{b,c}$\thanks{jw@cmsa.fas.harvard.edu}\quad\quad\quad  \\[10pt]
		$^{a}${\it\small International Centre for Theoretical Sciences, Tata Institute of Fundamental Research,} \\
	{\it\small Shivakote, Hesaraghatta, Bangalore 560089, India} \\ [5pt]
	$^{b}${\it\small Center of Mathematical Sciences and Applications, Harvard University,  Cambridge, MA 02138, USA} \\
	$^{c}${\it\small {School of Natural Sciences, Institute for Advanced Study, Einstein Drive, Princeton, NJ 08540, USA}} \\	
}

\begin{document}

\date{}
\maketitle

\thispagestyle{empty}

\begin{abstract}
 We show how 1+1-dimensional fermionic symmetry-protected topological states (SPTs, i.e. nontrivial short-range entangled gapped phases of quantum matter whose 
	 boundary exhibits 't Hooft anomaly and
	whose bulk cannot be deformed into a trivial tensor product state
	under finite-depth local unitary transformations
	only in the presence of global symmetries), 
	indeed can be \emph{unwound} to a trivial state by enlarging the Hilbert space
	via adding extra degrees of freedom and suitably extending the global symmetries. 
	The extended 
	projective global symmetry on the boundary can become supersymmetric in a specific sense, 
	i.e., it contains group elements that do \emph{not} commute with the fermion number parity $(-1)^F$, while the	
	anti-unitary time-reversal symmetry becomes fractionalized. 
	This also means we can uplift and remove certain exotic fermionic anomalies 
	(e.g., ``parity'' anomaly in time-reversal or reflection symmetry)
	via
	appropriate \emph{supersymmetry extensions} in terms of \emph{group extensions}. 
	{We work out explicit examples for multi-layers of 1+1d Majorana fermion chains, then 
	comment on models with Sachdev-Ye-Kitaev (SYK) interactions, 
	{intrinsic fermionic \emph{gapless} SPTs protected by supersymmetry,}
	and	generalizations to higher spacetime dimensions via a cobordism theory.}
\end{abstract}


\newpage

 \pagenumbering{arabic}
   \setcounter{page}{2}

\tableofcontents


\section{Overview and summary of results}

\subsection{Classification of gapped phases of matter}
\label{sec:classification}
A central goal of condensed matter physics is to enumerate and understand the properties of various phases of matter. The systems of interest are generally many-body quantum systems in the {infinite system size} 
 limit described by local interactions and typically invariant under some group of global symmetries. Long-range ordering is one common way a system can transit to a non-trivial phase. The paramagnet-ferromagnet transition of magnets and the superfluid transition via Bose-Einstein condensation are prominent examples of such an ordering and are well described by the Ginzburg-Landau framework based on \emph{spontaneous symmetry breaking} (SSB) and characterized by local order parameters. The discovery of 
 quantum Hall phases, which were beyond this framework 
 heralded a new era in condensed matter and was followed by intense theoretical and experimental research attempting to get a sense of the the various possible phases of matter (classification) and their distinguishing properties (characterization) which continues to this day. This program has been particularly successful in the case of \emph{gapped phases} of matter i.e. when the systems of interest have a non-zero spectral gap separating the lowest energy states from the excited states. 

Gapped phases can be divided into two categories- \emph{non-invertible} and \emph{invertible} phases. Loosely speaking, a physical system $\scr{P}$ belongs to an invertible phase if there exists an inverse system, $\scr{P}^{\text{inv}}$ such that the composite system produced by \emph{stacking} $\scr{P}$ and $\scr{P}^{\text{inv}}$, belongs to the \emph{trivial} phase i.e. can be connected, without encountering a phase transition, to a trivial system $\scr{P}_0$ which has always has a unique trivial ground state i.e. product state for bosons or slater determinant state for fermions. The collection of invertible phases forms an \emph{Abelian group}, with group multiplication arising from stacking and the identity element being the trivial phase. Examples of invertible phases include paramagnets, integer quantum Hall states, insulators and superconductors. On the other hand, any phase that is not invertible is called a non-invertible phase. Examples of non-invertible phases include all gapped phases with long range order arising through SSB like ferromagnets as well as fractional quantum Hall phases and spin liquids. In this work, we will be concerned with a \emph{specific} class of invertible phases in the presence of global symmetries.  

Given a physical system with some global symmetries $G$ belonging to a non-trivial invertible phase, if we explicitly break global symmetries, one of two things can happen- the phase can become trivial (eg: topological insulators), or the phase can continue to remain non-trivial (eg: integer quantum Hall phases). The latter phases form a \emph{subgroup} of the group of invertible phases and are called Symmetry-Protected-Topological states or Symmetry-Protected-Trivial states or simply SPT states (SPTs) 
\cite{ChenGuLiuWen_GroupChomology_PhysRevB.87.155114, Senthil1405.4015} and will be the focus of this work. 

Let us state the above in the language of groups. Let $\scr{I}_G$ be the Abelian group of invertible phases. Explicitly breaking symmetries maps each element of $\scr{I}_G$ to $\scr{I}_0$, the group of invertible phases with \emph{no symmetries}. It is easy to see that breaking symmetry induces a \emph{surjective group homomorphism}, $\pi$
\begin{eqnarray}
\scr{I}_G \xrightarrow{\pi} \scr{I}_0.
\end{eqnarray}
SPT phases are those elements of $\scr{I}_G$ that land on the trivial element of $\scr{I}_0$ under the symmetry breaking map i.e. $\pi$. In other words, SPT phases (which we will call $\mathcal{S}_G$) correspond to the \emph{kernel} of $\pi$. 
\begin{equation}
\mathcal{S}_G = \ker \pi
\end{equation}
These simple arguments reveals a rich structure in the space of gapped physical systems- for any dimension, for either bosons or fermions with global symmetry $G$, $\mathcal{S}_G$, $\scr{I}_G$, and $\scr{I}_0$ can be arranged in the following \emph{short exact sequence}~\footnote{An exact sequence is a sequence of homomorphisms $\{\phi_n\}$ between groups $	\{A_n\}$ of the form 
	\begin{equation}
		\ldots A_n \xrightarrow{\phi_n} A_{n+1} \xrightarrow{\phi_{n+1} } \ldots \nonumber
	\end{equation} such that  $ \text{im} \phi_n =  \ker\phi_{n+1}$. A short exact sequence is an exact sequence of the form
\begin{equation}
1 \rightarrow C \xrightarrow{i} B \xrightarrow{\pi} A \rightarrow 1 \nonumber
\end{equation} 
where $i$ is injective and $\pi$ is surjective as a consequence of exactness.
}
\begin{equation}
1 \rightarrow \mathcal{S}_G \xrightarrow{i} \scr{I}_G \xrightarrow{\pi} \scr{I}_0 \rightarrow 1
\end{equation}

\begin{figure}[!htbp]
	\centering	
	\includegraphics[width=0.75\textwidth]{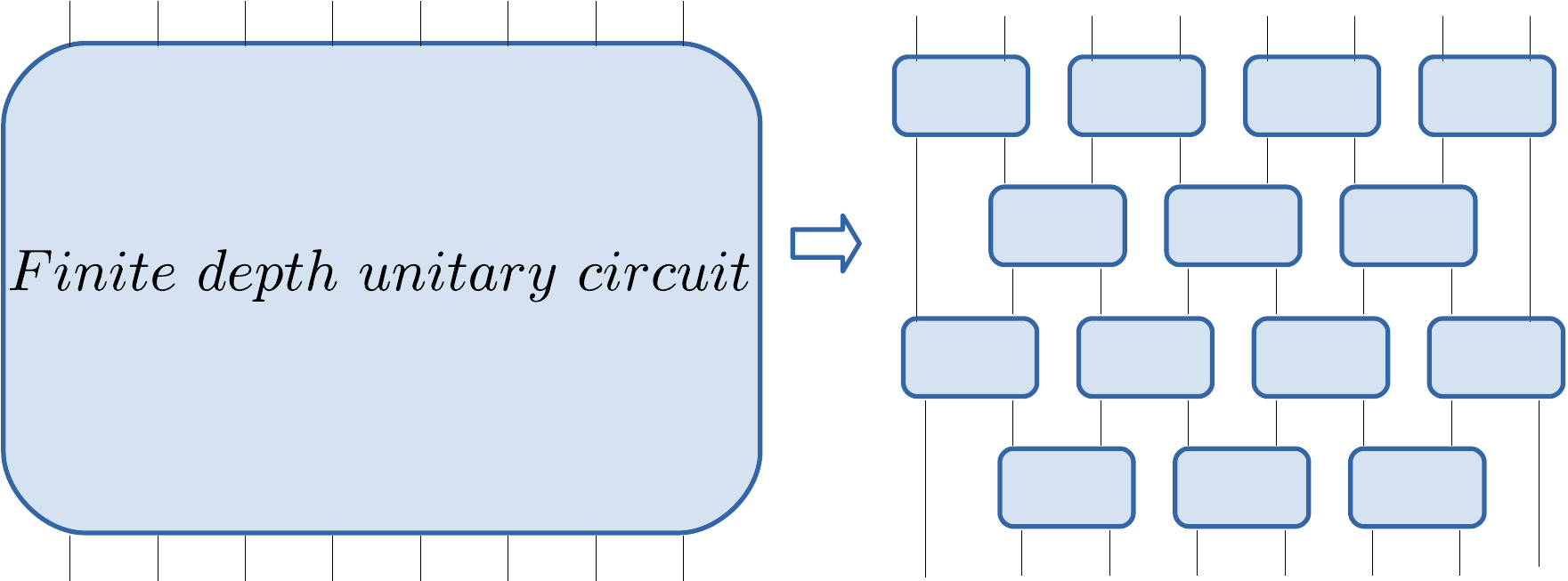}
	\caption{Finite depth unitary circuit (FDUC). The operator can be expressed as a produce of finite number of layers (vertically separated in the figure, indicating the time direction). Each layer is a product of unitary operators with strictly local support (square blobs in the figure) i.e. acts on the Hilbert space of a finite number of spins. \label{fig:FDUC}}
\end{figure}

Let us briefly review some important properties of SPT phases:

\begin{enumerate}
	\item If symmetries are disregarded, all SPT phases become trivial. This means that we can adiabatically connect the ground state to a tensor product state using a finite time-evolution operator generated by a local Hamiltonian which can be represented as a finite depth unitary circuit (FDUC). Following the terminology of our previous work~\cite{AP_Unwinding_PRB2018} based on the method of symmetry extension \cite{WangWenWitten_SymmetricGapped_PRX2018}, 
	we call this \emph{unwinding}.\footnote{A more mathematical way to interpret our approach on unwinding of SPTs 
	\cite{WangWenWitten_SymmetricGapped_PRX2018, AP_Unwinding_PRB2018} can be that
	pulling back trivialization of fiber bundle via fibrations \cite{Witten1605.02391}. 
	These have physical consequences of (1) removing 't Hooft anomalies or (2) constructing symmetry-extended topological boundaries, 
	via symmetry extensions \cite{WangWenWitten_SymmetricGapped_PRX2018, AP_Unwinding_PRB2018}. 
	The 't Hooft anomalies are the obstruction of gauging global symmetries; here we concern the global symmetries including
	 time-reversal or reflection symmetry, in additional to spacetime or internal symmetries.
	 These anomalies are generalized ``parity'' anomalies -- more accurately involving time-reversal $\Z_2^T$ or reflection symmetry $\Z_2$ instead of the spatial parity. 
	In fact,  \Refe{Witten1605.02391, Tachikawa1712.09542} 
	provide further mathematical insights of the problems dealt in  \cite{WangWenWitten_SymmetricGapped_PRX2018, AP_Unwinding_PRB2018}.
	} 
	An FDUC is a locality preserving unitary operator that be written as a product of a finite number of layers of \emph{ultra-local unitaries} schematically shown in Fig~\ref{fig:FDUC}.  
	\item Boundaries of SPT phases are always non-trivial i.e. are gapless or have some form of long-range order if gapped. This \emph{persistent ordering}~\cite{Komargodski_SciPostPhys.6.1.003_2019} is due to the presence of an \emph{'t Hooft anomaly} for the global symmetry $G$ that forbids a trivial phase.  
\end{enumerate} 

Although not immediately obvious, the above two properties are intimately connected --- the different gapped boundary terminations of SPT phases can tell us how to construct FDUC to unwind an SPT phase. Let us focus on symmetry breaking boundary terminations of SPT phases. If a $G$-SPT phase can be terminated by spontaneously breaking $G$ down to one of its subgroups, $G_{\text{sub}}$ on the boundary, this means that the same SPT phase can be unwound by a FDUC that is invariant under $G_{\text{sub}}$. Note that breaking all symmetries is one specific case of the above paradigm corresponding to $G_{\text{sub}}$ being the trivial group. 

A second boundary termination, particularly insightful for 3+1d bulk systems,\footnote{{Our convention on the dimensions
is that the one, two, three spatial dimensions are written as 1d, 2d, 3d, etc. 
We denote their spacetime dimensions as 1+1d, 2+1d, 3+1d, etc.} \label{ft:dim}} 
is when no global symmetry is broken but the boundary is topologically ordered. The anomalous nature of the boundary symmetry is seen in the way it acts on the anyons of the theory which is impossible in a purely 2+1d system 
\cite{VishwanathSentil_surfaceTI_PhysRevX.3.011016,LukaszAshvin_TSC_PRX2013} (see the review \cite{Senthil1405.4015} and references within for an overview). 
For a specific class of bosonic SPT phases (those classified by group cohomology), the authors of~\cite{WangWenWitten_SymmetricGapped_PRX2018} were able to place the understanding of symmetric gapped boundary terminations in a systematic framework. They prove that given a system belonging to a non-trivial $d+1$-dimensional bosonic $G$-SPT 
phase with criteria: 
\begin{enumerate}
\item  bosonic SPTs classified by group cohomology. 
\item $G$ is a finite group, including both unitary and/or anti-unitary symmetries.
\item $d \geq 1$ for $d$ is the spatial dimension. (Namely the spacetime dimension $d+1 \geq 2$.)
\end{enumerate}
there always exists a larger group, $\tilde{G}$ such that, starting with a $(d-1)+1$
dimensional $\tilde{G}$-symmetric invariant theory and gauging a particular subgroup, $K$ satisfying $\tilde{G}/K \cong G$, we are left with a $K$ gauge theory with $G$ symmetry realized anomalously and can form a boundary termination for the G-SPT. This in turn also teaches us something else very interesting- \emph{the G-SPT phase can be unwound, not by breaking symmetry but by extending to $\tilde{G}$!} In other words, the 't Hooft anomaly of $G$-symmetry
of a boundary theory
(in a $d-1$ dimensional space and 1-dimensional time)
can be trivialized by extending the system to $\tilde{G}$-symmetry.
In \Refe{AP_Unwinding_PRB2018}, this was demonstrated explicitly in the case of 1+1d for various groups $G$ by constructing FDUCs invariant under an appropriately identified $\tilde{G}$. 

This naturally brings about the following questions --- \emph{can there be a systematic framework to understand symmetric gapped boundary terminations for SPT phases not classified by group cohomology?} and \emph{can these be obtained by symmetry extension?} In particular, this includes all \emph{intrinsically fermionic} SPT phases i.e. which cannot be constructed by stacking a non-trivial bosonic SPT phase with a trivial fermionic phase. In this work, for 1+1d fermionic SPT phases, we provide evidence that the answer to both the above italicized questions is in the affirmative.

\subsection{Summary of results and outline}
We focus only on on-site, unitary and anti-unitary symmetries (including various time-reversal symmetries)  in this article. We find that just like bosonic SPT phases, 1+1d fermionic SPT phases can be unwound by an FDUC invariant under an extended symmetry. For intrinsically fermionic SPT phases, we find that the extended symmetry has an interesting property --- the fermion parity is not at the center of the symmetry group, i.e. there exists elements of the symmetry group that \emph{do not commute with fermion parity} locally. We term this as supersymmetry extension. This is our main result which we demonstrate using several examples. Our results are demonstrated using exactly solvable lattice models and constructing explicit FDUCs. 

This article is organized as follows: we warm up by considering a bosonic example in Sec.~\ref{eq:Haldane} which sets the language and the general ideas that are detailed in \Refe{AP_Unwinding_PRB2018}. In Sec.~\ref{sec:free fermions}, we consider non-trivial free-fermion models of SPT phases in 1+1d corresponding to classes BDI, AIII and DIII and demonstrate how they can be unwound to a trivial phase by symmetry extension. In Sec.~\ref{sec:interacting BDI}, we study an interacting example of a fermionic SPT phase with 
time-reversal symmetry and demonstrate its unwinding. {In \Sec{sec:SUSYQM}, we justify our usage of the term \emph{supersymmetry extension} and show that the extended symmetry has an intimate correspondence to the supersymmetric quantum mechanics. In Sec.~\ref{sec:conclusion1}, we make comments about generalization in the continuum, suitable for quantum field theory (QFT) formulation. 
In Sec.~\ref{sec:conclusion}, we explore generalizations to
higher dimensions and relations to a cobordism theory (from \Sec{sec:1+1d} to \Sec{sec:DIFF-TOP}) and the implications to {intrinsically fermionic \emph{gapless} SPTs (\Sec{sec:gaplessSPT}),}
and finally conclude with possible future directions in (\Sec{sec:Finalcomments}). Additional details are relegated to the Appendices.}

\section{Warm up: a bosonic example --- the Haldane phase}
\label{eq:Haldane}
In this section, we warm-up by considering a bosonic example to set the language as well as list the results that we generalize to fermionic systems in this work. Additional examples as well as general results for bosonic SPTs can be found in \cite{AP_Unwinding_PRB2018}. 
\begin{figure}[!htbp]
	\centering	
	\includegraphics[width=100mm]{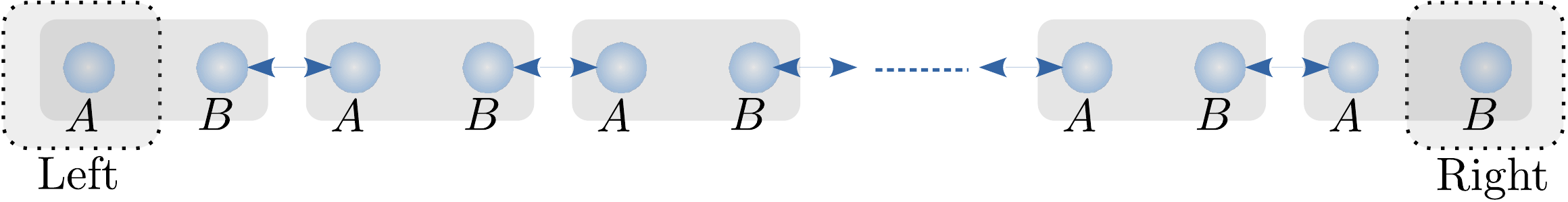}
	\caption{Model Hamiltonian belonging to the Haldane phase with open boundary conditions. The 
	SU(2) doublet written as the SU(2) representation $J=\frac{1}{2}$ qubits at both ends are indicated. }
\end{figure}
\subsection{The Hamiltonian, symmetry fractionalization, and classification}
\label{sec:Haldane-symmetry-fractionalize}
We consider a {1+1d} bosonic spin chain belonging to the Haldane phase protected by an on-site  SO(3) symmetry. The Hilbert space on each site is a pair of qubits each of which transforms as the $J=\frac{1}{2}$ representation of  the SU(2) group. Together, each site forms a $\frac{1}{2} \otimes \frac{1}{2} \cong 1 \oplus 0$ reducible representation that is faithful to SO(3). The Hamiltonian with $\ell$ sites and SO(3)  symmetry operators are 
\begin{align}
H &= - \sum_{j=1}^{\ell} \vec{\sigma}_{B,j}.\vec{\sigma}_{A,j+1}, \\
U(\hat{n},\theta) &= \prod_{j=1}^{\ell} \exp{\left(\ii \hat{n}.\vec{\sigma} \frac{\theta}{2}\right)_{A,j}}~ \exp{\left(\ii \hat{n}.\vec{\sigma} \frac{\theta}{2}\right)_{B,j}}.
\label{eq:AKLT}
\end{align}
For \eq{eq:AKLT}, two SU(2) symmetry operators (two of $J=\frac{1}{2}$) act on the same site $j$ makes the 
SO(3) symmetry operator (for $J=0 \oplus 1$).
The model by Affleck, Kennedy, Lieb and Tasaki (AKLT) is obtained by projecting out the $J=0$ sector on each site and only retaining $J=1$~\cite{AKLT_PhysRevLett.59.799}. We will choose not to perform this projection and work with the exactly solvable model of \cref{eq:AKLT}. With periodic boundary conditions, \cref{eq:AKLT} has the following unique ground state which has SO(3) singlets on each bond connecting the sites.  
\begin{align}
\ket{{\rm GS}} = \prod_{j=1}^{\ell} \frac{\ket{\uparrow_{B,j} \downarrow_{{A,(j|\ell)+1}}}-\ket{\downarrow_{B,j} \uparrow_{A,(j|\ell)+1}}}{\sqrt{2}},
\end{align}
where, $(j|\ell) \equiv j \mod \ell$. With any symmetry preserving open boundary conditions, the ground state degeneracy is four (GSD = 4), i.e. any ground state can be written as a vector in a four-dimensional Hilbert subspace of ground states: 
\begin{align}
\ket{{\rm GS}} &= \left(\prod_{j=1}^{\ell} \frac{\ket{\uparrow_{B,j} \downarrow_{A,j+1}}-\ket{\downarrow_{B,j} \uparrow_{A,j+1}}}{\sqrt{2}}\right) \otimes \ket{\psi},\\
\ket{\psi} &= \sum_{\alpha = \uparrow, \downarrow} \sum_{\beta = \uparrow, \downarrow} \psi_{\alpha,\beta} \ket{\alpha_{A,1}} \ket{\beta_{B,\ell}}, ~~~ \sum_{\alpha = \uparrow, \downarrow} \sum_{\beta = \uparrow, \downarrow} |\psi_{\alpha,\beta}|^2=1.
\end{align} 
We can define a Hermitian  projection operator $\PGS =  \PGS^\dagger$  onto the ground space that has support on the ends of the chain
\begin{align}
\PGS &= \mathbf{P}_{{\rm left}}\otimes\outerproduct{s}{s} \otimes \mathbf{P}_{{\rm right}} \\
\mathbf{P}_{{\rm left}} &=  \sum_{\alpha = \uparrow, \downarrow} \outerproduct{\alpha_{A,1}}{\alpha_{A,1}},~~~\mathbf{P}_{{\rm right}} =  \sum_{\alpha = \uparrow, \downarrow} \outerproduct{\alpha_{B,\ell}}{\alpha_{B,\ell}}, \\
\ket{s} &= \left(\prod_{j=1}^{\ell} \frac{\ket{\uparrow_{B,j} \downarrow_{A,j+1}}-\ket{\downarrow_{B,j} \uparrow_{A,j+1}}}{\sqrt{2}}\right).
\end{align}
$\PGS$ has support on the boundaries of the chain. This allows us to study the \emph{fractionalization} of the SO(3) symmetry on the boundaries at low energies.
\begin{align}
\PGS U(\hat{n},\theta) \PGS &=  \exp{\left(\ii \hat{n}.\vec{\sigma} \frac{\theta}{2}\right)_{A,1}} \otimes ~ \exp{\left(\ii \hat{n}.\vec{\sigma} \frac{\theta}{2}\right)_{B,\ell}} \nonumber \\ &= V(\hat{n},\theta)_{{\rm left}} \otimes V(\hat{n},\theta)_{{\rm right}}.
\end{align}
The matrix symmetry representation of each of the boundary modes corresponds to $J=\frac{1}{2}$ which is a fraction of the quantum numbers allowed by the bulk symmetry group SO(3). Furthermore, the representation is faithful to the group SU(2) which is a non-trivial extension of SO(3) by $\ztwo$. This is an example demonstration of the emergence of extended symmetries on the boundaries of SPT phases~\cite{WangWenWitten_SymmetricGapped_PRX2018}. 

\begin{figure}[!htbp]
	\centering	
	\includegraphics[width=0.6\textwidth]{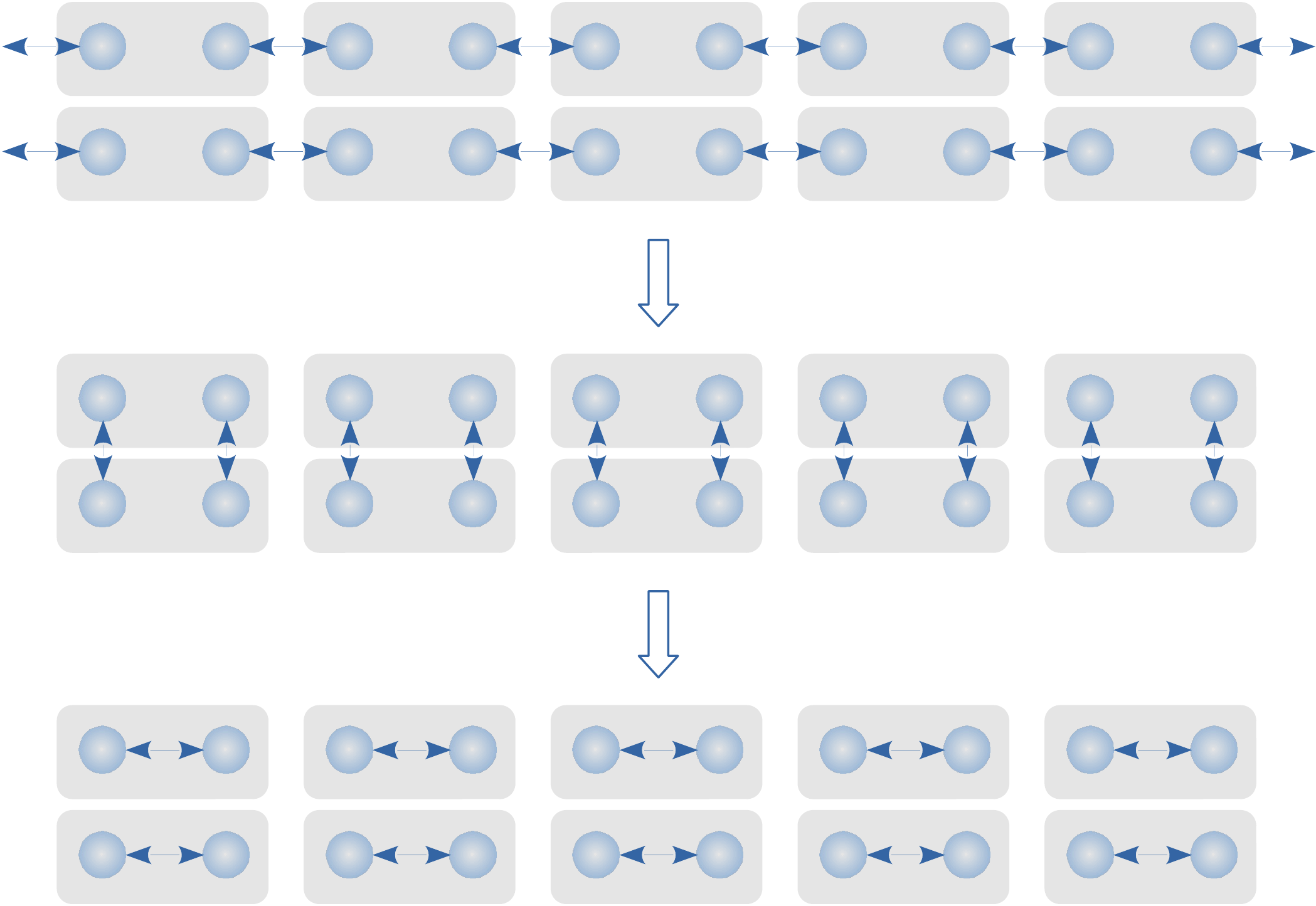}
	\caption{Unwinding two-layers of~\cref{eq:AKLT2}
	via the two-layer FDUC $W = W_2 W_1$ in \eq{eq:W1W2}. \label{fig:AKLT_unwinding}}
\end{figure}

The non-trivial nature of the Haldane phase comes from the fact that there does not exist an SO(3) invariant FDUC that can map the ground state to a product state. If we consider two layers of the Hamiltonian~\cref{eq:AKLT} however, 
\begin{align}
H^{\rm II} &= - \sum_{j=1}^{\ell} \vec{\sigma}_{B,j}.\vec{\sigma}_{A,j+1} - \sum_{j=1}^{\ell} \vec{\tau}_{B,j}.\vec{\tau}_{A,j+1} , \label{eq:AKLT2} \\
U(\hat{n},\theta) &= \prod_{j=1}^{\ell} \exp{\left(\ii \hat{n}.\vec{\sigma} \frac{\theta}{2}\right)_{j,A}} \exp{\left(\ii \hat{n}.\vec{\sigma} \frac{\theta}{2}\right)_{j,B}}  \exp{\left(\ii \hat{n}.\vec{\tau} \frac{\theta}{2}\right)_{j,A}}\exp{\left(\ii \hat{n}.\vec{\tau} \frac{\theta}{2}\right)_{j,B}}.
\end{align}
it is easy to see that it can be \emph{unwound} to one whose ground state is a product state as shown in \Fig{fig:AKLT_unwinding} using a two-layer FDUC $W=W_2 W_1$:
\begin{align}
W H^{\rm II} W^\dagger &= - \sum_{j=1}^{\ell} \vec{\sigma}_{A,j}.\vec{\sigma}_{B,j} - \sum_{j=1}^{\ell} \vec{\tau}_{A,j}.\vec{\tau}_{B,j} , \\
W &= W_2 W_1, \\
W_1 &= \prod_{j=1}^{\ell} \frac{1}{2} \left(\mathds{1} + \vec{\sigma}_{B,j}.\vec{\tau}_{A,j+1}\right),~~~W_2 = \prod_{j=1}^{\ell} \frac{1}{2} \left(\mathds{1} + \vec{\sigma}_{A,j}.\vec{\tau}_{B,j}\right).
\label{eq:W1W2}
\end{align}
The bosonic SWAP operator $\frac{1}{2}\left(\mathds{1} + \vec{\sigma}.\vec{\tau}\right)$ swaps the basis states of the $\sigma$ and $\tau$ qubits. What we have reproduced is the well-known $\ztwo$ classification of the Haldane phase~\cite{ChenGuLiuWen_GroupChomology_PhysRevB.87.155114}.

\subsection{Unwinding by symmetry extension}
\label{sec:Haldane-symmetry-extension} \label{sec: Unwind Cluster bulk}
\begin{figure}[!htbp]
	\centering	
	\includegraphics[width=100mm]{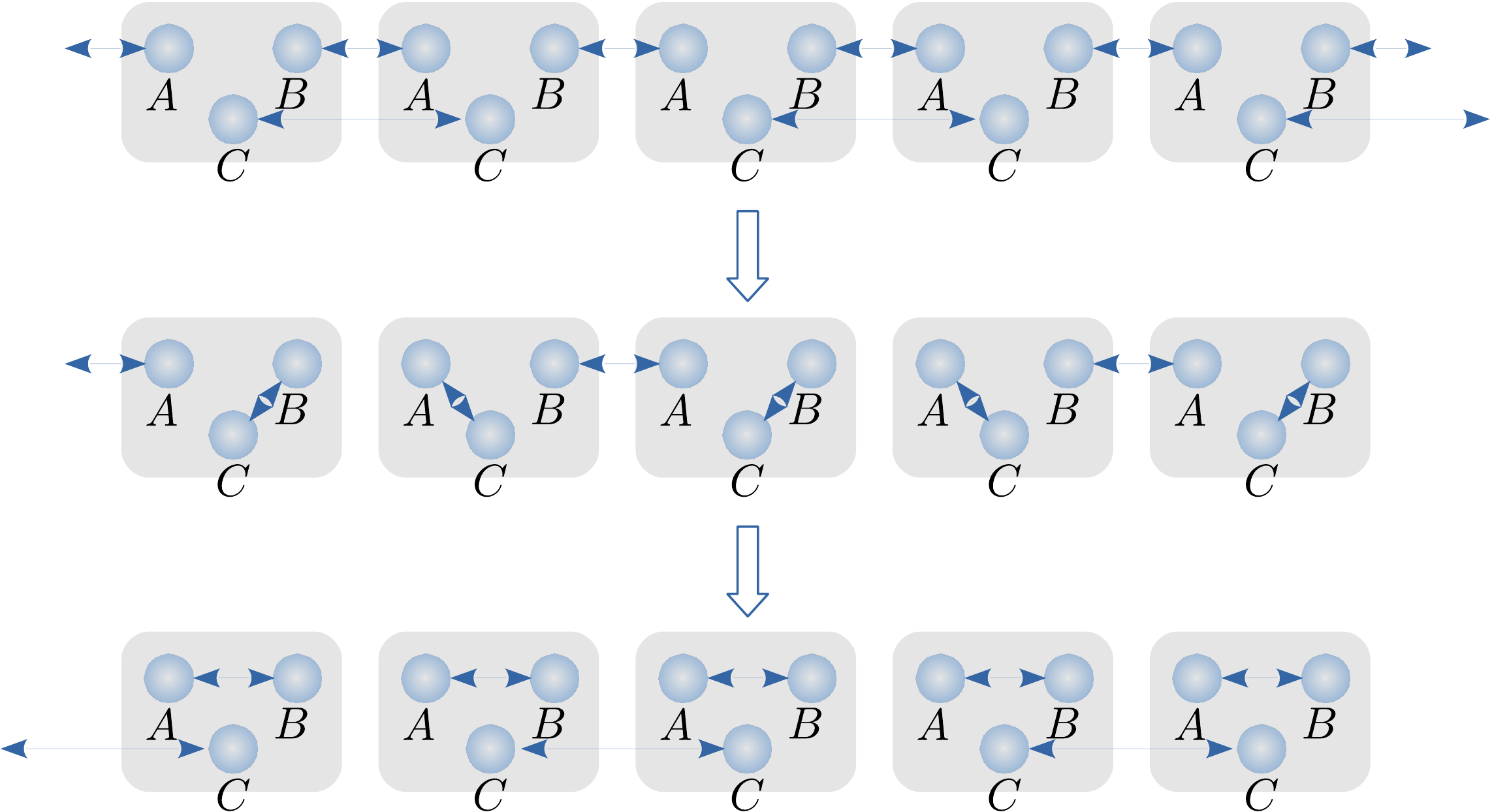}
	\caption{Unwinding the extended Hamiltonian~\cref{eq:AKLT_extended_symmetry} state by extending the bulk on-site symmetry to SU(2),
	via $W = W_2 W_1$ in two steps: 
	$W_1$ and $W_2$ in \eq{eq:W1W2-AKLT-extend}.
	\label{fig:AKLT_unwinding_extension}}
\end{figure}
We now demonstrate that by elevating the extended boundary symmetry (SU(2)) to the bulk, we can unwind the single layer Hamiltonian~\cref{eq:AKLT}. To do this, we first add an extra $J=\frac{1}{2}$ qubit on each site (which we label ``C''). We also add a trivial Hamiltonian that dimerizes the $C$ qubits in the ground state~\footnote{\label{ft:LSM}
An aside: When the on-site representation is half-odd-integer as in \cref{eq:AKLT_extended_symmetry}, the Hamiltonian has to explicitly break the lattice translation symmetry if we want to have a unique ground state. This is a consequence of the Lieb-Schultz-Mattis theorem~\cite{LiebSchultzMattis1961fr,AP_LSM}}. 
\begin{equation}
H = - \sum_{j=1}^{\ell} \vec{\sigma}_{B,j}.\vec{\sigma}_{A,j+1} - \sum_{\text{odd j}} \vec{\sigma}_{C,j}.\vec{\sigma}_{C,j+1}.
\label{eq:AKLT_extended_symmetry}
\end{equation}
This Hamiltonian can be unwound as shown in \cref{fig:AKLT_unwinding_extension} using a two-layer FDUC
\begin{align}
W H W^\dagger &= - \sum_{j=1}^{\ell} \vec{\sigma}_{A,j}.\vec{\sigma}_{B,j} - \sum_{\text{even j}} \vec{\sigma}_{C,j}.\vec{\sigma}_{C,j+1},\\
W &= W_2 W_1\\
W_1 &= \prod_{\text{odd j}}\frac{1}{2} \left(\mathds{1} + \vec{\sigma}_{C,j}.\vec{\sigma}_{A,j+1}\right),~~ W_2 = \prod_{\text{odd j}}\frac{1}{2} \left(\mathds{1} + \vec{\sigma}_{A,j}.\vec{\sigma}_{C,j}\right).
\label{eq:W1W2-AKLT-extend}
\end{align}

\subsection{Boundary Hamiltonians and protected degeneracy for the Haldane chain:
Sachdev-Ye interaction}
\begin{figure}[!htbp]
	\centering	
	\includegraphics[width=0.74\textwidth]{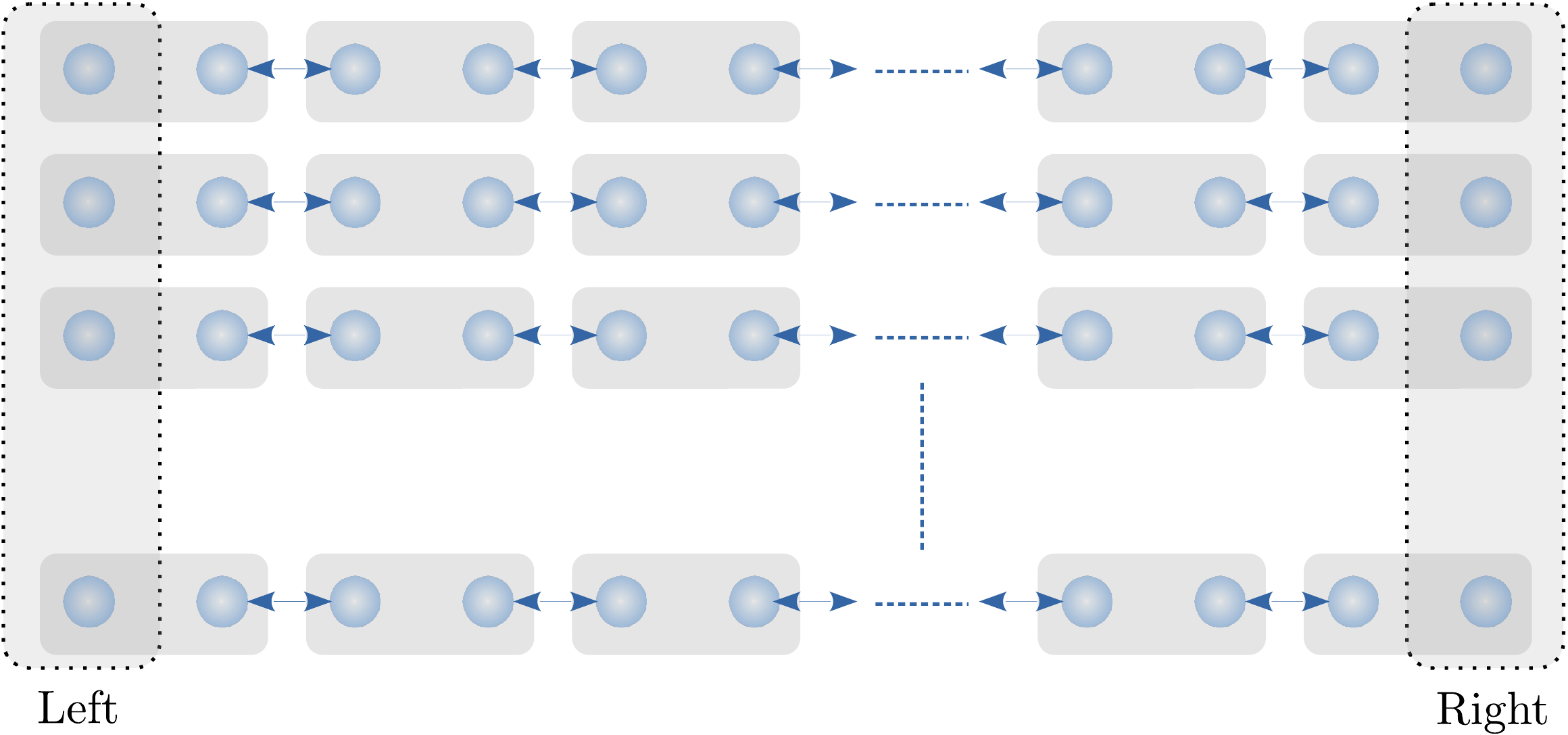}
	\caption{Multiple copies of the Hamiltonian~\cref{eq:AKLT} \label{fig:AKLT_layers}}
\end{figure}
The classification of the Haldane phase tells us that even copies of Hamiltonian~\cref{eq:AKLT} belong to the trivial phase and odd number of copies belong to the Haldane phase. This result can also be obtained by attempting to lift the boundary degeneracy by introducing interactions. Let us consider $N$ copies of the Hamiltonian~\cref{eq:AKLT}. The boundary Hilbert space consists of $N$ of bosonic $J=\frac{1}{2}$ qubits which we shall label $m=1, \ldots, N$. The most general symmetry preserving Hamiltonian we can write down is as follows
\begin{align}
H_{{\rm bdry}} = \sum_{a,b=1}^N J_{ab}~ \vec{\sigma}_a.\vec{\sigma}_b + \sum_{a,b,c,d=1}^N K_{abcd}~  \left(\vec{\sigma}_a.\vec{\sigma}_b\right)  \left(\vec{\sigma}_c.\vec{\sigma}_d\right) + \ldots 
\label{eq:AKLT_boundary}
\end{align}
The leading term in the above Hamiltonian with $J_{ab}$ drawn from a random distribution is the model by Sachdev and Ye~\cite{Sachdev1992fkSYK9212030}. We are concerned with the question of when there is a possibility that \cref{eq:AKLT_boundary} can have a unique ground state. This can be answered from some basic representation theoretic facts of SU(2). The total Hilbert space of $N$ spins forms a reducible representation of SU(2) corresponding to the Clebsch-Gordan (CG) decomposition 
\begin{align}
\frac{1}{2} \otimes \frac{1}{2} \otimes \frac{1}{2} \ldots \otimes \frac{1}{2} ~~ (\text{N terms}).
\end{align}
The only way to obtain a symmetric unique ground state is if the Hilbert space contains a $J=0$ irreducible representation (irrep) in the above CG decomposition. If $N$ is even, this is possible. For example, consider $N=2$:
\begin{eqnarray}
\frac{1}{2} \otimes \frac{1}{2} \cong 1 \oplus 0.
\end{eqnarray}
If $N$ is odd however, we only obtain half-odd-integer values of $J$ in the CG decomposition. For example, for $N=3$, 
\begin{eqnarray}
\frac{1}{2} \otimes \frac{1}{2} \otimes \frac{1}{2} \cong \frac{3}{2} \oplus \frac{1}{2} \oplus \frac{1}{2}.
\end{eqnarray}
This recovers the $\ztwo$ classification.

\subsection{Formal explanation}
We now provide a formal way of understanding the above results following for the mathematically inclined reader \cite{WangWenWitten_SymmetricGapped_PRX2018}.
In \Sec{sec:Haldane-symmetry-fractionalize}, 
the boundary spin $J=\frac{1}{2}$ is a projective representation of SO(3), 
which is in fact the faithful representation of SU(2). 
This can also be interpreted as the group extension as a short exact sequence:
\bea \label{eq:SU2-SO3}
1  \longrightarrow \ztwo   \overset{ }{\longrightarrow} 
{\SU(2)}\vert_{\text{boundary}} \overset{ r}{\longrightarrow} {\SO(3)} \vert_{\text{bulk}} \longrightarrow 1, 
\eea
as the boundary-bulk symmetry extension,
with the following implications:
\begin{enumerate}
\item \emph{Topological invariant}:
The 1+1d Haldane chain is characterized by a 1+1d topological invariant given by a nontrivial group element of the cohomology group
$\H^2(\B\SO(3),\U(1)) =\Z_2$ \cite{ChenGuLiuWen_GroupChomology_PhysRevB.87.155114}
also the cobordism group
$\Omega_2^{\SO}(\B\SO(3)) =\Z_2$ (see Sec.~5.5.2 of  \cite{WanWang1812.11967}. )
Precisely the 1+1d topological invariant contains the 
second Stiefel-Whitney (SW) class $w_2$ of the associated vector bundle $V$ of SO(3), as
\bea \label{eq:w2VSO3}
\exp(\ii \pi \int_{M^2} w_2 (V_{\SO(3)}) )
\eea
(which is known as the group cocycle in a group cohomology, or the cobordism invariant in a cobordism theory).
The \eq{eq:w2VSO3} is a partition function of the low energy physics of Haldane chain on a 1+1d spacetime manifold $M^2$.

For example, we can construct a nontrivial manifold generator: take the base manifold $M^2 = S^2$ as a 2-sphere, 
the map $S^2 \to \B\SO(3)$ determines an SO(3) bundle over $S^2$. 
Recall the homotopy group $\pi_2(\B\SO(3)) = \pi_1(\SO(3)) = \Z_2$.
The homotopy class of the map $S^2 \to \B\SO(3) \to \B^2 \Z_2$ is 
given by a nontrivial generator exactly $w_2 (V_{\SO(3)})$ that can be pull back to the base manifold $S^2$.

\item \emph{Trivialization of topological invariant}:
This group extension \eq{eq:SU2-SO3} actually solves a problem about the trivialization of topological invariant via the 
symmetry extension ---
once we pull back the
SO(3) bundle to the SU(2) bundle via the pull back $r^*$, then the
1+1d topological invariant of Haldane chain
$r^* \exp(\ii \pi \int_{} w_2 (V_{\SO(3)})) \sim 1$ becomes trivialized 
because $r^* w_2 (V_{\SO(3)}) =w_2 (V_{\SU(2)})=0$ 
(the trivialization means that it is a group coboundary as a trivial element in the cohomology group
(in $\H^2(\B\SU(2),\U(1)) =0$),
also a trivial identity invariant in the cobordism theory (in $\Omega_2^{\SO}(\B\SU(2))  =0$)). 

This implies that we can in fact add an extra SU(2) doublet $J=\frac{1}{2}$ on each of the boundary of Haldane chain,
then we can add SU(2)-invariant interaction terms to the boundary Hamiltonian to gap the degenerate $\GSD = 4$ states, left with only a single unique ground state ($\GSD=1$)
preserving the extended SU(2) symmetry (but not preserving the SO(3) symmetry), see  \cite{AP_Unwinding_PRB2018}. 

\item \emph{Trivialization of 't Hooft anomaly}:
 The 1+1d bulk topological invariant from 
 $\H^2(\B\SO(3),\U(1))$ $=$ $\Omega_2^{\SO}(\B\SO(3))$ $=\Z_2$ classifies its 0+1d boundary 't Hooft anomaly   
 of the SO(3) global symmetry. The $w_2 (V_{\SO(3)})$ regarded as the SO(3) background field, once turned on, 
 can detect the 't Hooft anomaly. 
 
It is shown that the \eq{eq:SU2-SO3}, with the suitable condition mentioned above, says that the
 0+1d 't Hooft anomaly of the SO(3) global symmetry becomes anomaly-free once we pull back the theory to the SU(2) symmetry in  0+1d \cite{WangWenWitten_SymmetricGapped_PRX2018}.

\item   \emph{Bulk trivialization}: 
If we consider the trivialization of topological invariant not only at the boundary but also at the full bulk via the unwinding method in \Sec{sec:Haldane-symmetry-extension}. 
Then  we also write the group extension as the bulk-bulk symmetry extension:
\bea \label{eq:SU2-SO3}
1  \longrightarrow \ztwo   \overset{ }{\longrightarrow} 
{\SU(2)}\vert_{\text{bulk}} \overset{ r}{\longrightarrow} {\SO(3)} \vert_{\text{bulk}} \longrightarrow 1. 
\eea
\end{enumerate}
The above bosonic SPTs serve as a guiding principle, for the readers to follow.
In the following sections, we will explore the generalization to fermionic SPTs counterpart.
We provide many examples in \Sec{sec:free fermions} and \Sec{sec:interacting BDI}, then we provide a formal explanation of the fermionic examples in
\Sec{sec:conclusion1}.

\section{Unwinding SPT phases of fermions}
\label{sec:free fermions}
In this section, we present the key results of the paper. Fermionic invertible phases as well as SPT phases were originally discovered in \emph{non-interacting} systems i.e. Hamiltonians are quadratic in fermion creation and annihilation operators. Prominent examples are the integer quantum Hall states, topological insulators and superconductors with various symmetries. In a monumental series of works~\cite{SchnyderRyu_classification_doi:10.1063/1.3149481,Kitaev_periodic_doi:10.1063/1.3149495}, all free-fermion invertible phases have been classified in all dimensions. The program of extending the classification to interacting fermions in any dimension is under progress, 
{although so far the best known mathematical framework is known to be  the generalization of group supercohomology 
\cite{Gu1201.2648, Kitaev2015, Kapustin1701.08264,WangGu1703.10937, GaiottoJohnson-Freyd1712.07950, WangGu1811.00536},
and the cobordism group classification 
\cite{Kapustin1406.7329, FreedHopkins1604.06527, 1711.11587GPW, GuoJW1812.11959} and \cite{WanWang1812.11967}} 

In the coming sections, we will demonstrate unwinding 1+1 d fermionic SPT phases by extended symmetries similar to~\cref{sec:Haldane-symmetry-extension} and ref.\cite{AP_Unwinding_PRB2018} with examples of models from fermion SPT phases with a \emph{free-limit} i.e. have representatives which are quadratic Hamiltonians and whose symmetries can be placed in one of the symmetry classes (see ref. \cite{ChiuRyuSchnyderRyu_RevModPhys.88.035005} for an accessible review on symmetry classes). This is only for convenience.
We focus on  the following five symmetry classes   --- 
D, BDI, DIII, AII, and CII and pick a particular representative symmetry for each class.
\footnote{The A and C symmetry classes have no SPT phases in 1+1d.
The AI and CI have also SPT phases but they are only bosonic phases 
(in fact, Haldane spin-1 chain) 
up to stacking gapped fermionic product states. For bosonic SPTs, we can unwind by the approach of 
\cite{WangWenWitten_SymmetricGapped_PRX2018, AP_Unwinding_PRB2018}. 
See the later \Table{table:web} for the systematic web of SPTs related by a symmetry group embedding.} In Table~\ref{tab:free fermion phases}, we list the group of invertible phases and the subgroup of SPT phases. Following Sec.~\ref{sec:classification}, we use the  convention: 
\begin{table}[!tb]
	\centering
	\begin{tabular}{@{}lll@{}}
		\toprule
		$
		\begin{array}{l}		
		\text{Symmetry}\\
		\text{Class}
		  \end{array}
		  $
		& $\invertibleG$ & $\SPT$ \\ \midrule
		D     & $\bZ_2$ & $0$ \\
		BDI   & $\bZ$   & $2 \bZ$ \\	
		DIII  &	$\bZ_2$ & $\bZ_2$ \\ 	
		AIII  & $\bZ$   & $\bZ$  \\
		CII   & $\bZ$   & $\bZ$ \\
		\bottomrule	
	\end{tabular}
	\quad
	\begin{tabular}{@{}llll@{}}
		\toprule
		$
		\begin{array}{l}		
		\text{Internal}\\
		\text{Symmetry}
		  \end{array}
		  $
		  &
		  $
		\begin{array}{l}		
		\text{Spacetime-Internal}\\
		\text{Symmetry $G$}
		  \end{array}
		  $		  
		  & $\TP_2(G)$ & $\frac{\TP_2(G)}{\TP_2(\Spin)}$ \\ \midrule
		  $\Z_2^F$     & Spin & $\bZ_2$ & $0$ \\
		  $\Z_2^T \times \Z_2^F$  & Pin$^-$ & $\bZ_8$   & $\bZ_4$ \\		
		  $\Z_4^{TF}$   & Pin$^+$ &	$\bZ_2$ & $\bZ_2$ \\
		{${\U(1)^{F}_{} \times \Z^{{T}}_{2}}$}  & Pin$^c$ & $\bZ_4$   & $\bZ_4$  \\
		$\SU(2)^F \times \Z_2^T$   &  Pin$^- \times_{\Z_2^F} \SU(2)$ & $\bZ_2$   & $\bZ_2$ \\
		\bottomrule	
	\end{tabular}
	\caption{ 
	{For 1+1d fermionic systems, on the left-hand side (l.h.s) table, we show 
	the free fermion classification of invertible phases, $\invertible_G$ and the subgroup of SPT phases, $\SPT$ of a given symmetry class.
	On the right-hand side (r.h.s) table,
	we show the corresponding symmetry group $G$ suitable for the interacting systems (interacting analogous of symmetry class),
	their interacting fermion classification of invertible topological phases (TP) denoted as a version of cobordism group 
	${\TP_2(G)}$ in Freed-Hopkins's \cite{FreedHopkins1604.06527},
	and  interacting fermion classification of SPT phases denoted as $\frac{\TP_2(G)}{\TP_2(\Spin)}$.
	The $\frac{\TP_2(G)}{\TP_2(\Spin)}$ means that the invertible topological phases (TP) of symmetry group $G$ mod out 
	the invertible fermionic topological order ${\TP_2(\Spin)}$ that is long-range entangled (LRE). 
	Thus the $\frac{\TP_2(G)}{\TP_2(\Spin)}$ is the short-range entangled (SRE) fermionic SPT phases.
	Note that {${\U(1)^{F}_{} \times \Z^{{T}}_{2}}$} is equivalent to $\frac{\U(1)^{F}_{} \times \Z^{{T}}_{4}}{\Z_2^F}$
	via a redefinition of the time reversal symmetry (see \Sec{sec:free AIII}).
	See also \Table{table:web} for a systematic web of these 1+1d fermionic SPTs.
	The cobordism group data is explained in more details in \Sec{sec:formal-classification}.
	}}
	\label{tab:free fermion phases}
\end{table}
{
\begin{enumerate}[leftmargin=2.mm, label=\textcolor{blue}{(\arabic*)}., ref={(\arabic*)}]
\item {For the free fermion symmetry classes and their group classification of phases,} 
we denote the group of invertible phases as $\invertibleG$ and the group of SPT phases, which forms a subgroup of $\invertibleG$, as $\SPT$.
\item
For the interacting fermion systems and their group classification of phases, 
we can obtain the group classification from a cobordism group 
$\TP_d(G)$ defined in \cite{FreedHopkins1604.06527} for invertible topological phases with a $G$-global symmetry.
Here $G$ is the full spacetime-internal global symmetry in the continuum limit, see examples of 
$G$ listed in Table~\ref{tab:free fermion phases}.
{The low energy physics of
 invertible topological phases are described by the invertible topological field theory (iTQFT) \cite{FreedHopkins1604.06527} 
 which is also known as SPT invariants \cite{1405.7689}.}
 Based on Wen's definition \cite{Wen2016ddy1610.03911}, 
 part of the invertible topological phases are long-ranged entangled invertible topological orders,
while another part are short-ranged entangled invertible SPTs.
For example, any layer of Kitaev chains is an invertible topological phase:
 a single Kitaev chain is a \emph{long-ranged entangled} invertible topological order, 
while two or more layers of Kitaev chains are \emph{short-ranged entangled} invertible SPTs.
\end{enumerate}
}

We can generate a representative of each 1+1d free-fermion invertible phase for every class by considering an appropriate number of layers of Kitaev's Majorana chains~\cite{Kitaev_majorana_2001} and identifying an appropriate symmetry belonging to the symmetry class in consideration. SPT phases on the other hand turn out to be those invertible phases which can be represented by an even number of Majorana chains.\footnote{{By Wen's definition \cite{WangWenWitten_SymmetricGapped_PRX2018},  
 a single layer of Kitaev chain is protected by no symmetry except by the fermion parity, this means that
in fact  
$$\hspace{-6mm}
\text{an odd number of Kitaev chains is a 1+1d \emph{long-range entangled state} as a 1+1d \emph{fermionic invertible topological order}.}$$
Based on the Wen's definition using the local unitary quantum circuit, 
Kitaev chain is long-range entangled (LRE) but not short-range entangled (SRE) because it cannot be deformed to a trivial product state by breaking all symmetries 
(but maintaining the fermion parity $\Z_2^F$ as a fermionic system). In contrast,
$$\hspace{-6mm}
\text{an even number of Kitaev chains is a 1+1d \emph{short-range entangled state} as a 1+1d \emph{fermionic invertible SPT state}.}$$
because it can be deformed to a trivial product state by breaking all symmetries (but maintaining the fermion parity $\Z_2^F$  as a fermionic system).
}
}
We relegate 
a {mathematical and formal explanation on the classifications of invertible phases vs SPT phases}
via the cobordism group data
in the end of section in \Sec{sec:formal-classification}.

\subsection{Toy model Hamiltonian of Majorana fermionic chains}
\label{sec:free D}

Let us start by writing down the Hamiltonian for the Kitaev's Majorana chain \cite{Kitaev_majorana_2001}. The Hilbert space is of one {complex} fermion per unit site on a 1d lattice which can be represented using two Majorana operators, $\gamma,~\bar{\gamma}$ satisfying $\gamma_i^\dagger = \gamma_i$, $\bar{\gamma}_i^\dagger =\bar{\gamma}_i$, $\{\gamma_i, \gamma_j\} = \{\bar{\gamma}_i, \bar{\gamma}_j\} = 2 \delta_{ij}$, and $\{\gamma_i, \bar{\gamma}_j\} = 0$.
{Majorana operators, which are Hermitian, can be defined in terms of creation and annihilation operators of the complex fermions, 
$\psi_i$ and $\psi_i^\dagger$ as \footnote{{Throughout the article, we denote the imaginary number as $\ii$ with $\ii^2=-1$.
The complex fermion operators obey $\{\psi_i, \psi_j^\dagger \}=\delta_{ij}$, 
$\{\psi_i, \psi_j \}=0$,
and $\{\psi_i^\dagger, \psi_j^\dagger \}=0$. We have:
$$
\psi_i=\frac{1}{2}({\gamma}_i+\ii \bar{\gamma}_i), \quad
\psi_i^\dagger=\frac{1}{2}({\gamma}_i-\ii \bar{\gamma}_i).
$$}}
\begin{equation} \label{eq:complex-Majorana}
{\gamma}_i =   ( \psi_i^\dagger + \psi_i  ), ~~~ \bar{\gamma}_i = {\ii} (\psi_i^\dagger - \psi_i  ).
\end{equation}
Kitaev's Majorana chain~\cite{Kitaev_majorana_2001} has the Hamiltonian:
}
\begin{equation}
\label{eq:Majorana chain}
H_{\text{Kitaev}}  = -\ii \sum_j \gamma_j \bar{\gamma}_{j+1}.
\end{equation}
\begin{figure}[!h] 
	\centering	
	\includegraphics[width=100mm]{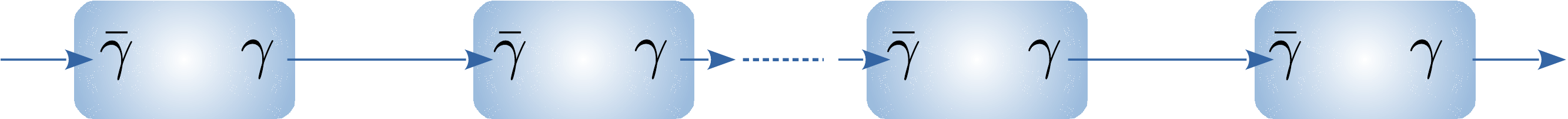}
	\caption{Graphical representation of Kitaev's 
	Majorana chain of \eq{eq:Majorana chain}.\label{fig:kitaev}}
\end{figure}

Let us consider two layers of Majorana chains {with their flavor indices $\varsigma=\uparrow,\downarrow$}:
\begin{equation}
\label{eq:2 layer Kitaev}
H^{\text{II}}_{\text{Kitaev}}  = -\ii \sum_{\varsigma = \uparrow, \downarrow} \sum_j \gamma_{\varsigma,j} \bar{\gamma}_{\varsigma, j+1}.
\end{equation}
\begin{figure}[!h] 
	\centering	
	\includegraphics[width=100mm]{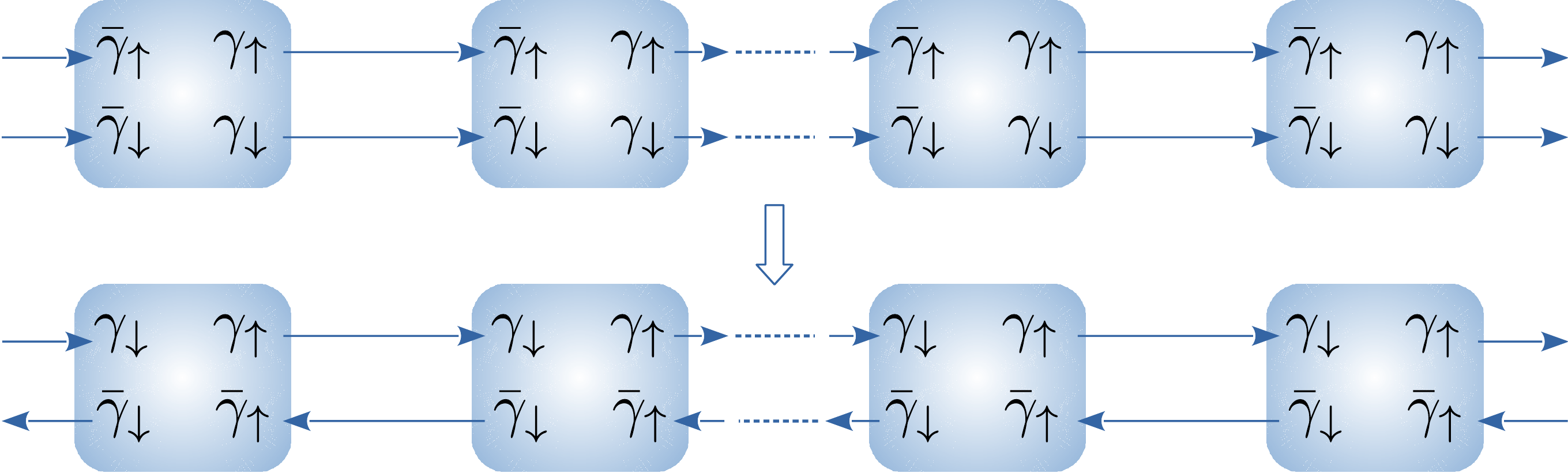}
	\caption{Fermionic SWAP operator
	and two layers of Kitaev's Majorana chains: 
	Two layers of Majorana chains before and after change of basis {by 
	the fermionic SWAP operator
	$\SWAP_f=
	M= \prod_{j} \frac{1 }{\sqrt{2}}\left( 1 + \gamma_{\downarrow,j} \bar{\gamma}_{\uparrow,j} \right) $ in \Eqn{eq:M}.
	The $M $ is the fermionic analog of the SWAP operator of bosonic spin systems (i.e.,
	the bosonic SWAP operator in \Fig{fig:AKLT_unwinding} and in \Refe{AP_Unwinding_PRB2018}'s Fig.~7).
	The Hamiltonian before  the basis changing is \Eq{eq:2 layer Kitaev}, 
	after the basis changing is 
	$H_{\text{gen}} =M H^{\text{II}}_{\text{Kitaev}}  M^\dagger  = -\ii \sum_{j} \left(\gamma_{\uparrow,j} \gamma_{\downarrow,j+1}- \bar{\gamma}_{\uparrow,j} \bar{\gamma}_{\downarrow,j+1}\right)$ 
	in \Eq{eq:2layerKitaev_basischanged}.
        }
	\label{fig:2layer_kitaev_basischanged}}
\end{figure}
For convenience, we now perform a change-of-basis on the Hamiltonian of eq~(\ref{eq:2 layer Kitaev}) using the unitary operator $M$,
\begin{equation} \label{eq:M}
{M =\prod_{j} \exp\left(\frac{\pi}{4}  \gamma_{\downarrow,j} \bar{\gamma}_{\uparrow,j} \right)}= \prod_{j} \frac{1 }{\sqrt{2}}\left( 1 + \gamma_{\downarrow,j} \bar{\gamma}_{\uparrow,j} \right). 
\end{equation}
{The new Hamiltonian $H_{\text{gen}}$ under the change of basis (see \Fig{fig:2layer_kitaev_basischanged}) 
becomes:\footnote{{In
terms of complex fermion bases \eq{eq:complex-Majorana}, we have
$H^{\text{II}}_{\text{Kitaev}}  = + \sum_{\varsigma = \uparrow, \downarrow} \sum_j 
( \psi_{\varsigma,j}^\dagger + \psi_{\varsigma,j}  )(\psi_{\varsigma,j+1}^\dagger - \psi_{\varsigma,j+1}  ),$
while
\\
$
H_{\text{gen}}  =
M H^{\text{II}}_{\text{Kitaev}}  M^\dagger = 
 -\ii \sum_{j} 
 \left(
( \psi_{\uparrow,j}^\dagger + \psi_{\uparrow,j}  )
( \psi_{\downarrow,j+1}^\dagger + \psi_{\downarrow,j+1}  )
 +
(\psi_{\uparrow,j}^\dagger - \psi_{\uparrow,j}  )
(\psi_{\downarrow,j+1}^\dagger - \psi_{\downarrow,j+1}  )
\right).$
}}
\bea
 H_{\text{gen}}  &=&M H^{\text{II}}_{\text{Kitaev}}  M^\dagger = 
 -\ii \sum_{j} \left(\gamma_{\uparrow,j} \gamma_{\downarrow,j+1}- \bar{\gamma}_{\uparrow,j} \bar{\gamma}_{\downarrow,j+1}\right). 
 \label{eq:2layerKitaev_basischanged}
\eea
}
        {The $M$  in \Eqn{eq:M} and in \Fig{fig:2layer_kitaev_basischanged} is the fermionic analog of the SWAP operator of bosonic spin systems (such as
	the bosonic SWAP operator in \Refe{AP_Unwinding_PRB2018}'s Fig.~7).}\footnote{{The bosonic SWAP operator
	$\SWAP_b \equiv \frac{1}{2} \left({1} + \vec{\sigma}_{A}\cdot \vec{\sigma}_{B}\right)$ 
	exchanges the basis states $\ket{\uparrow},~\ket{\downarrow}$ 
	of two qubit Hilbert spaces, 
	$A$ and $B$, and satisfies $(\SWAP_b)^2 =1$.
	On the other hand, the fermionic SWAP operator 
	$$\SWAP_f \equiv\frac{1 }{\sqrt{2}}\left( 1 + \gamma_{A} \gamma_{B} \right)  =\exp(\frac{\pi}{4} \gamma_{A} \gamma_{B})$$ 
	acts on two Majorana operators  
	$(\gamma_A , \gamma_B)$ to give 
	$$\SWAP_f \cdot (\gamma_A , \gamma_B) \cdot \SWAP_f^{-1}=(-\gamma_B,  \gamma_A).$$}
	Note that
	$(\SWAP_f)^2=
	 \gamma_{A} \gamma_{B},$
	$(\SWAP_f)^4=-1,$ and $(\SWAP_f)^8=1.$
	}
The Hamiltonian  $H_{\text{gen}}$ of eq~(\ref{eq:2layerKitaev_basischanged}) 
is the proper form of generator of two layers of Kitaev chains,
which will be a primary object of focus of this section and throughout our work.

\subsection{Class BDI: $\ztwo^T \times \ztwo^F$ symmetry or Pin$^-$}
\label{sec:free BDI}

We start with class BDI with representative global symmetry $\ztwo^T \times \ztwo^F$ where $\ztwo^T$ is generated by the anti-unitary operator $\mathcal{T}$ satisfying $\mathcal{T}^2 = 1$ and $\ztwo^F$ is the fermion parity group generated by  $P_F=(-1)^F$ where $F$ is the total fermion number. 
In the absence of interactions, it is known that there are an infinite number of invertible phases~\cite{Kitaev_periodic_doi:10.1063/1.3149495} indexed by an integer $k \in \bZ$ representing the number of `dangling' Majorana modes on each end i.e $\invertibleG \cong \bZ$. This reduces to $\bZ_8$ in the presence of interactions~\cite{FidkowskiKitaev_PhysRevB.81.134509_2010,FidkowskiKitaev_PhysRevB.83.075103_2011}. A representative of the generator of these phases is the Majorana chain of eq~(\ref{eq:Majorana chain}) with symmetry operator 
\bea
\mathcal{T} = \mathcal{K}
\eea 
which is a complex conjugation which acts as  
\begin{equation}
\label{eq:complex conjugation Majorana}
\mathcal{K} \gamma \mathcal{K} = \gamma,~~\quad\quad \mathcal{K} \bar{\gamma} \mathcal{K} =- \bar{\gamma},~~\quad\quad\mathcal{K} \ii \mathcal{K} = -\ii,
\end{equation}
and 
\begin{equation}
P_F = (-1)^{F}= \prod_j \ii \bar{\gamma}_j \gamma_j. 
\end{equation}
The subgroup of SPT phases are those corresponding to even members $k \in 2 \bZ$ and are non-trivial only in the presence of symmetries $\ztwo^T \times \ztwo^F$. 
{The SPT phases also correspond to also the even members $k=0,2,4,6 \in \Z_8$ in the interacting SPT classification (See
Table \ref{tab:free fermion phases}).}
A generator of the SPT phases (corresponding to $k=2$) is two layers of the Majorana chains, eq~(\ref{eq:2 layer Kitaev}) or its equivalent representation in eq~(\ref{eq:2layerKitaev_basischanged}) 
{with the new time-reversal $\ztwo^T$ symmetry operators $\tilde{\mathcal{T}}$ under the \eq{eq:M} basis-change via a unitary $M$: 
\bea \label{eq:T}
 \tilde{\mathcal{T}}  &\equiv&  M \mathcal{T} M^\dagger =  M \mathcal{K} M^\dagger = 
 \left(\prod_{j} 
 (\gamma_{\downarrow,j} \bar{\gamma}_{\uparrow,j}) \right) \mathcal{K}.
\eea
The form of fermion parity $P_F$ stays unchanged under basis change $M {P_F}=  {P_F} M$.}

\subsubsection{Gapping the boundary:  Extension to a non-Abelian $\bD_8^{F,T}$} 
\label{subsec:gap-bdry-BDI}

\begin{figure}[!htbp]
	\centering	
	\includegraphics[width=100mm]{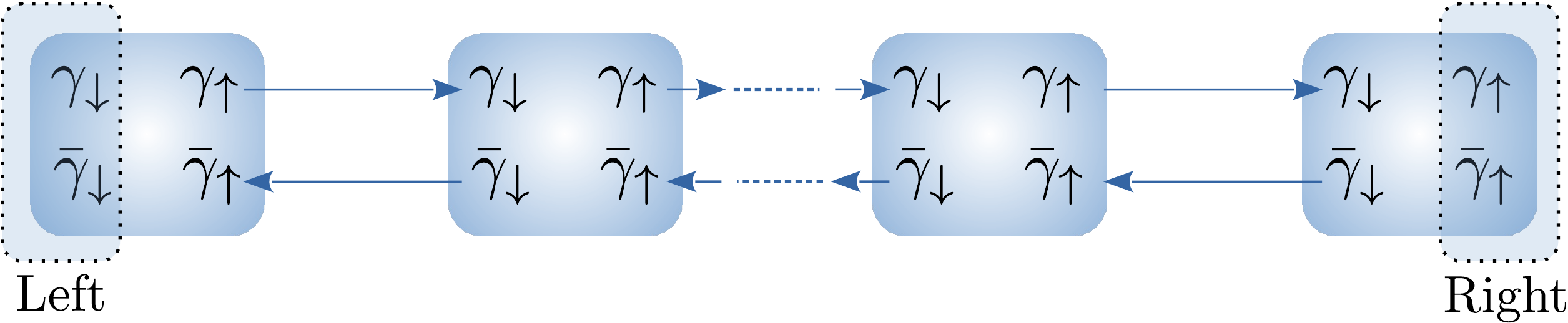}
	\caption{
	{Basis-changed 2-layer Majorana chain with left ($L$) and right ($R$) open boundaries, 
	with a symmetry as in \eq{eq:T}, we have four fold ground state degeneracy (GSD = 4) 
	with two degenerate Majorana modes on each end, emphasized in the box frames.}
	\label{fig:bdry-sym}}
\end{figure}

Let us see how we can gap the boundary when we have open boundary conditions for Hamiltonian~(\ref{eq:2layerKitaev_basischanged}). {\Fig{fig:bdry-sym} shows that the boundary 
  has four fold ground state degeneracy (GSD = 4) coming from the two dimensional Hilbert space from 
  two Majorana modes (i.e., one complex fermion) on each boundary site, $L$ and $R$}

\begin{figure}[!htbp]
	\centering	
	\includegraphics[width=100mm]{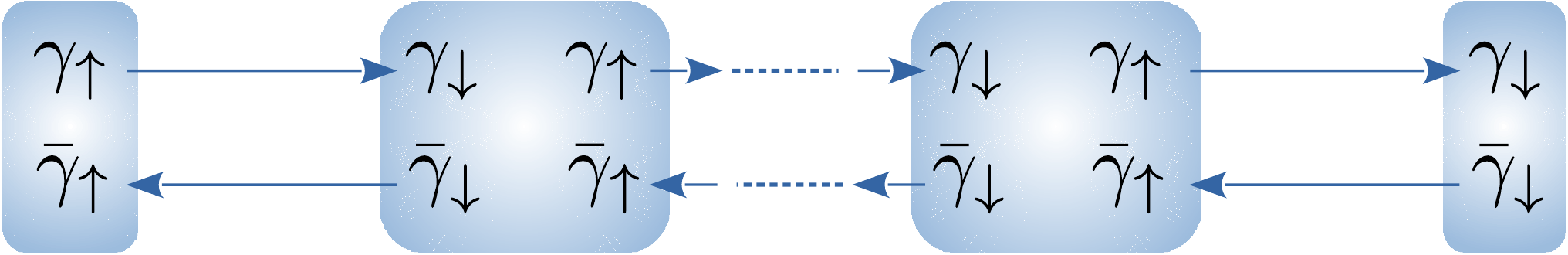}
	\caption{{Basis-changed 2-layer Majorana chain from \Fig{fig:bdry-sym} with two open boundaries,
	but now each of the boundary couples to additional two Majorana modes on both the left $L$ and right $R$ 
	(the $j=1$ site couples to a new site $j=0$,
	  $j={\ell}$ site  couples to a new site $j={\ell}+1$ respectively). 
	We can modify $H_{\text{gen}}$ to $H_{\text{gen}}+H_{\text{gen,bdry}}$ by adding local Hamiltonian terms to 
	 two boundaries as $H_{\text{gen,bdry}}=
	 -\ii  \left(\gamma_{\uparrow,0} \gamma_{\downarrow,1}- \bar{\gamma}_{\uparrow,0} \bar{\gamma}_{\downarrow,1}\right)
	 -\ii  \left(\gamma_{\uparrow,{\ell}} \gamma_{\downarrow,{\ell}+1}- \bar{\gamma}_{\uparrow,{\ell}} \bar{\gamma}_{\downarrow,{\ell}+1}\right).$
	 The whole system becomes gapped but with only a unique ground state (GSD = 1).
	 \label{fig:2layer_kitaev_boundary_basischanged}}}
\end{figure}
{\Fig{fig:2layer_kitaev_boundary_basischanged} shows another boundary termination of the Hamiltonian of eq.~\ref{eq:2layerKitaev_basischanged} 
{by adding two extra Majorana modes on each end ($L$ and $R$)}
leading to a unique ground state.}
From \Fig{fig:2layer_kitaev_boundary_basischanged}, we can read-off the effective \emph{extended} 
boundary symmetry operators {on the left ($L$) and right ($R$)} boundaries 
{(on the site $j=0$ and $j={\ell}+1$) omitting site indices below}:
\begin{eqnarray}
\mathcal{T}_L &=& \bar{\gamma}_{\uparrow} \mathcal{K},\quad~P_{F,L} = \ii \bar{\gamma}_{\uparrow} \gamma_{\uparrow},\quad~\mathcal{T}_L^2 = -1, P_{F,L}^2 = 1, 
\quad\mathcal{T}_L P_{F,L} = -P_{F,L}  \mathcal{T}_L. \label{eq:L-sym}\\
\mathcal{T}_R &=& \gamma_{\downarrow} \mathcal{K},\quad~P_{F,R} = \ii \bar{\gamma}_{\downarrow} \gamma_{\downarrow}, \quad~\mathcal{T}_R^2 = P_{F,R}^2 = 1, 
\quad\quad\quad \mathcal{T}_L P_{F,R} = -P_{F,R}  \mathcal{T}_R. \label{eq:R-sym}
\end{eqnarray}
We can also show:
\bea
&&\mathcal{T}_L  {\gamma}_{\uparrow}  \mathcal{T}_L ^{-1} 
= -{\gamma}_{\uparrow},
\quad
\mathcal{T}_L  \bar{\gamma}_{\uparrow}  \mathcal{T}_L ^{-1}
=-\bar{\gamma}_{\uparrow}. \\
&&\mathcal{T}_R  {\gamma}_{\downarrow}  \mathcal{T}_R ^{-1} 
=\gamma_{\downarrow},
\quad\;\;
\mathcal{T}_R  \bar{\gamma}_{\downarrow}  \mathcal{T}_R ^{-1} 
= \bar{\gamma}_{\downarrow}
\eea
It can be easily checked that the boundary symmetry operators generate a faithful representation of 
{the dihedral group of order 8 known as} $\bD_8$ which is a group with two generators satisfying the following relation: 
\begin{eqnarray}\label{eq:D8}
\bD_8\equiv \Z_4 \rtimes \Z_2\equiv\innerproduct{a,x}{a^4 = x^2 = 1, xax = a^{-1}}. 
\end{eqnarray}
On the  left ($L$) end, $a = \mathcal{T}_L,~x = P_{F,L}$, giving us
$$
\bD_8^{F,T} \vert_{\text{boundary}} = \Z_4^T \rtimes \Z_2^F \equiv\innerproduct{ \mathcal{T}_L, P_{F,L}}{\mathcal{T}_L^4 = P_{F,L}^2 = 1, 
P_{F,L} \mathcal{T}_L P_{F,L} = \mathcal{T}_L^{-1}}. 
$$
On the right ($R$) end, $a = \mathcal{T}_R P_{F,R},~x = P_{F,R}$, giving us
$$
\bD_8^{F,T}\vert_{\text{boundary}} 
= \Z_4^T \rtimes \Z_2^F
\equiv\innerproduct{ (\mathcal{T}_R P_{F,R}), P_{F,R}}{( \mathcal{T}_R P_{F,R})^4 = (P_{F,R})^2 = 1, P_{F,R} ( \mathcal{T}_R P_{F,R}) P_{F,R} = ( \mathcal{T}_R P_{F,R})^{-1}}. 
$$
Importantly, we observe that the fermion parity ($P_{F,L}$ or $P_{F,R}$) is \emph{not} in the center of 
$\bD_8^{F,T} = \Z_4^T \rtimes \Z_2^F$ with $\Z_4^T$ generated by $a$ and $\Z_2^F$ generated by $x$.
This is the extended symmetry that will help us to unwind 
the bulk.\footnote{{Some comments about our conventions and notations:
\begin{enumerate}
\item  Here the $\mathbb{D}_8^{F,T}$ 
in our conventions is the left ($L$) or right ($R$) boundary versions of $\bD_8$ in \Eq{eq:D8}.
The upper indices (${F,T}$) emphasize that fermion parity and time reversal are included, but they do \emph{not} commute in the extended groups.
\item  In general, when we write a group
$G^{F,T}$, the fermion parity ($P_F$) and time-reversal ($T$) do not necessarily commute in $G$.
\item  
$G^{TF}$ means that $G$ contains time-reversal ($T$) which commutes with and generates fermion parity  $\cT^2 =(-1)^F$ .
\item  
$G^{TB}$ means $G$ contains the time-reversal ($T$) which satisfies 
$\cT^2 = (-\mathds{1})$ .
\end{enumerate}
}
\label{ft:D8Q8}} 

In particular, given the bulk onsite symmetry $\ztwo^T \times \ztwo^F$ (where time reversal $\ztwo^T$ and fermion parity $\ztwo^F$ commutes),
we can extend the symmetry transformation on the boundary to $\mathbb{D}_8^{F,T}$ written as a short exact sequence: 
\bea
1  \longrightarrow \ztwo   \overset{ }{\longrightarrow} \mathbb{D}_8^{F,T}\vert_{\text{boundary}} \overset{ }{\longrightarrow}  \ztwo^T \times \ztwo^F  \vert_{\text{bulk}} \longrightarrow 1. 
\eea
See footnote \ref{ft:D8Q8} for our conventions.
The normal subgroup $N=\ztwo=\{1,-1\}$ is at the center of $\bD_8$ on both ends (left and right), while the quotient group
$\ztwo^T \times \ztwo^F$ contains four commutative elements: $N\{1, a, x, ax\}$ for $\mathbb{D}_8^{F,T}$.

Some more comments:
\begin{enumerate}[leftmargin=0.mm, label=\textcolor{blue}{\arabic*}., ref={\arabic*}]
\item We can 
explicitly check that the fully gapped system in \Fig{fig:2layer_kitaev_boundary_basischanged} preserves the global symmetry.
Not only the bulk Hamiltonian term $H_{\text{gen}}$ 
respects the bulk $\ztwo^T \times \ztwo^F$ symmetry, but 
the boundary Hamiltonian 
$H_{\text{gen,bdry}}=
H_{\text{gen,bdry},L}
+
H_{\text{gen,bdry},R}
=
-\ii  \left(\gamma_{\uparrow,0} \gamma_{\downarrow,1}- \bar{\gamma}_{\uparrow,0} \bar{\gamma}_{\downarrow,1}\right)$
	 $-\ii  \left(\gamma_{\uparrow,\ell} \gamma_{\downarrow,{\ell}+1}- \bar{\gamma}_{\uparrow,{\ell}} \bar{\gamma}_{\downarrow,\ell+1}\right)$
also respects the boundary $\mathbb{D}_8^{F,T}$ symmetry. More precisely,
for the time reversal symmetry on the boundary Hamiltonian:

On the left end $L$,
we have the symmetry operator from $\mathcal{T}_L$ on the site 0 and
$\tilde{\mathcal{T}}$ on the site 1, which together act on the left end 
$H_{\text{gen,bdry},L}$.

On the right end $R$,
we have the  symmetry operator from $\mathcal{T}_R$ on the site $\ell+1$ and
$\tilde{\mathcal{T}}$ on the site $\ell$,
which together act on the 
right end $H_{\text{gen,bdry},R}$. 

The system of \Fig{fig:2layer_kitaev_boundary_basischanged}
is fully gapped with a unique ground state (GSD = 1),
preserving 
the bulk $\ztwo^T \times \ztwo^F$ symmetry
and the boundary $\mathbb{D}_8^{F,T}$ symmetry.

\item \label{BDI-remark2}
We should emphasize that 
there are at least three different ways to look at $\mathcal{T}$ symmetries on the boundary depending on (1) the boundary truncation types of quantum systems and (2) the Hilbert space projection $\PGS$ to ground state subspace
(Below we omit the boundary site $j$ labels, $j=1$ for the left $L$ and $j=\ell$ for the right  $R$, throughout):
\begin{enumerate}
\item On boundaries of \Fig{fig:bdry-sym}, if we look at
    $\mathcal{T}_L$ and $\mathcal{T}_R$ acting on whole boundary sites 
    of $j=1$ and $j=\ell$ respectively 
    without $\PGS$ projection, then we have:
    $$\mathcal{T}_L = \gamma_{\downarrow}\bar{\gamma}_{\uparrow} \mathcal{K},\quad \mathcal{T}_R = \gamma_{\downarrow}\bar{\gamma}_{\uparrow} \mathcal{K}.$$
\item On boundaries of 
    \Fig{fig:2layer_kitaev_boundary_basischanged},
    if we look at
    $\mathcal{T}_L$ and $\mathcal{T}_R$ acting on extra boundary sites 
    of $j=0$ and $j=\ell+1$ respectively (with or without $\PGS$ projection), then we have:
    $$\mathcal{T}_L = \bar{\gamma}_{\uparrow} \mathcal{K},\quad \mathcal{T}_R = \gamma_{\downarrow}\mathcal{K}.$$
    \item On boundaries of \Fig{fig:bdry-sym},
    if we look at
    $\mathcal{T}_L$ and $\mathcal{T}_R$ acting on whole boundary sites 
    of $j=1$ and $j=\ell$ respectively 
    with $\PGS$ projection to dangling zero energy modes, then we have:
    $$\mathcal{T}_L = \gamma_{\downarrow} \mathcal{K},\quad \mathcal{T}_R = \bar{\gamma}_{\uparrow} \mathcal{K}.$$

\end{enumerate} 
\item  \label{BDI-remark3}
There is a special \emph{supersymmetric} feature for a 2 layer of Kitaev chains which holds when the layer number $N= 2 \mod 4$ only. 
\begin{itemize}           
\item  When the layer number $N= 2 \mod 4$, 
          we have two Majorana zero modes (with two operators $\gamma, \bar{\gamma}$) 
          with GSD = 2 on each end of
           system (with the left and right ends combined to give GSD = 4).
           In fact, for this GSD = 2, on one end, 
           the Hilbert space $\cH$ 
           must have one bosonic ground state $| B \rangle$
           and one fermionic ground state $| F \rangle$.
           We denote the GSD = 2 ground state subspace 
           as $\cH =\{ | B \rangle, | F \rangle\}  = \cH_B \oplus \cH_F$, say $| B \rangle = \begin{pmatrix} 1\\ 0 \end{pmatrix}$
           and $| F \rangle = \begin{pmatrix} 0 \\ 1 \end{pmatrix}$.
           Consider the $L$ boundary in \Fig{fig:bdry-sym}, we can write down the operators acting on the two-dimensional Hilbert space $\cH$ explicitly:
           \bea
           &&\gamma = \begin{pmatrix}  0 & 1\\ 1 & 0  \end{pmatrix} = \sigma_x , \quad
            \bar{\gamma} = \begin{pmatrix}  0 & -\ii\\ \ii & 0  \end{pmatrix} = \sigma_y , \quad 
            P_F= -\ii \gamma \bar{\gamma}  = \begin{pmatrix}  1 & 0\\ 0 & -1  \end{pmatrix} = \sigma_z. \cr
           &&\psi=\frac{1}{2}({\gamma}+\ii \bar{\gamma}) = \begin{pmatrix}  0 & 1\\ 0 & 0  \end{pmatrix} , \quad
           \psi^\dagger=\frac{1}{2}({\gamma}-\ii \bar{\gamma}) = \begin{pmatrix}  0 & 0\\ 1 & 0  \end{pmatrix}, \quad
           \mathcal{T} = \bar{\gamma} \mathcal{K} = \sigma_y \mathcal{K}.
           \eea
           The bosonic $| B \rangle$ and fermionic $| F \rangle$ have the distinct fermion parity $P_F$ (even vs odd).
           
           The GSD = 2 on each end is protected by time-reversal $\mathcal{T}$ symmetry: because to gap       
           any ground state, the only possible Hamiltonian is  
           \bea
           H'\propto - \ii \gamma \bar{\gamma}  = P_F
           \eea 
           which is disallowed by a global symmetry $\mathcal{T}$ due to the fact
           $H'$ does not preserve $\mathcal{T}$.    
           (i.e., $\mathcal{T}$ does not commute with $P_F$, 
           namely $\mathcal{T} P_F \mathcal{T}^{-1} = - P_F$ or   $P_F \mathcal{T} P_F =  -\mathcal{T}$.)
      
           We will revisit  the underlying supersymmetric quantum mechanics in \Sec{sec:SUSYQM} in depth.
           
\item   When the layer number $N = 4$ (or a nonzero $N= 0 \mod 4$), 
           we  have four Majorana zero modes (with four operators $\gamma_1, \bar{\gamma}_1, \gamma_2, \bar{\gamma}_2$)
           with
           GSD = 4 on one end (with the left and right ends combined to give GSD = 16).
           However, we can introduce a four Majorana interaction with a $\mathcal{T}$-symmetric Hamiltonian 
           \bea
           H' = \alpha \gamma_1 \bar{\gamma}_1 \gamma_2 \bar{\gamma}_2 =  - \alpha (P_{F,1})(P_{F,2}) =   - \alpha  P_F
           \eea
           where $P_{F,i} = - \ii \gamma_i \bar{\gamma}_i$ is the fermion parity for the $i$-th complex fermion $\psi_i$ sector.
           This $H'$ is  $\mathcal{T}$-symmetric because  $\mathcal{T} (P_{F,1}P_{F,2}) \mathcal{T}^{-1}  = (-P_{F,1})(-P_{F,2})=(P_{F,1}P_{F,2})$.
            Overall, we can gap two ground states out of GSD = 4 to leave with only GSD = 2 on one end. 
            By turning on $H' = \alpha \gamma_1 \bar{\gamma}_1 \gamma_2 \bar{\gamma}_2$, depend on the sign of $\alpha$,
            we are left with only either \emph{two bosonic states} ($\alpha > 0$, so ground states with $P_F =+ 1$) or \emph{two fermionic states} 
            ($\alpha < 0$, so ground states with $P_F = -1$). 
            
\end{itemize}

            In summary, the $N= 0 \mod 4$ may be left with two ground states but always with the same fermion parity. 
            In contrast, only $N= 2 \mod 4$, we have two ground states with the different fermion parity, hinting the supersymmetric quantum mechanics to be revealed in \Sec{sec:SUSYQM}.

\end{enumerate}
 \color{black}

\subsubsection{Unwinding the bulk:  Extension to a non-Abelian $\bD_8^{F,T}$} 

In order to unwind {the bulk}, we follow the similar steps as in 
\Sec{sec:Haldane-symmetry-extension}, 
see Fig~\ref{fig:unwinding_BDI}: 
\begin{enumerate}[leftmargin=2.mm, label=\textcolor{blue}{\arabic*}., ref={\arabic*}]
\item The first step, 
we extend the Hilbert space by adding one {complex} fermion per site 
({which corresponds} to Majorana operators $\gamma$ and $\bar{\gamma}$) with dynamics given by a dimerizing Hamiltonian: 
\begin{equation}
H_o = \ii \sum_{\text{odd} ~j} (\gamma_j \gamma_{j+1} - \bar{\gamma}_j \bar{\gamma}_{j+1}). 
\end{equation}
The extended Hamiltonian and symmetry operators are (with $H_{\text{gen}}$ in \Eq{eq:2layerKitaev_basischanged})
\begin{eqnarray}
H_{\text{ext} } &=& H_{\text{gen}}  + H_o.  \label{eq:Hext}\\
\cT_{\text{ext} } &=& \Big(\prod_{\text{odd} ~j} (\gamma_{\downarrow,j} \bar{\gamma}_{\uparrow,j}) \gamma_j\Big) 
\Big(\prod_{\text{even} ~j} (\gamma_{\downarrow,j} \bar{\gamma}_{\uparrow,j}) \bar{\gamma}_j \Big) \cK. \label{eq:Text-D8}\\
P_{F}&=& \prod_j 
\left(\ii \bar{\gamma}_{\downarrow,j} \gamma_{\downarrow,j}\right) 
\left(\ii \bar{\gamma}_{\uparrow,j} \gamma_{\uparrow,j}\right) 
\left(\ii \bar{\gamma}_{j} \gamma_{j}\right).
\end{eqnarray}
{In the extended 1+1d bulk, we have the extended $\bD_8^{F,T}$ symmetry on-site in the bulk:}  
\bea
\bD_8^{F,T} \vert_{\text{bulk}}  
\equiv\innerproduct{ \cT_{\text{ext} }, P_{F}}{\cT_{\text{ext} }^4 = P_{F}^2 = 1, 
P_{F} \mathcal{T}_{\text{ext} } P_{F} = \mathcal{T}_{\text{ext} }^{-1} {=- \mathcal{T}_{\text{ext} }}}.
\eea
\begin{figure}[!htbp] 
	\centering	
	\includegraphics[width=100mm]{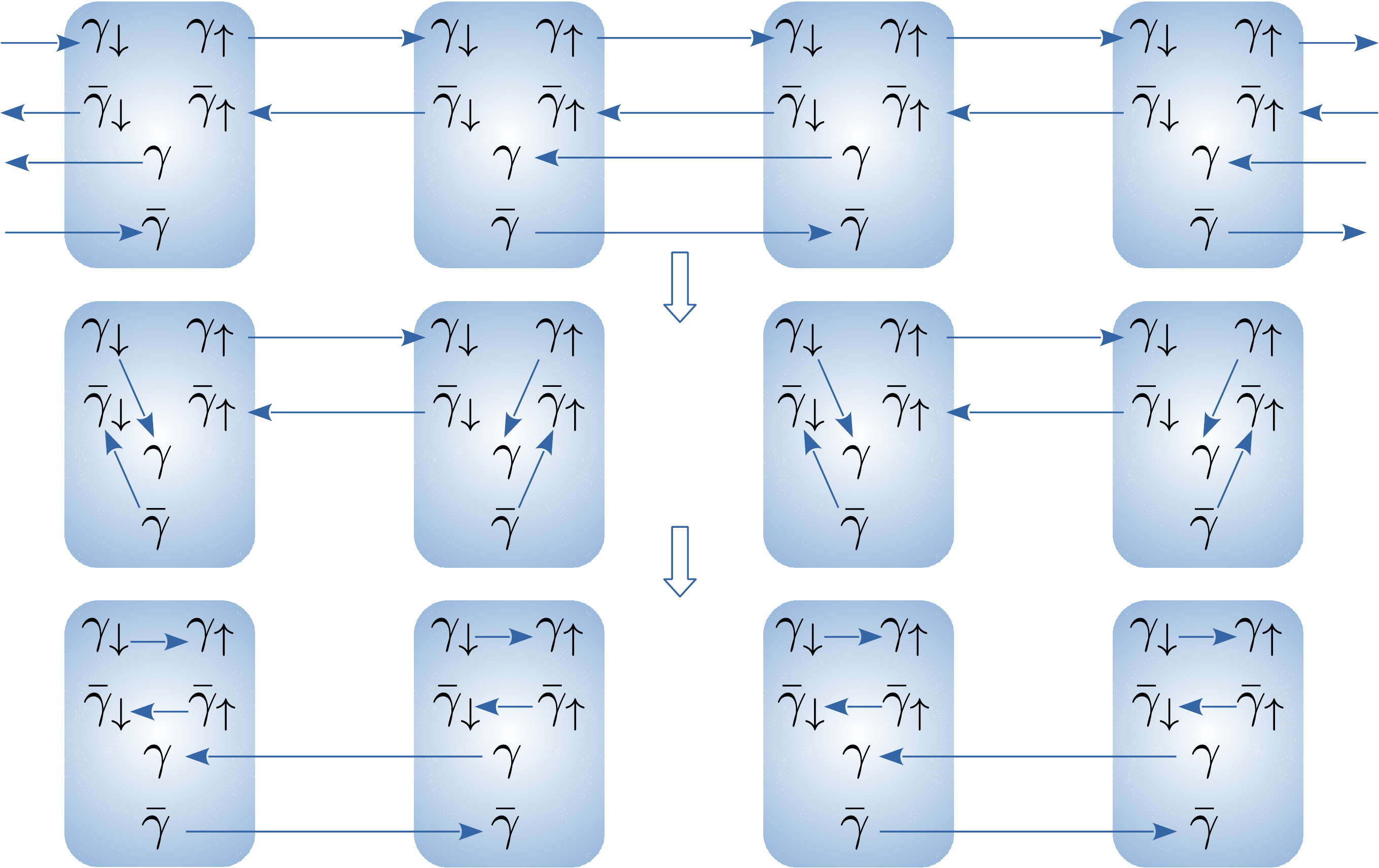}
	\caption{Unwinding the Hamiltonian 
	of the
	basis-changed 2-layer Majorana chain
	(\eq{eq:Hext} for BDI class) by three steps.
	 {We show four sites (as even, odd, even, and odd sites) from the left to the right.
	Step 1: we add one extra complex fermion per site to extend the Hilbert space. 
	Step 2: we perform a local unitary transformation $W_1$ on each of two neighbor sites (a neighbor pair of odd and even sites).
	Step 3: we perform a local unitary transformation $W_2$ on each of sites, while the odd and even sites have different transformations.
	The final form of Hamiltonian becomes \eq{eq:Unwinding_BDI3}, 
	while the ground state becomes a tensor product state, which is 
	a trivial tensor product state respect to effective sites (with each effective site combining two neighbor [an even and an odd] sites).}
	See \ft{ft:LSM}: We have the final ground state  
	breaking the translation symmetry with a double unit cell due to the LSM constraint.
	\label{fig:unwinding_BDI}}
\end{figure}
\item The second and third steps, we show
the Hamiltonian can be unwound as shown in Fig~\ref{fig:unwinding_BDI} using a 2-layer $\bD_8$-symmetry invariant FDUC $W = W_2 W_1$ as
\begin{equation}
\label{eq:Unwinding_BDI3}
W H_{\text{ext} } W^\dagger =
(W_2 W_1) H_{\text{ext} } (W_2 W_1)^\dagger = -\ii \sum_{j} \left(\gamma_{\downarrow,j} \gamma_{\uparrow,j}- \bar{\gamma}_{\downarrow,j} \bar{\gamma}_{\uparrow,j}\right) + H_e,
\end{equation}
{such that $\mathcal{T}_{\text{ext} }W_{l}\mathcal{T}_{\text{ext} }^{-1} =W_{l}$ 
and  $P_{F} W_{l}P_{F}^{-1} =W_{l}$ with $l=1,2$,} 
where,
\begin{eqnarray} \label{eq:unwinding_BDI1}
W_1 &=& \prod_{\text{odd} ~j} \left(\frac{1+\gamma_j \gamma_{\downarrow,j+1}}{\sqrt{2}}\right) \left(\frac{1+\bar{\gamma}_j \bar{\gamma}_{\downarrow,j+1}}{\sqrt{2}}\right). \\
W_2 &=&  \prod_{\text{odd} ~j} \left(\frac{1+\gamma_j \gamma_{\downarrow,j}}{\sqrt{2}}\right) \left(\frac{1+\bar{\gamma}_j \bar{\gamma}_{\downarrow,j}}{\sqrt{2}}\right) \prod_{\text{even} ~j} \left(\frac{1+ \gamma_{\uparrow,j} \gamma_j }{\sqrt{2}}\right) \left(\frac{1+ \bar{\gamma}_{\uparrow,j} \bar{\gamma}_j }{\sqrt{2}}\right). 
 \label{eq:Unwinding_BDI2} \\
H_e &=& \ii \sum_{\text{even} ~j} (\gamma_j \gamma_{j+1} - \bar{\gamma}_j \bar{\gamma}_{j+1}).
 \label{eq:unwinding_He}
\end{eqnarray}
\end{enumerate}
The deformed Hamiltonian \eq{eq:Unwinding_BDI3} under FDUC is exactly what we aim for,
which shows that under the \emph{enlarged} Hilbert space and \emph{extended} global symmetry, 
its ground state can be deformed to a trivial tensor product state (See the outcome state in \Fig{fig:unwinding_BDI})!
In short, the bulk unwinding follows also the symmetry extension:
\bea
1  \longrightarrow \ztwo   \overset{ }{\longrightarrow} \mathbb{D}_8^{F,T}\vert_{\text{bulk}} \overset{ }{\longrightarrow}  \ztwo^T \times \ztwo^F  \vert_{\text{bulk}} \longrightarrow 1. 
\eea
\subsection{Class DIII: $\bZ_4^{TF}$ symmetry or Pin$^+$}
\label{sec:free DIII}
We now consider 1+1d superconductors with a \emph{conventional} $\bZ_4^{TF}$ time-reversal invariance, {which is
known as the Class DIII for the free fermion systems}.
We work with the same Hilbert space as the previous subsection i.e. two species of fermions per unit site. 
{The 2-layer Majorana chain with a Hamiltonian 
$H^{\text{II}}_{\text{Kitaev}}$ \Eq{eq:2 layer Kitaev} is in fact the generator of $\nu =1 \in \Z_2$ class of
the interacting SPTs with $\bZ_4^{TF}$-symmetry (see \Table{tab:free fermion phases}).
The $\cT$ transformation generating  $\bZ_4^{TF}$ obey
\bea
\cT &=& \prod_j \left(\frac{1+\gamma_{\uparrow,j}\gamma_{\downarrow,j}}{\sqrt{2}}\right) \left(\frac{1+\bar{\gamma}_{\uparrow,j}\bar{\gamma}_{\downarrow,j}}{\sqrt{2}}\right) \cK
=\prod_j 
\e^{\frac{\pi}{4} \gamma_{\uparrow,j}\gamma_{\downarrow,j}}
\e^{\frac{\pi}{4} \bar\gamma_{\uparrow,j}\bar\gamma_{\downarrow,j}}
\cK.\\
\cT^2 &=& P_F=(-1)^{F} {=   
\prod_j \prod_{\varsigma = \uparrow, \downarrow}   \ii \bar{\gamma}_{\varsigma, j} \gamma_{\varsigma, j}}.
 \quad\quad\quad \cT^4=1.
\eea
This means $\Z_2^F \subset  \Z_4^{TF}$, the fermion parity generated by $P_F$ is a normal subgroup of time reversal $\Z_4^{TF}$, thus
these \emph{bulk} global symmetries form a short exact sequence:
\bea \label{eq:Z4TF}
1\to \Z_2^F \to \Z_4^{TF} \to \Z_2^T \to 1.
\eea}
It can easily be checked that the Hamiltonian of eq~(\ref{eq:2 layer Kitaev}) is invariant under the time-reversal symmetry defined above. Once again, we use the basis change 
via \eq{eq:M}'s $M$ to get the Hamiltonian $H_{\text{gen}}$
of eq~(\ref{eq:2layerKitaev_basischanged}) and the following basis-changed time-reversal operator:
\bea
{ \tilde{\mathcal{T}} } &\equiv&  M \cT M^\dagger 
=\prod_{j} \gamma_{\downarrow,j} 
\left(\frac{1- \gamma_{\downarrow,j}\bar{\gamma}_{\downarrow,j}}{\sqrt{2}}\right) 
\bar{\gamma}_{\uparrow,j} 
\left(\frac{1+\gamma_{\uparrow,j}\bar{\gamma}_{\uparrow,j}}{\sqrt{2}}\right)  \cK.
\eea
{The fermion parity $\tilde{P_F}  = M {P_F} M^\dagger=P_F$ stays unchanged.}

\subsubsection{Gapping the boundary:  Extension to a non-Abelian $\bM^{F,T}_{16}$}

To find the symmetry extension, let us now look at the symmetry transformation on the edges. We have
\begin{eqnarray}
\cT_L &=& \bar{\gamma}_{\uparrow} \left(\frac{1+\gamma_{\uparrow}\bar{\gamma}_{\uparrow}}{\sqrt{2}}\right)   \cK
 {=  \bar{\gamma}_{\uparrow}\e^{\frac{\pi}{4} \gamma_{\uparrow}\bar\gamma_{\uparrow}} \cK},
~~~~~~~~ \,
P_{F,L} = \ii \bar{\gamma}_{\uparrow} \gamma_{\uparrow}.\\
\cT_{R} &=& \gamma_{\downarrow} \left(\frac{1-\gamma_{\downarrow}\bar{\gamma}_{\downarrow}}{\sqrt{2}}\right) \cK
{= \gamma_{\downarrow}\e^{-\frac{\pi}{4} \gamma_{\downarrow}\bar\gamma_{\downarrow}} \cK},
~~~~~~~ \,
P_{F,R} = \ii \bar{\gamma}_{\downarrow} \gamma_{\downarrow} .
\end{eqnarray}
{We can also show:
\bea
&&\mathcal{T}_L  {\gamma}_{\uparrow}  \mathcal{T}_L ^{-1} 
=-\bar\gamma_{\uparrow}
 ,
\quad
\mathcal{T}_L  \bar{\gamma}_{\uparrow}  \mathcal{T}_L ^{-1}
={\gamma}_{\uparrow}. \\
&&\mathcal{T}_R  {\gamma}_{\downarrow}  \mathcal{T}_R ^{-1} 
=-\bar\gamma_{\downarrow}
,
\quad\;\;
\mathcal{T}_R  \bar{\gamma}_{\downarrow}  \mathcal{T}_R ^{-1} 
={\gamma}_{\downarrow}.
\eea}
The 0+1d edge symmetry operators form a finite non-Abelian group $\bM_{16}$~\cite{M16} of order 16:\footnote{{Note that $\cK^2=1$, 
we can show
$\cT_{L}^2=\gamma_{\uparrow}\bar{\gamma}_{\uparrow} = - \ii P_{F,L}$,
\; 
$\cT_{R}^2=\gamma_{\downarrow }\bar{\gamma}_{\downarrow} = - \ii P_{F,R}$,
\; $\cT_{L}^4=\cT_{R}^4=-1$,
and $\cT_{L}^8=\cT_{R}^8=1$,
while 
$P_{F,L/R} \cT_{L/R} P_{F,L/R} =- \cT_{L/R}= \cT_{L/R}^5$}
}
{\begin{equation}
\bM^{F,T}_{16}  \vert_{\text{boundary}}  \equiv \langle {\cT_{L/R}}, {P_{F,L/R}} \mid \cT_{L/R}^4 =-1, \;
\cT_{L/R}^8 = {P^2_{F,L/R}} = 1, \;
P_{F,L/R} \cT_{L/R} P_{F,L/R} = \cT_{L/R}^5 \rangle .
\end{equation}
which again has the fermion parity ${P_{F,L/R}}$  not at its center $Z(\bM^{F,T}_{16})\cong \Z_4$
$=\{1,$ ${\cT^2_{L/R}},$ ${\cT^4_{L/R}},$ ${\cT^6_{L/R}}\}$.}
{There is a short exact sequence for $\bM^{}_{16}$ as a central extension
\bea  \label{eq:M16-extension-0}
1  \longrightarrow \Big(Z(\bM^{}_{16})\cong \Z_4 \Big) \longrightarrow
(\bM^{}_{16}) \longrightarrow (\Z_2)^2\longrightarrow 1,
\eea
where the quotient group can be regarded as 
$(\Z_2)^2=Z(\bM^{}_{16}) \{1, \; {\cT_{L/R}}, \;{P_{F,L/R}},\; {\cT_{L/R}}{P_{F,L/R}}  \}$.
But the short exact sequence that relates the extended symmetry on the boundary $\bM_{16}^{F,T} \vert_{\text{boundary}}$ 
and the original symmetry group $\bZ_4^{TF} \vert_{\text{bulk}}$, is the following
\begin{equation} \label{eq:M16-extension}
1  \longrightarrow \bZ_4   \overset{ }{\longrightarrow} \bM_{16}^{F,T} \vert_{\text{boundary}} \overset{ }{\longrightarrow}  \bZ_4^{TF} \vert_{\text{bulk}}   \longrightarrow 1, 
\end{equation}
where the normal subgroup $N=\Z_4 
=\{1,$ ${\cT^2_{L/R}} {P_{F,L/R}},$ ${\cT^4_{L/R}},$ ${\cT^6_{L/R}}{P_{F,L/R}}\}$ is not the center $Z(\bM^{F,T}_{16})$.
Thus \eq{eq:M16-extension} is the non-central extension. 
The quotient group can be regarded as $\bZ_4^{TF}=$$N \{1, \; {\cT_{L/R}}, \;{\cT^2_{L/R}},\; {\cT^3_{L/R}} \}$.}

In fact the boundary fractionalized symmetry with $\cT^4=-1$ and $\cT^8=+1$ has been noticed by a remarkable work by Gu in \Refe{Gu1308.2488}.
We further point out that the full boundary fractionalized symmetry group forms a non-Abelian $\bM^{F,T}_{16}$.
\subsubsection{Unwinding the bulk:  Extension to a non-Abelian $\bM^{F,T}_{16}$}
The DIII Hamiltonian can be unwound using the \emph{same} Hilbert space extension and the \emph{same} 2-layer FDUC, 
$$W = W_2 W_1$$ as for class BDI (\eq{eq:Unwinding_BDI3}-\eq{eq:Unwinding_BDI2}). It is easy to check that $W$ is also invariant under the extended $\bM_{16}^{F,T}$ symmetry generator $\cT_{\text{ext} }$ {and the new $P_{F}$} defined below
\begin{eqnarray}
\cT_{\text{ext} } &=& \prod_{\text{odd} ~j} \gamma_{\downarrow,j} \left(\frac{1-\gamma_{\downarrow,j}\bar{\gamma}_{\downarrow,j}}{\sqrt{2}}\right) \bar{\gamma}_{\uparrow,j} \left(\frac{1+\gamma_{\uparrow,j}\bar{\gamma}_{\uparrow,j} }{\sqrt{2}}\right) \gamma_{j} \left(\frac{1-\gamma_{j}\bar{\gamma}_{j}}{\sqrt{2}}\right) \nonumber \\
& & \cdot \prod_{\text{even} ~j} \gamma_{\downarrow,j} \left(\frac{1-\gamma_{\downarrow,j}\bar{\gamma}_{\downarrow,j}}{\sqrt{2}}\right) \bar{\gamma}_{\uparrow,j} \left(\frac{1+\gamma_{\uparrow,j}\bar{\gamma}_{\uparrow,j}}{\sqrt{2}}\right) \bar{\gamma}_{j} \left(\frac{1+\gamma_{j}\bar{\gamma}_{j}}{\sqrt{2}}\right)   \cK.\\
P_{F}&=& \prod_j 
\left(\ii \bar{\gamma}_{\downarrow,j} \gamma_{\downarrow,j}\right) 
\left(\ii \bar{\gamma}_{\uparrow,j} \gamma_{\uparrow,j}\right) 
\left(\ii \bar{\gamma}_{j} \gamma_{j}\right).
\end{eqnarray}
{The extended bulk symmetry operators form again a non-Abelian $\bM_{16}$~\cite{M16} of order 16:\footnote{{Note that $\cK^2=1$, 
we can show
$\cT_{\text{ext} }^2=\prod_j 
\left( \gamma_{\downarrow,j}  \bar{\gamma}_{\downarrow,j}\right)
\left(\gamma_{\uparrow,j}\bar{\gamma}_{\uparrow,j} \right)
(\gamma_{j}\bar{\gamma}_{j})
$,
\; $\cT_{\text{ext} }^4=-1$,
and $\cT_{\text{ext} }^8=1$,
also $P_{F} \cT_{\text{ext} } P_{F} = -\cT_{\text{ext} } = \cT_{\text{ext} }^5 $.
}
}
\begin{equation}
\bM^{F,T}_{16}  \vert_{\text{bulk}}  \equiv \langle {\cT_{\text{ext} }}, {P_{F}} \mid
{ \cT_{\text{ext} }^4 =-1,} \;
 \cT_{\text{ext} }^8 = {P^2_{F}} = 1, \;
 P_{F} \cT_{\text{ext} } P_{F} = \cT_{\text{ext} }^5 \rangle .
\end{equation}
The short exact sequence that relates the extended and original symmetry groups in the bulk again is the following
\begin{equation}
1  \longrightarrow \bZ_4   \overset{ }{\longrightarrow} \bM_{16}^{F,T} \vert_{\text{bulk}}  \overset{ }{\longrightarrow}  \bZ_4^{TF} \vert_{\text{bulk}}   \longrightarrow 1. 
\end{equation}
where the normal subgroup $\Z_4 
=\{1,$ ${\cT^2_{\text{ext} }} {P_{F}},$ ${\cT^4_{\text{ext} }},$ ${\cT^6_{\text{ext} }}{P_{F}}\}$ is not the center $Z(\bM^{F,T}_{16})$.
The quotient group $\bZ_4^{TF}$ contains $\{1, \; {\cT_{\text{ext} }}, \;{\cT^2_{\text{ext} }},\; {\cT^3_{\text{ext} }} \}$.
}

\subsection{Class AIII: $\U(1)^{F} \times \ztwo^T$ or $\frac{\U(1)^{F}_{} \times \Z^{{T}}_{4}}{\Z_2^F}$ symmetry or Pin$^c$}
\label{sec:free AIII}

Let us look at {another example for free fermion Hamiltonian AIII class, which can be chosen to be
$\U(1)^{F} \times \ztwo^T$ or $\frac{\U(1)^{F}_{} \times \Z^{{T}}_{4}}{\Z_2^F}$ symmetry, see Table \ref{tab:free fermion phases}.}\footnote{{The time reversal 
$\cT$ with $\cT^2=+1$ in $\U(1)^{F} \times \ztwo^T$ can be redefined as a
new time reversal 
$\cT'=\exp(\ii \pi/2) \cT =\ii  \cT$ 
with
${\cT'}^2=-1$ and ${\cT'}^4=+1$ in $\frac{\U(1)^{F}_{} \times \Z^{{T}}_{4}}{\Z_2^F}$. 
Here the $\exp(\ii \pi/2)=\ii$ is a $\pi/2$ rotation of the U(1) symmetry transformation which is also
related to the  $\pi$ rotation of fermion since the $2 \pi$ rotation of fermion gives a spin-statistics $(-1)$ from ${\Z_2^F}$.}}
{For $\U(1)^{F} \times \ztwo^T$ symmetry}, a representative of class AIII systems is a superconductor with time reversal $\cT = \cK $ and a spin-rotation $\U(1)$ symmetry
\begin{equation}
V(\theta) = \prod_j \exp \Big(\frac{\theta}{2} \left( \gamma_{\downarrow,j} \gamma_{\uparrow,j} + \bar{\gamma}_{\downarrow,j} \bar{\gamma}_{\uparrow,j} \right)\Big).
\end{equation}
It can be easily checked that the Hamiltonian $H^{\text{II}}_{\text{Kitaev}}$
of eq~(\ref{eq:2 layer Kitaev}) is invariant under the symmetries defined above. Once again, we use the basis change to 
get the Hamiltonian $H_{\text{gen}}$ of eq~(\ref{eq:2layerKitaev_basischanged}) and the following basis-changed symmetry operators:
\begin{eqnarray}
{ \tilde{\mathcal{T}} } &\equiv&  M \cT M^\dagger =  M \cK M^\dagger 
 =  \left(\prod_{j} 
 (\gamma_{\downarrow,j} \bar{\gamma}_{\uparrow,j}) \right) \mathcal{K}. \\
 \label{eq:tildeV}
{  \tilde{V}{(\theta)} } &\equiv& M V(\theta) M^\dagger = \prod_j \exp 
\Big(\frac{\theta}{2} \left( \bar{\gamma}_{\downarrow,j} \gamma_{\downarrow,j}  -  \bar{\gamma}_{\uparrow,j} \gamma_{\uparrow,j}  \right)\Big)
{= \prod_j \exp 
\Big(\ii  {\theta} \big( (N_{F,\downarrow})_{j}  - (N_{F,\uparrow})_{j} \big)\Big).} \quad\quad \quad
\end{eqnarray}
{Here we use the fact for the flavor $\uparrow$ or $\downarrow$ (abbreviated as $\uparrow/\downarrow$),\footnote{{We remark 
that the left-handed side expression fermion number $(N_{F,\uparrow/\downarrow})_{j}$ has eigenvalues in $\{0,1\}$
for states with a complex fermion empty or filled.
In contrast, $ \ii \bar{\gamma}_{\uparrow/\downarrow,j}\gamma_{\uparrow/\downarrow,j} =(P_{F,\uparrow/\downarrow})_j$ is the fermion parity on the site $j$ for the 
flavor $\uparrow$ or $\downarrow$,
which $(P_{F,\uparrow/\downarrow})_j$ has eigenvalues in $\{1,-1\}$. So the right-handed side 
$\frac{1- (P_{F,\uparrow/\downarrow})_j }{2}$ also indeed has eigenvalues in $\{0,1\}$ exactly matching $(N_{F,\uparrow/\downarrow})_{j}$.}} 
$$(N_{F,\uparrow/\downarrow})_{j}\equiv \psi^\dagger_{\uparrow/\downarrow,j} \psi_{\uparrow/\downarrow,j}  
=\frac{1- \ii  \bar{\gamma}_{\uparrow/\downarrow,j} \gamma_{\uparrow/\downarrow,j}}{2}
=\frac{1 -(P_{F,\uparrow/\downarrow})_j }{2}.$$
Note that \eq{eq:tildeV} is in the form of $S_z$ spin-rotational U(1) symmetry generator $\prod_j \exp 
(-\ii {\theta} (S_z)_j )$ of the spin-1/2 system (a fundamental representation {\bf 2} of SU(2)), with a periodicity $\theta \sim \theta + 2\pi$,
so $\theta  \in [0,2 \pi)$.}
\subsubsection{Gapping the boundary:  Extension to a non-Abelian 
$\frac{\U(1) \times \bD_8^{F,T}}{\ztwo}$} 

Let us now look at the symmetry transformation on the edges, 
 {on the left ($L$) and right ($R$) boundaries}: 
\begin{eqnarray}
V_L(\theta) &=& \exp\Big( \frac{\theta}{2} \bar{\gamma}_\downarrow \gamma_\downarrow\Big),~~~~~~~~~ 
V_R(\theta) = \exp\Big(-\frac{\theta}{2} \bar{\gamma}_\uparrow \gamma_\uparrow \Big),  \label{eq:tildeVbdry}\\
\cT_{L} &=& \gamma_{\downarrow}  \cK, ~~~~~~~~~~~~~~~~~~~~ \cT_R = \bar{\gamma}_{\uparrow}    \cK, \\
P_{F,L} &=& \ii \bar{\gamma}_{\downarrow} \gamma_{\downarrow} ,~~~~~~~~~~~~~~~~P_{F,R} = \ii \bar{\gamma}_{\uparrow} \gamma_{\uparrow}.
\label{eq:PF}
\end{eqnarray}
{Some comments are in order:
\begin{itemize}
\item The bulk $\U(1)^F$ symmetry \eq{eq:tildeV} acting on the two complex fermions per site has
a $2 \pi$ periodicity: $\theta  \in [0,2 \pi)$.
Note that
\eq{eq:tildeV} at $\theta = \pi$, we have effectively 
${  \tilde{V}{(\theta = \pi)} }=P_F$;
this generates a $\Z_2^F=  \{ \tilde{V}{(\theta =0)}, \tilde{V}{(\theta = \pi)} \} = \{ 1, P_F\}$,
which is a normal subgroup of $\U(1)^F$.
\item The boundary U(1) symmetry \eq{eq:tildeVbdry} acting on the one complex fermion per site has
a $4 \pi$ periodicity: $\theta  \in [0,4 \pi)$. Note that
this boundary U(1) does \emph{not} contain $\Z_2^F= \{ 1, P_F\}$,
since \eq{eq:tildeVbdry}'s generator 
$\exp(\frac{\theta}{2} \bar{\gamma}{\gamma})=\cos(\frac{\theta}{2})+  \sin(\frac{\theta}{2}) \bar{\gamma}{\gamma}$
can \emph{not} contain \eq{eq:PF}'s $P_{F} = \ii \bar{\gamma} \gamma$.
Instead \eq{eq:tildeVbdry} of the $L$/$R$ boundary
contains a $\Z_2= \{1,V_{L/R}(\theta = 2 \pi) \}=\{1,-1\}$.
\item On the boundary, still there is a $ \bD_8^{F,T}$ finite group symmetry generated by 
$\cT_{L/R}$ and $P_{F,L/R}$, 
similar to \Sec{subsec:gap-bdry-BDI}.
There is a normal subgroup $\Z_2=\{1,-1\}$
shared by both the boundary symmetry 
$\U(1)$ and 
$\bD_8^{F,T} \equiv  \ztwo^F \ltimes  \Z_4^T$ 
with the new fractionalized $\Z_4^T$.
We can explicitly check the boundary U(1) commutes with other generators of $\mathbb{D}_8^{F,T}$.
\end{itemize}
In summary of the above, so the edge symmetry operators form the boundary extended symmetry group
$ \tilde{G} \vert_{\text{boundary}}$,
which can be written into the short exact sequence:\footnote{{To check some of the above mentioned equalities, we use
the fact 
$\exp(\ii \theta X)=\cos(\theta X) + \ii \sin(\theta X)$ for a matrix operator $X$. 
For example, we have
$\exp(\frac{\theta}{2} \bar{\gamma}{\gamma})=\cos(\frac{\theta}{2})+  \sin(\frac{\theta}{2}) \bar{\gamma}{\gamma}$.
\\
$\bullet$ Note
that for \eq{eq:tildeV}  of the bulk, 
$\tilde{V}{(\theta = \pi)}=
\prod_j \exp 
(\ii \pi (   (N_{F,\uparrow})_{j} + (N_{F,\downarrow})_{j}  \big) ) 
=(-1)^{(   (N_{F,\uparrow})_{j} + (N_{F,\downarrow})_{j} )}= P_F= 
\prod_j 
\left(\ii \bar{\gamma}_{\uparrow,j} \gamma_{\uparrow,j}\right)
\left(\ii \bar{\gamma}_{\downarrow,j} \gamma_{\downarrow,j}\right) 
$
generates the fermion parity $\Z_2^F$ of the bulk system.\\
$\bullet$ Note
that for \eq{eq:tildeVbdry} of the $L$ or $R$ boundary, 
$V_{L/R}(\theta = 2 \pi) = \exp\Big( {\pi} \bar{\gamma}_{\downarrow/\uparrow} \gamma_{\downarrow/\uparrow}\Big)= 
(-1)^{\ii (  \bar{\gamma}_{\downarrow/\uparrow} \gamma_{\downarrow/\uparrow})
}
= (-1)^{P_{F,L/R}}=-1$,
generates a $\Z_2$ but \emph{not} the fermion parity $\Z_2^F$ of the effective boundary.}} 
\bea
&&1\to \Z_2 \to  \underset{\quad \supseteq \mathbb{D}_8^{F,T} =  \ztwo^F \ltimes  \Z_4^T}{\frac{\U(1) \times \bD_8^{F,T}}{\ztwo}}\vert_{\text{boundary}}  \to 
 \underset{\quad \supseteq  \ztwo^F \times  \ztwo^T }{\U(1)^{F} \times \Z_2^T}  
\vert_{\text{bulk}   
} \to 1.  
\eea
In the underset of groups $ \tilde{G} \vert_{\text{boundary}}$ and $G\vert_{\text{bulk}}$, we indicate that they contain the subgroups ($\supseteq \dots$) 
established in \Sec{subsec:gap-bdry-BDI}.
}

\subsubsection{Unwinding the bulk:  Extension to a non-Abelian $\frac{\U(1) \times \bD_8^{F,T}}{\ztwo}$}
The AIII Hamiltonian can be unwound using the same Hilbert space extension and 2-layer FDUC, $W = W_2 W_1$ as for class BDI 
(\eq{eq:Unwinding_BDI3}-\eq{eq:unwinding_He}).  
It is easy to check that $W$ is also invariant under the extended $\frac{\U(1) \times \bD_8^{F,T}}{\ztwo}$ symmetry generators defined below
\begin{eqnarray}
V_{\text{ext} }(\theta) &=& \prod_{\text{odd} ~j} 
\exp\Big( \frac{\theta}{2} \left( \bar{\gamma}_{\downarrow,j} \gamma_{\downarrow,j}  -  \bar{\gamma}_{\uparrow,j} \gamma_{\uparrow,j} + \bar{\gamma}_{j} \gamma_{j}  \right)\Big) \prod_{\text{even} ~j} \exp\Big( \frac{\theta}{2} \left( \bar{\gamma}_{\downarrow,j} \gamma_{\downarrow,j}  -  \bar{\gamma}_{\uparrow,j} \gamma_{\uparrow,j} - \bar{\gamma}_{j} \gamma_{j}  \right)\Big), \quad\quad\;\\
\cT_{\text{ext} } &=& \prod_{\text{odd} ~j} \gamma_{\downarrow,j} \bar{\gamma}_{\uparrow,j}  \gamma_{j} \prod_{\text{even} ~j} \gamma_{\downarrow,j}  \bar{\gamma}_{\uparrow,j} \bar{\gamma}_{j}   \cK,\\
P_{F}&=& \prod_j \left(\ii \bar{\gamma}_{\downarrow,j} \gamma_{\downarrow,j}\right) 
\left(\ii \bar{\gamma}_{\uparrow,j} \gamma_{\uparrow,j}\right) 
\left(\ii \bar{\gamma}_{j} \gamma_{j}\right).
\end{eqnarray}
{In short, the bulk unwinding follows also the symmetry extension:
\bea
1  \longrightarrow \ztwo   \overset{ }{\longrightarrow} 
\frac{\U(1) \times \bD_8^{F,T}}{\ztwo}\vert_{\text{bulk}} \overset{ }{\longrightarrow}  \U(1)^{F} \times \Z_2^T  \vert_{\text{bulk}} \longrightarrow 1. 
\eea}
\subsection{Class CII: $\SU(2)^F \times \ztwo^T$ symmetry or Pin$^- \times_{\Z_2^F} \SU(2)^F$}
\label{sec:CII}
The 1+1d CII fermion phase has 
{a nontrivial class
$\nu_{\text{CII}}\in \Z_2$ which} actually corresponds to a bosonic SPT phase with a $\Z_2^T$ time reversal symmetry
{tensor product with a trivial fermionic gapped phase,}
and can be unwound by an extended symmetry that has a fermion parity 
{at the center, see the detailed discussions in \cite{AP_Unwinding_PRB2018}.}
In terms of the CII class symmetry realization, other than $\SU(2)^F \times \ztwo^T$, it can also be realized as
$\frac{({\U}(1) \rtimes {\Z}_4^C)}{{\Z}_2^F}$ symmetry \cite{1711.11587GPW, AP_Unwinding_PRB2018}.
Let us interpret how to gap the boundary and 
unwind the bulk via the following symmetry extensions.

If we extend the $\mathbb{Z}_2^T$ via a new species of fermions with a fermion parity $\Z_2^{F_+}$ to $ \mathbb{Z}_4^{T{F_+}}$ with $\cT^2=(-1)^{F_+}$, we have the symmetry extensions:
\bea \label{eq:CII-1ext}
\begin{array}{ccccccccc}
1& \longrightarrow&	\Z_2^{F_+} & \longrightarrow &\SU(2)^F  \times \mathbb{Z}_4^{T{F_+}} &\longrightarrow& \SU(2)^F \times \ztwo^T &   \longrightarrow& 1.\\
1 &\longrightarrow& \Z_2^{F_+} &\longrightarrow &\frac{({\U}(1) \rtimes {\Z}_4^C)}{{\Z}_2^F} \times \mathbb{Z}_4^{T{F_+}}& \longrightarrow & \frac{({\U}(1) \rtimes {\Z}_4^C)}{{\Z}_2^F} \times \mathbb{Z}_2^T&  \longrightarrow 1. \\
1&\longrightarrow &	\Z_2^{F_+} & \longrightarrow &\mathbb{Z}_4^{T{F_+}} & \longrightarrow &  \mathbb{Z}_2^T &   \longrightarrow & 1.
\end{array}
\eea

If we extend the $\mathbb{Z}_2^T$ via a bosonic $\Z_2^{B}$ to $ \mathbb{Z}_4^{T{B}}$ symmetry with $\cT^2=-1$ as a bosonic
$(-1)$, we have the symmetry extensions:
\bea  \label{eq:CII-2ext}
\begin{array}{ccccccccc}
1& \longrightarrow&	\Z_2^B & \longrightarrow &\SU(2)^F  \times \mathbb{Z}_4^{TB} &\longrightarrow& \SU(2)^F \times \ztwo^T &   \longrightarrow& 1.\\
1 &\longrightarrow& \Z_2^B &\longrightarrow &\frac{({\U}(1) \rtimes {\Z}_4^C)}{{\Z}_2^F} \times \mathbb{Z}_4^{TB} & \longrightarrow & \frac{({\U}(1) \rtimes {\Z}_4^C)}{{\Z}_2^F} \times \mathbb{Z}_2^T&  \longrightarrow 1. \\
1&\longrightarrow &	\Z_2^B & \longrightarrow &\mathbb{Z}_4^{TB} & \longrightarrow &  \mathbb{Z}_2^T &   \longrightarrow & 1.
\end{array}
\eea
The quantum mechanical lattice constructions of the above are done in \cite{AP_Unwinding_PRB2018}. 
Formally, we can also understand it from trivializing the 2d topological term in the bordism group 
$\Omega_2^{\Pin^- \times_{\Z_2^F} \SU(2)^F}$, 
which will be explained in \Sec{sec:conclusion1}.

\subsection{Formal explanation on the classifications of invertible phases vs SPT phases}
\label{sec:formal-classification}
\begin{itemize}[leftmargin=2.mm]
\item
\emph{Long-ranged entangled (LRE) invertible fermionic topological orders} (protected by no other symmetry, except the fermion parity $\Z_2^F$ essential for fermionic systems):
To obtain its classification,
we can read from $\TP_d(\Spin)$ with the group $G=\Spin(d)$ symmetry.
Note that $\Spin(d)/\SO(d)=\Z_2^F$ with the $\SO(d)$ is the rotational symmetry for the continuum limit of the tangent manifold of spacetime manifold.
\item \emph{Short-ranged entangled (SRE) invertible SPTs}:
To obtain its classification, we can consider the quotient group
$\frac{\TP_d(G)}{\TP_d(\Spin)}$.
Namely, the invertible fermionic phases classified by $\TP_d(G)$ mod out the
LRE invertible fermionic topological orders in $\TP_d(\Spin)$, indeed are left with only SRE invertible SPTs \cite{GuoJW1812.11959}.
In fact, for interacting systems, 
for the finite group subclasses of the SRE SPT classification  
(known as the torsion [tors] classes in mathematics),
we can rewrite the SRE fermionic SPT finite-subgroup-part classification following \cite{GuoJW1812.11959} as
 the torsion part of the quotient group $\frac{\TP_d(G)}{\TP_d(\Spin)}$:
\bea \label{eq:cobordism}
\big[\frac{\TP_d(G)}{\TP_d(\Spin)} \big]_{{\rm tors}}=\frac{[\TP_d(G)]_{\rm tors}}{[\TP_d(\Spin)]_{\rm tors}} 
= \frac{[\Omega_d^G]_{\rm tors}}{[\Omega_d^\Spin(pt)]_{\rm tors}}.
\eea
Here we use the fact that the torsion (finite subgroup) part of cobordism and bordism groups are the same
${[\TP_d(G)]_{\rm tors}}={[\Omega_d^G]_{\rm tors}}$; 
this \eqn{eq:cobordism} relates to the 
$d$-th bordism group $\Omega_d$.
{{This expression in \Eq{eq:cobordism}
left out other infinite group $\Z$ classes: \\
$\bullet$ Some of $\Z$ classes are known as the free classes in mathematics correspond to LRE invertible topological orders by Wen's definition \cite{Wen2016ddy1610.03911},
captured by 
\bea
\big[{\TP_d(\Spin)} \big]_{{\rm free}}.
\eea
$\bullet$  Other $\Z$ classes are captured by the free part of quotient group $\frac{\TP_d(G)}{\TP_d(\Spin)}$ as
\bea
\big[\frac{\TP_d(G)}{\TP_d(\Spin)} \big]_{{\rm free}},
\eea
which happen especially in an odd dimensional spacetime and correspond to the $\Z$ classes of SRE invertible SPTs whose boundary exhibits the 
\emph{perturbative local anomalies}. For example, in 2+1d, there is a $\Z$ class of SRE invertible SPTs protected by U(1) symmetry, for both bosonic or fermionic systems \cite{WanWang1812.11967}.
\\
$\bullet$  In contrast, the other $\Z_n$ classes (the torsion) from $\big[{\TP_d(G)} \big]_{{\rm tors}}$
correspond to invertible topological phases whose boundary exhibits the \emph{non-perturbative global anomalies}.
This $\big[{\TP_d(G)} \big]_{{\rm tors}}$ includes both 
the $\big[{\TP_d(\Spin)} \big]_{{\rm tors}}$ as classes of LRE invertible topological orders
and the $\big[\frac{\TP_d(G)}{\TP_d(\Spin)} \big]_{{\rm tors}}$ as SRE invertible SPTs.}
}
The $pt$ is a point as a certain classifying space.
\end{itemize}

\section{Interacting models for even $\nu \in \Z_8$ classes with $\ztwo^T \times \ztwo^F$ symmetry  or Pin$^-$}
\label{sec:interacting BDI}
\begin{figure}[!h] 
	\centering	
	\includegraphics[width=140mm]{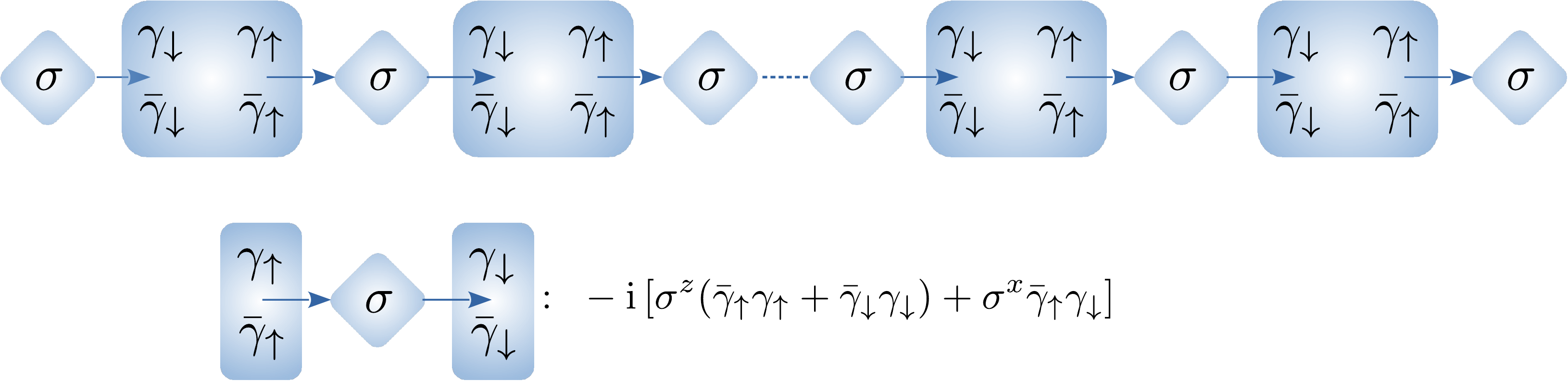}
	\caption{The interacting $\ztwo^T \times \ztwo^F$ invariant SPT Hamiltonian of \cref{eq:Nat}. The squares consist of two complex fermions, labeled $\uparrow$ and $\downarrow$ and reside on the sites of the lattice. The diamonds represent the qubits $\sigma$ and reside on the links of the lattice. \label{fig:BDI_interacting}}
\end{figure}

We now consider an interacting model of time-reversal invariant fermions with $\ztwo^T \times \ztwo^F$ symmetry belonging to the same $k=2$ phase as the model in \cref{sec:free BDI} of the $\bZ_8$ classification. This is a basis-changed version of the model introduced by Tantivasadakarn and Vishwanath~\cite{NatAshvin_PhysRevB.98.165104} (see Appendix~\ref{app:Nat_basis_change} for details of the basis-change operations). The Hilbert space for the model consists of two species of fermions 
{($ \varsigma = \uparrow, \downarrow$: each complex fermions beget two Majorana fermions, ${\gamma}$ and $\bar{\gamma}$)} on the sites and qubits ($\sigma$) on the links of a one-dimensional lattice. Using the Majorana operators as defined in the previous sections, the Hamiltonian is written as
\begin{equation}
H = -\ii \sum_j \left[\sigma^z_{j,j+1} \left( \bar{\gamma}_{\uparrow,j}\gamma_{\uparrow,j} +  \bar{\gamma}_{\downarrow,j+1}\gamma_{\downarrow,j+1}\right) +   
 {\sigma^x_{j,j+1}  \bar{\gamma}_{\uparrow,j}\gamma_{\downarrow,j+1}} \right]. \label{eq:Nat}
\end{equation}
The above Hamiltonian is also represented graphically in Fig~\ref{fig:BDI_interacting}.

The time-reversal and fermion parity operators are
\begin{equation}
\cT = \left(\prod_j \left(\ii \gamma_{\downarrow,j} \bar{\gamma}_{\uparrow,j}\right)  \sigma^x_{j,j+1}  \right)~\cK, ~~~~ P_F = \prod_j \prod_{\varsigma = \uparrow, \downarrow}  \ii \bar{\gamma}_{\varsigma,j} \gamma_{\varsigma,j},
\end{equation}
where $\cK$ is the complex conjugation operation as before. The boundary symmetry operators are as follows:
\begin{eqnarray}
\cT_L &=& \gamma_\downarrow \cK, ~~~~~~~~~~ \cT_R = \ii \bar{\gamma}_\uparrow \cK \\
P_{F,L} &=& \ii \bar{\gamma}_\downarrow \gamma_\downarrow, ~~~~~~ P_{F,R} = \ii \bar{\gamma}_\uparrow \gamma_\uparrow
\end{eqnarray} 
On both ends, the symmetry operators generate the group $\bD_8^{F,T}$ which does not have fermion parity at the center
\begin{equation}
{\bD_8^{F,T} \equiv \langle (\cT_{L/R} P_{F,L/R}), P_{F,L/R} \;\vert\; \cT_{L/R}^2 = P_{F,L/R}^2 = 1, P_{F,L/R} \cT_{L/R} P_{F,L/R} = -\cT_{L/R} \rangle .}
\end{equation}
 We now proceed to extend the Hilbert space and unwind the Hamiltonian using a 
 $\bD_8^{F,T}$ invariant quantum circuit. The extended Hilbert space we choose consists of an additional qubit per link ($\tau_{j,j+1}$), an additional qubit per site ($\sigma_j$) and an additional fermion per site ($\gamma_j, \bar{\gamma_j}$). The extended Hamiltonian we consider is 
 \begin{equation}
 H_{\text{ext} } = H - \sum_j \sigma^x_j -\ii \sum_{\text{odd} ~j} \left[\tau^z_{j,j+1} \left( \bar{\gamma}_{j}\gamma_{j} +  \bar{\gamma}_{j+1}\gamma_{j+1}\right)  + 
{ \tau^x_{j,j+1}   \gamma_{j}  \bar{\gamma}_{j+1}} \right] - \sum_{\text{even} ~j} \tau^x_{j,j+1}. \label{eq:Nat_extend}
 \end{equation}

\begin{figure}[!htbp]
	\centering	
	\includegraphics[width=140mm]{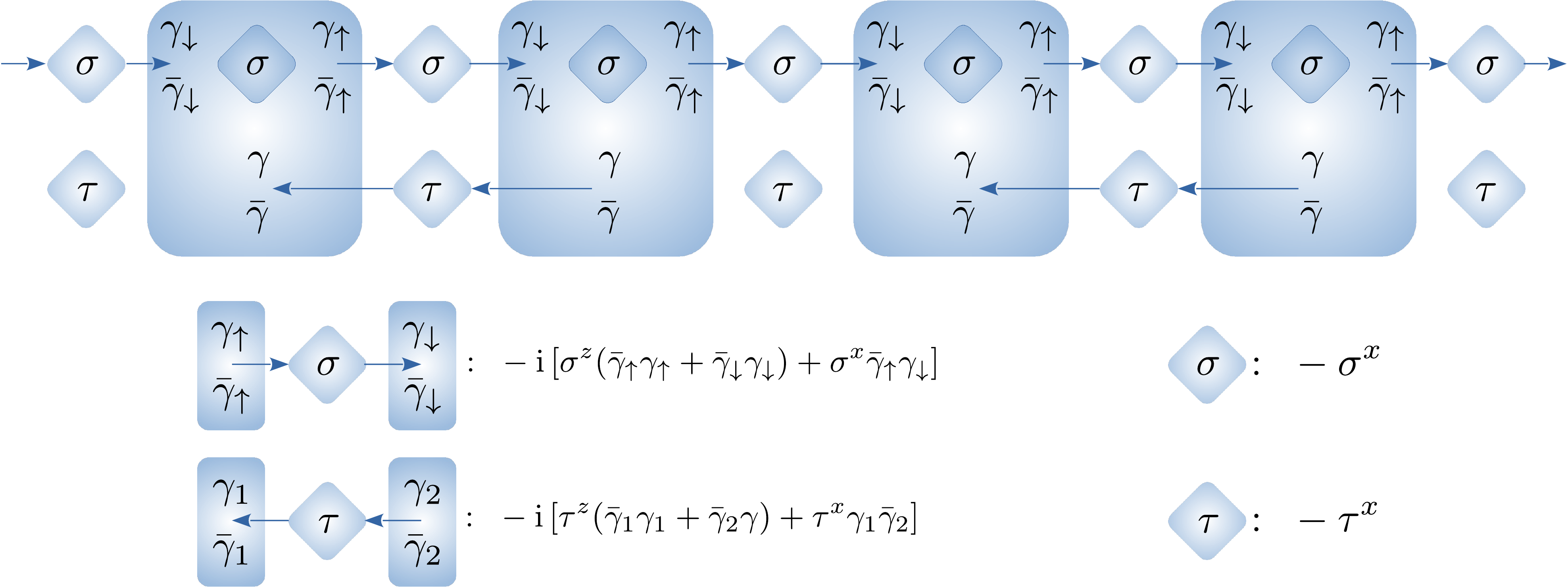}
	\caption{Extended $\bD_8 \equiv \bD_8^{F,T}$ invariant SPT Hamiltonian of \cref{eq:Nat_extend}. The squares consist of three complex fermions and one qubit and reside on the sites of the lattice. Two other qubits (diamonds not enclosed in squares) reside on the links of the lattice.  \label{fig:BDI_interacting_extended}}
\end{figure}

\begin{figure}[!htbp]
	\centering	
	\includegraphics[width=140mm]{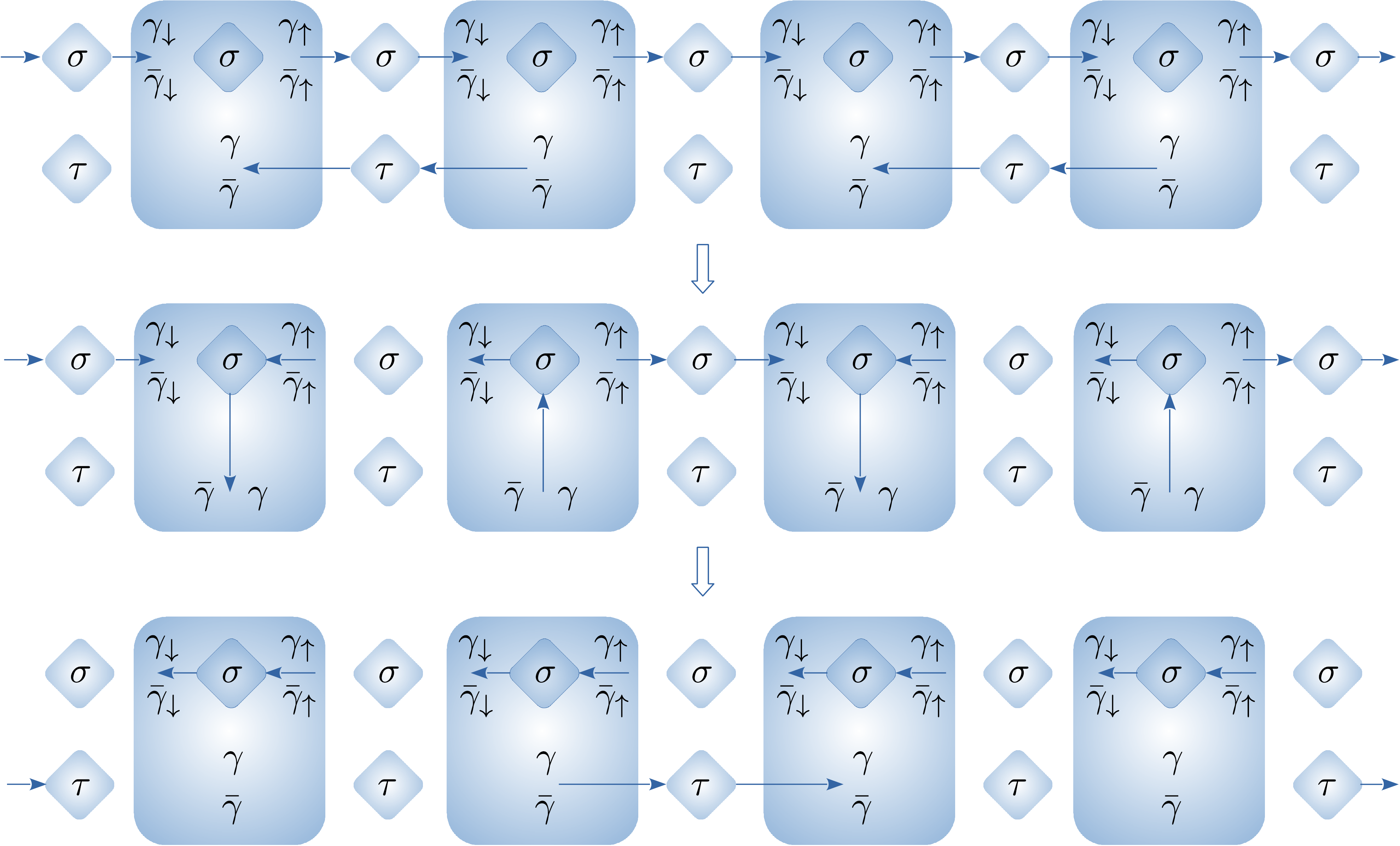}
	\caption{Unwinding the interacting $\ztwo^T \times \ztwo^F$ invariant SPT Hamiltonian \cref{eq:Nat_extend}. The unwinding is done in two steps using a series of bosonic and fermionic swap operators which commute with the extended symmetry. \label{fig:BDI_interacting_unwinding}}
\end{figure}

This extended Hamiltonian is schematically shown in Fig~\ref{fig:BDI_interacting_extended}. 
The Hamiltonian $H_{\text{ext} }$ is unwound to $H_{\text{triv} }$ as shown in Fig~\ref{fig:BDI_interacting_unwinding} using the FDUC, $W$ as follows
 \begin{eqnarray}
WH_{\text{ext} } W^\dagger &=& H_{\text{triv} }. \\
H_{\text{triv} } &=& - \sum_j  \left[\sigma^z_j \Big(\sum_{\varsigma {= \uparrow, \downarrow}}  \ii \bar{\gamma}_{\varsigma,j}\gamma_{\varsigma,j} \Big)+\ii \sigma^x_j \gamma_{\downarrow,j} \bar{\gamma}_{\uparrow,j}  + \sigma^x_{j,j+1}\right] \nonumber \\
&& -\ii \sum_{\text{odd} ~j} \left[ \tau^z_{j,j+1} \left( \bar{\gamma}_j \gamma_j + \bar{\gamma}_{j+1} \gamma_{j+1} \right) + \tau^x_{j,j+1} \bar{\gamma}_j \gamma_{j+1} \right] -\sum_{\text{even} ~j} \tau^x_{j,j+1}. \\
W &=& W_{2} W_1. \\
W_1 &=& \prod_{\text{odd} ~j} \SWAP\left(\sigma_j,\sigma_{j,j+1}\right) \SWAP\left(\tau_{j,j+1},\sigma_{j+1}\right)  \left(\frac{1+ \gamma_j \gamma_{\downarrow,j+1}}{\sqrt{2}}\right) \left(\frac{1+ \bar{\gamma}_j \bar{\gamma}_{\downarrow,j+1}}{\sqrt{2}}\right). \nonumber \\
W_2 &=& \prod_{\text{even} ~j} \left[ \SWAP \left(\sigma_{j,j+1}, \tau_{j,j+1}\right)  \left(\frac{1+ \gamma_j \gamma_{\uparrow,j}}{\sqrt{2}}\right) \left(\frac{1+ \bar{\gamma}_j \bar{\gamma}_{\uparrow,j}}{\sqrt{2}}\right)\right] \nonumber \\ &&~~~~~~~~~ \prod_{\text{odd} ~j} \left[\left(\frac{1+ \gamma_j \gamma_{\downarrow,j}}{\sqrt{2}}\right) \left(\frac{1+ \bar{\gamma}_j \bar{\gamma}_{\downarrow,j}}{\sqrt{2}}\right)\right]. \nonumber
 \end{eqnarray}
 
 It can be verified that the extended Hamiltonian as well as the unwinding FDUC are invariant under the extended symmetry operator transformation 
 \begin{eqnarray}
 \cT_{\text{ext} } &=&  \prod_{\text{odd} ~j} \left(\ii \gamma_{\downarrow,j} \bar{\gamma}_{\uparrow,j} \gamma_j ~\sigma^x_j  \sigma^x_{j,j+1}  \tau^x_{j,j+1}    \right)  \prod_{\text{even} ~j} \left(\ii \gamma_{\downarrow,j} \bar{\gamma}_{\uparrow,j} \ii \bar{\gamma}_j ~ \sigma^x_j  \sigma^x_{j,j+1}  \tau^x_{j,j+1}    \right) \cK\cr
 &=& { \prod_{\text{all} ~j} \left(\ii \gamma_{\downarrow,j} \bar{\gamma}_{\uparrow,j}~\sigma^x_j  \sigma^x_{j,j+1}  \tau^x_{j,j+1}    \right)
 ( \prod_{\text{odd} ~j}  \gamma_j ) ( \prod_{\text{even} ~j} \ii \bar{\gamma}_j)
 \cK.}
 \end{eqnarray}

\newpage

\section{Supersymmetric quantum mechanics}
\label{sec:SUSYQM}
In this section, we detail the relationship between the extended symmetries and supersymmetric quantum mechanics. More details can be found in ref~\cite{APJW-PRL}. Supersymmetric systems are characterized by the presence of \emph{supercharges}- symmetry operators which are fermionic i.e. anti-commute with fermion parity and square to the generators of space-time translations~\cite{Witten_IntroSUSY_1983}. The simplest of supersymmetric systems are supersymmetric quantum-mechanical (SUSY QM) systems where supercharges only square to the generator of time-translations i.e. the Hamiltonian. More specifically, a Hamiltonian $H$ is a SUSY QM system if there exists $\alpha = 1 \ldots \cN$ Hermitian supercharges $Q_\alpha$ satisfying the following algebra~\cite{CooperKhareSukhatme_SUSYQM_1995,WITTEN_SUSYQM_1981513,WITTEN_SUSYQM_1982253,WittenSUSY1982}
\begin{align}
    \{Q_\alpha, Q_\beta \} = 2 H \delta_{\alpha,\beta},~ [Q_\alpha,H] = \{Q_\alpha,P_F \} = 0. \label{eq:SUSY_algebra}
\end{align}
If we consider the \emph{fractionalized} time-reversal symmetry on the boundaries,
\begin{itemize}
\item  from $\Z_2^T$ ($\cT^2=+1$ for BDI)  to 
a fractionalized $\Z_4^T$ ($\cT^4=+1$ in  \Sec{sec:free BDI}), and
\item
from $\Z_4^{TF}$ ($\cT^4=+1$ for DIII) to a fractionalized 
$\Z_8^T$ ($\cT^8=+1$ in \Sec{sec:free DIII}),
\end{itemize} 
their anti-commutation with the fermion parity $P_F$ operator, 
$$
\{\cT , P_F \}  \equiv \cT   P_F + P_F \cT  =0,
$$
will result in any compatible Hamiltonian being a SUSY QM system. As a result, we refer to the presence of any symmetry operators which do not commute with fermion parity as supersymmetric. 

\subsection{General setting for SUSY quantum mechanics}
\label{sec:General}
Let us see the emergence of SUSY QM from global symmetries. Consider a fermionic system described by some Hamiltonian operator $H$. This means that the Hamiltonian commutes with a fermion parity $\Z_2^F$ symmetry, generated by 
$P_F \equiv (-1)^{F}$,
so $P_F H P_F^{-1}= H$. Let $\{V_g\}$ be a collection of symmetry operators, including $P_F$ that commute with $H$. Recall that $P_F$ also grades states and operators as bosonic and fermionic as follows
\begin{align}
    P_f \ket{b} &=+ \ket{b},~~ P_F O_b P_F = +O_b\\
    P_f \ket{f} &=+ \ket{f},~~ P_F O_f P_F = -O_f   
\end{align}
We show that if there exist symmetry operators which are fermionic i.e. anti-commutes with $P_F$,  $H$ is a SUSY QM system with at least $\cN=2$ supercharges.
Let $\cT$ be a such a symmetry of $H$ i.e. $\cT H \cT^{-1}= H$, and $P_F \cT P_F = -\cT$. This implies that any eigenstate with non-zero eigenvalue (ensured by a constant shift of $H$) is degenerate and contains an even number of bosonic and fermionic states. The diagonalized Hamiltonian can be schematically written, with $\mu$ the eigenvalue labels, as
\begin{equation}
H = \sum_{\mu} \sum_{a_\mu} E_\mu (\outerproduct{\mu,a_\mu,+}{\mu,a_\mu,+} + \outerproduct{\mu,a_\mu,-}{\mu,a_\mu,-}).
\end{equation}
{The $+$ or $-$ specifies the $(-1)^F$ eigenvalue which correspond to bosonic or fermionic sector and $a_\mu$ keeps track of additional degeneracies.}
Let us consider the minimal setting when there are no other degeneracies in the system. For concreteness, let us also take $\cT^2 = 1$ 
{so we have a  $\Z_2^\cT$ symmetry 
(in fact, $\Z_2^\cT$ can be more general, 
\emph{either unitary or anti-unitary}
in this section)}. We assume that the Hamiltonian is shifted so as to ensure that all energies are positive-definite. The eigenstates are organized as
\begin{equation}
H \ket{\mu,\pm} = E_\mu \ket{\mu,\pm}, ~~~P_F \ket{\mu,\pm} = \pm \ket{\mu,\pm}.
\end{equation}
 It is clear that $\cT$ switches the parity
\begin{equation}
\cT \ket{\mu,\pm} = \ket{\mu,\mp}.
\end{equation}
Following ref~\cite{BehrendsBeri_SUSYSYK_PhysRevLett.124.236804}, we can now define the following operators
\begin{equation}
Q_\mu \equiv \sqrt{E_\mu } \outerproduct{\mu,+}{\mu,-},
\quad Q^\dagger_\mu \equiv \sqrt{E_\mu } \outerproduct{\mu,-}{\mu,+},
\end{equation}
using which we can define the following two Hermitian \emph{supercharges}
\begin{align}
Q_{+} &\equiv \sum_{\mu} Q_\mu + Q^\dagger_\mu. \\
Q_{-} &\equiv\sum_{\mu} {-} \ii\left( Q_\mu - Q^\dagger_\mu \right).
\end{align}
{We also have the fermion parity $P_F$ and identity $\mathds{1}$ operators as:
\bea
P_F \equiv \sum_{\mu} \outerproduct{\mu,+}{\mu,+} -  \outerproduct{\mu,-}{\mu,-}, \quad
\mathds{1} \equiv \sum_{\mu} \outerproduct{\mu,+}{\mu,+} +  \outerproduct{\mu,-}{\mu,-}.
\eea
}
It can be checked that the supercharges satisfy the algebra of \cref{eq:SUSY_algebra} {(below $\alpha, \beta \in \{\pm\}$)} \cite{WittenSUSY1982,CooperKhareSukhatme_SUSYQM_1995} i.e.
\begin{enumerate}
\item Supersymmetry anti-commutes with the  $\Z_2^F$ fermion parity symmetry, generated by $P_F \equiv (-1)^{F}$:
\begin{equation}
\{Q_\alpha, P_F \}  \equiv Q_\alpha P_F + P_F Q_\alpha =0.
\end{equation}
\item Supercharges commute with Hamiltonian thus supersymmetry is a global symmetry:
\begin{equation}
[Q_\alpha, H]  \equiv Q_\alpha H -H Q_\alpha  =0.
\end{equation}
\item
The anti-commutator for supersymmetric quantum mechanics obeys:
\begin{equation}
\{Q_\alpha,Q_\beta\} \equiv Q_\alpha Q_\beta + Q_\alpha Q_\beta= 2H \delta_{\alpha \beta}.
\end{equation}
\end{enumerate}
This means that we have an $\cN = 2$ SUSY QM system at hand. The value of $\cN$ could be larger depending on the additional symmetries but some SUSY is guaranteed. The symmetry 
$\cT$ acts on the supercharges as follows.\footnote{{In general, for each energy $E_\mu$ sector, we have a 2-dimensional Hilbert space
$\cH_{\mu}=\cH_{B,{\mu}} \oplus \cH_{F,{\mu}}$ where
the bosonic sector $\cH_{B,{\mu}}$ is spanned by the vector
 $\{ \ket{\mu,+}\}$,
 the fermionic sector $\cH_{F,{\mu}}$ is spanned by the vector
 $\{ \ket{\mu,-}\}$.
Overall, the total Hilbert space is even dimensional:
\bea
\cH =\cH_{B} \oplus \cH_{F} = \bigoplus_{\mu} \cH_{\mu}=\bigoplus_{\mu}  \big(\cH_{B,{\mu}} \oplus \cH_{F,{\mu}}\big).
\eea
We can write all the aforementioned operators 
$H, P_F, \cT$, and $Q_{\pm}$
respect to the 2-dimensional Hilbert space $\cH_{\mu}$ associated to each energy $E_\mu$. We can denote the projection
of the $H, P_F, \cT$, and $Q_{\pm}$ to a 2-dimensional Hilbert space $\cH_{\mu} \equiv \{ \ket{\mu,+}, \ket{\mu,-}\}$ as
$(H)_{{\mu}}, (P_F)_{\mu}$, $(Q_{\pm})_{\mu}$,  and $(\cT)_{\mu}$. Then we have:
\bea
&&(H)_{{\mu}}= E_\mu \begin{pmatrix} 1& 0\\ 0  &1\end{pmatrix}_{\mu} = E_\mu (\sigma_0)_{\mu}, \;\;
(P_F)_{\mu} =\begin{pmatrix} 1& 0\\ 0  &-1\end{pmatrix}_{\mu} \equiv (\sigma_z)_{\mu}, \;\;
\cr
&&(Q_{+})_{\mu}= \begin{pmatrix}  0 & 1\\ 1 & 0  \end{pmatrix}_{\mu} \equiv (\sigma_x)_{\mu}, \;\;
(Q_{-})_{\mu}= \begin{pmatrix}  0 & -\ii\\ \ii & 0  \end{pmatrix}_{\mu} \equiv (\sigma_y)_{\mu}, \;\;
Q_{\mu}= \begin{pmatrix}  0 & 1\\ 0 & 0  \end{pmatrix}_{\mu} , \;\;
Q_{\mu}^\dagger= \begin{pmatrix}  0 & 0\\ 1 & 0  \end{pmatrix}_{\mu}. \cr
&&(\cT)_{\mu} =\begin{cases}
(n_x (Q_{+})_{\mu}+n_y (Q_{-})_{\mu} )=(n_x ( \sigma_x)_{\mu}+n_y (\sigma_y)_{\mu} ) ,&\text{ if $\cT$ is unitary}.\\
(n_x (Q_{+})_{\mu}+n_y (Q_{-})_{\mu} ) K
=(n_x (\sigma_x)_{\mu}+n_y (\sigma_y)_{\mu}) K
,&\text{ if $\cT$ is anti-unitary}.
\end{cases}
\eea
Here $(n_x,n_y) \in S^1 \subset\mathbb{R}^2$ is a unit vector so $n_x^2+n_y^2=1$,
and $K$ is a complex conjugation operator. 
It is amusing to notice such a $(\cT)_{\mu}$ is in fact
a linear combination $(n_x (Q_{+})_{\mu}+n_y (Q_{-})_{\mu} )$. 
In the case if $\cT$ is an anti-unitary time-reversal symmetry,
interestingly
we have $(\cT)_{\mu}$ proportional to SUSY charge linear combination 
$(Q_{+})_{\mu}$ and $(Q_{-})_{\mu}$ up to a 
complex conjugation $K$. In the main text, we choose
$(n_x,n_y)=(1,0)$
to derive \eq{eq:TQ1} and \eq{eq:TQ2}.
}} 
If {$\cT$} is a unitary $\ztwo$ symmetry, we {can choose}
\begin{equation} \label{eq:TQ1}
\cT: Q_{\pm} \mapsto \cT Q_{\pm}\cT^{-1}= \pm Q_\pm.
\end{equation} 
If the {$\cT$} is an anti-unitary $\ztwo$ symmetry, {we can choose} $\cT$ acts as an identity operator on the supercharges: 
{
\begin{equation} \label{eq:TQ2}
\cT: Q_{\pm} \mapsto \cT Q_{\pm}\cT^{-1}=  + Q_\pm.
\end{equation} 
}
Generally, the original symmetries acts by permuting the supercharges.

\subsection{$\cN=2$ SUSY on the boundary of 1+1d fermionic SPT phase with $\ztwo^T \times \ztwo^F$ symmetry:
Generalized Sachdev-Ye-Kitaev interactions}
\label{sec:SYK-SUSY-BDI}
Recall that one-dimensional SPT phases have symmetries that are realized \emph{projectively}. For \emph{intrinsically} fermionic SPT phases i.e phases that cannot be thought of as a stack of a bosonic SPT phase, {tensor product} with a trivial {gapped} fermionic system, some of the projective symmetry generators do not commute with fermion parity~\cite{KapustinTurzilloYou_PhysRevB.98.125101_2018}. Due to the arguments of the previous subsection, this leads to the boundary Hamiltonian being supersymmetric. Let us consider a specific phase for concreteness. We again consider the $\nu = 2 \mod 8$ phase of time-reversal invariant superconductors where $\cT$ which acts anti-unitarily and has the property $\cT^2 = 1$. This belongs to class BDI and (in the presence of interactions) has a $\bZ_8$ classification. We again consider the Hamiltonian in \eq{eq:2layerKitaev_basischanged}:
\begin{equation}
H = -\ii \sum_{j=1}^\ell \left(\gamma_{\uparrow,j} \gamma_{\downarrow,j+1}- \bar{\gamma}_{\uparrow,j} \bar{\gamma}_{\downarrow,j+1}\right) \label{eq:BDI}.
\end{equation}
Recall the symmetries are
\begin{align} \label{eq:T-whole-SYK}
	\cT  = 
	\left(\prod_{j} 
	(\gamma_{\downarrow,j} \bar{\gamma}_{\uparrow,j}) \right) \mathcal{K}.\quad
	\quad
	P_F = \left( \prod_j
	\prod_{\varsigma = \uparrow, \downarrow}   \ii \bar{\gamma}_{\varsigma, j} \gamma_{\varsigma, j} \right). 
\end{align}
$\cK$ is the complex conjugation which acts on the Majorana operators as
\begin{equation}
\label{eq:complex conjugation majorana}
\mathcal{K} \gamma \mathcal{K} = \gamma,~~\mathcal{K} \bar{\gamma} \mathcal{K} =- \bar{\gamma},~~\mathcal{K} \ii \mathcal{K} = -\ii.
\end{equation}
Let us now consider a boundary termination as shown in \Fig{fig:bdry-sym}. 
There are 2 dangling Majorana zero modes on each (left or right) end of 
 \Fig{fig:bdry-sym}, each end has a 2-dimensional Hilbert space.
{For  \Fig{fig:bdry-sym}, 
 the symmetries acting on the boundary sites are the same as the bulk sites.
However, if we only consider how 
the \emph{effective symmetries} acting on
the ground state subspace which associated with the boundary zero modes,
we can do the ground state (GS) projection $\PGS = \PGS^\dagger$ on
$\mathcal{T}$
which factorize \eq{eq:T-whole-SYK} to left and right boundary symmetry operators:
$$
\PGS \cT\PGS  \simeq
\mathcal{T}_L
\otimes
\mathcal{T}_R. 
$$
Here \emph{effective} $\mathcal{T}_L$ and $\mathcal{T}_R$ act only on the
left and right Majorana zero modes of \Fig{fig:bdry-sym} as:
}
\begin{eqnarray}
\mathcal{T}_L &=& \gamma_{\downarrow} \mathcal{K},~P_{F,L} = i \bar{\gamma}_{\downarrow} \gamma_{\downarrow},~\mathcal{T}_L^2 = P_{F,L}^2 = 1, \mathcal{T}_L P_{F,L} = -P_{F,L}  \mathcal{T}_L. \label{eq:TLP} \\
\mathcal{T}_R &=& \bar{\gamma}_{\uparrow} \mathcal{K},~P_{F,R} = i \bar{\gamma}_{\uparrow} \gamma_{\uparrow},~\mathcal{T}_R^2 = -1, P_{F,R}^2 = 1, \mathcal{T}_R P_{F,R} = -P_{F,R}  \mathcal{T}_R. \label{eq:TRP} 
\end{eqnarray}
It can be easily checked that the boundary symmetry operators generate a faithful representation of {dihedral group of order 8,} $\bD_8 \equiv \bD_8^{F, T}$, which is a group with two generators satisfying the following relation: $\innerproduct{a,x}{a^4 = x^2 = 1, xax = a^{-1}}$. On the left end, $a = \mathcal{T}_L P_{F,L},~x = P_{F,L}$ and on the right, $a = \mathcal{T}_R,~x = P_{F,R}$.\footnote{{Readers may be puzzled by why that
the boundary $\mathcal{T}_L$ and $\mathcal{T}_R$ symmetries act oppositely
in comparison between those of \eq{eq:L-sym} and \eq{eq:R-sym}
versus those of \eq{eq:TLP} and \eq{eq:TRP}. 
Recall \Sec{subsec:gap-bdry-BDI}'s remark \ref{BDI-remark2} 
that
there are three different ways to look at $\mathcal{T}$ symmetries on the boundary depending on (1) the boundary truncation types of quantum systems and (2) the Hilbert space projection $\PGS$ to ground state subspace.
Here we do the remark \ref{BDI-remark2}'s (c) case,
with the $\PGS$ projection to dangling zero energy modes.
}
}
Observe that fermion parity is not in the center of $\bD_8^{F,T}$. This is going to be crucial. From now on, we focus only on the left boundary and drop the $\downarrow$ index. The effective symmetry can be simply written as
\begin{equation}
\cT = \gamma \cK,~~~ P_F = \ii \bar{\gamma} \gamma.
\end{equation} 
The symmetry acts on the Majorana operators trivially:
\begin{equation}
\cT: \begin{pmatrix}
\gamma \\
\bar{\gamma}
\end{pmatrix} \mapsto  \begin{pmatrix}
\gamma \\
\bar{\gamma}
\end{pmatrix}.
\end{equation}
However, it is anti-linear $\cT: \ii \mapsto -\ii$. As a result, the only possible 
boundary Hamiltonian term in this Hilbert space,
\begin{equation}
H { \propto - \ii  \gamma \bar{\gamma} = P_F},
\end{equation}
is disallowed by $\cT$ symmetry. Therefore, the only allowed boundary Hamiltonian {is proportional to the identity operator},
\begin{equation}
    H = c \mathds{1}.
\end{equation}
As shown in \cite{APJW-PRL}, with $c$ being positive (with no loss of generality), this has an $\cN=2$ SUSY with supercharges 
\begin{equation}
    Q_+ = \sqrt{c} \gamma,~Q_- = \sqrt{c} \bar{\gamma}.
\end{equation}
\begin{figure}[!htbp]
	\centering	
	\includegraphics[width=100mm]{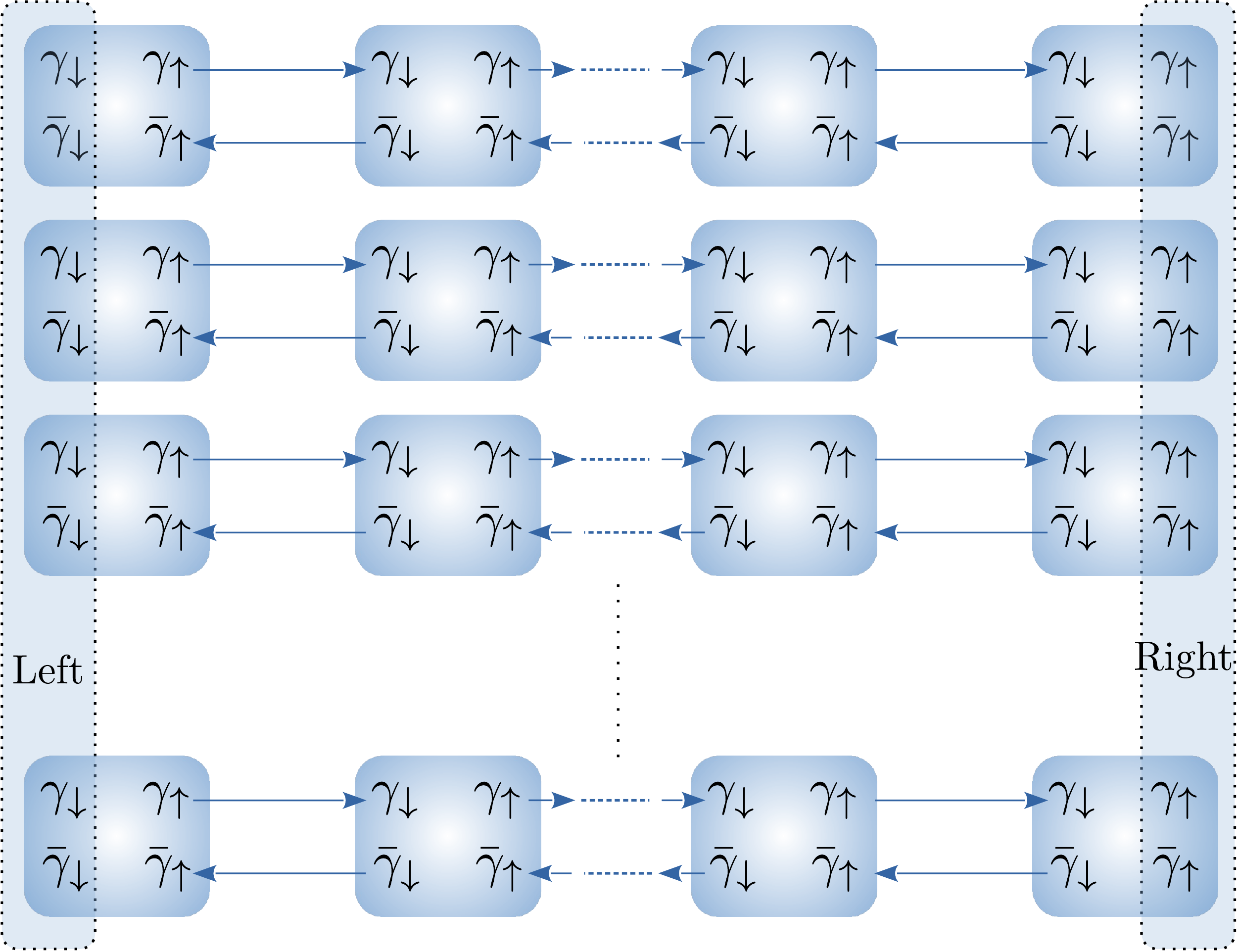}
	\caption{Representative of the $\nu=2$ phase with $\ztwo^T \times \ztwo^F$ symmetry consisting of multiple layers --- 
	{here $(4N+1)$ layers for the $\nu=2$ phase}.
	We can introduce SYK model Hamiltonian on the boundary, similar to the setup of \cite{YouLudwigXu_SYKSPT_PhysRevB.95.115150}.
	\label{fig:Layered_BDI}}
\end{figure}
We can increase the possible boundary dynamics by \emph{layering} trivial phases on the bulk and therefore increasing the boundary {by mod 8 layers of Kitaev chains}. Since the Hamiltonian \cref{eq:BDI} is the $\nu = 2 \mod 8$ member, this means that we can add multiples of four copies of the Hamiltonian~\cref{eq:BDI} to the bulk and preserve the same phase. The bulk Hamiltonian and symmetry operators correspond to this layered system shown in \Fig{fig:Layered_BDI} is 
\begin{align}
H &= -\ii \sum_{j=1}^{\ell} \sum_{a=1}^{4N + 1} \left(\gamma_{\uparrow,a,j} \gamma_{\downarrow,a,j+1}- \bar{\gamma}_{\uparrow,a,j} \bar{\gamma}_{\downarrow,a,j+1}\right) \label{eq:BDI_layer}. \\
\cT  &= 
\left(\prod_{a=1}^{4N+1}\prod_{j=1}^{\ell} 
(\gamma_{\downarrow,a,j} \bar{\gamma}_{\uparrow,a,j}) \right) \mathcal{K}.\\
P_F &= \left(\prod_{a=1}^{4N+1} \prod_{j=1}^{\ell}
\prod_{\varsigma = \uparrow, \downarrow}   \ii \bar{\gamma}_{\varsigma,a, j} \gamma_{\varsigma, j} \right). 
\end{align}
We now focus on the boundary. The Hilbert space can be acted by a total $8N+2$ Majorana operators: $\{\gamma_{a}, \bar{\gamma}_a\}$ for 
$a=1, \ldots, 4N+1$. The symmetry operators are
\begin{equation}
\cT = \prod_{a=1}^{4N+1} \gamma_a \cK, ~~~ P_F = \prod_{a=1}^{4N+1} \ii \bar{\gamma}_a \gamma_a.
\end{equation}
It can be checked that the operators again form a $\mathbb{D}_8^{F, T}$ group where the fermion parity is not at the center subgroup because $\cT$ anti-commutes with $P_F$. 
\begin{equation}
\cT^2 = P_F^2  = 1,~~~~ \cT P_F = - P_F \cT.
\end{equation} 
However, we can have a boundary Hamiltonian that is not quadratic, but can consist of Majorana operators coupled in multiples of 4:
\begin{multline}
H = \sum_{a,b,c,d}J^1_{a,b,c,d} \gamma_a \gamma_b \gamma_c \gamma_d + J^2_{a,b,c,d} \bar{\gamma}_a \gamma_b \gamma_c \gamma_d + J^3_{a,b,c,d} \bar{\gamma}_a \bar{\gamma}_b \gamma_c \gamma_d  + J^4_{a,b,c,d} \bar{\gamma}_a \bar{\gamma}_b \bar{\gamma}_c \gamma_d + J^5_{a,b,c,d} \bar{\gamma}_a \bar{\gamma}_b \bar{\gamma}_c \bar{\gamma}_d \\
{+K^1_{a,b,c,d,e,f,g,h} \gamma_a \gamma_b \gamma_c \gamma_d  \gamma_e \gamma_f \gamma_g \gamma_h + K^2_{a,b,c,d,e,f,g,h} \bar{\gamma}_a \gamma_b \gamma_c \gamma_d  \gamma_e \gamma_f \gamma_g \gamma_h +\dots}. \label{eq:SYK}
\end{multline}
{This Hamiltonian $H$ is time reversal invariant: $\cT H \cT^{-1} = H$.}
A shown in \cite{YouLudwigXu_SYKSPT_PhysRevB.95.115150}, if the couplings are restricted to the quartic operators and $J^{i}_{abcd}$ are drawn from a random distribution, this is the famous Sachdev-Ye-Kitaev (SYK) model 
(\cite{Sachdev1992fkSYK9212030, 
Kitaevtalk} 
also the later expositions in \cite{MaldacenaStanford1604.07818,KitaevSuh1711.08467}). 
The SUSY nature of the SYK model was shown by the authors of \cite{BehrendsBeri_SUSYSYK_PhysRevLett.124.236804}. 
{This SUSY nature still holds if we add any number of $4n$-Majorana operators (such as quartic, or octatonic as shown in \cref{eq:SYK})
to the Hamiltonian all of which preserve the $\cT$ and $P_F$ symmetry.} 
The logic is precisely what was sketched out in the previous subsection and detailed in \cite{BehrendsBeri_SUSYSYK_PhysRevLett.124.236804}: 
{once we have the anti-commutative $\cT P_F = - P_F \cT$, this implies that
we must have two-fold degenerate spectra from the equal number of bosonic sectors  $\{ \ket{\mu,+}\}$ and fermionic sectors  $\{ \ket{\mu,-}\}$
at every energy level $E_\mu$,
each with a distinct fermion parity $P_F=+1$ and $P_F=-1$.
We denote the number of states
\bea \label{eq:nBF}
n_{B,\mu} = n_{F,\mu} =1 
\eea
for each $E_\mu$.
}
By a constant shift, we can ensure that the Hamiltonian has positive definite energies and can be written as
\begin{equation}
H = \sum_{\mu} E_\mu (\outerproduct{\mu,+}{\mu,+} + \outerproduct{\mu,-}{\mu,-}),
\end{equation}
where
\begin{equation}
H \ket{\mu,\pm} = E_\mu \ket{\mu,\pm}, ~~~P_F \ket{\mu,\pm} = \pm \ket{\mu,\pm}.
\end{equation}
and $\cT$ switches the parity
\begin{equation}
\cT \ket{\mu,\pm} = \ket{\mu,\mp}.
\end{equation}
We can now define the two \emph{supercharges}
\begin{align}
Q_{+} &= \sum_{\mu} Q_\mu + Q^\dagger_\mu, ~~~~~Q_{-} =\sum_{\mu} {-} \ii \left( Q_\mu - Q^\dagger_\mu \right),
\end{align}
where $Q_\mu \equiv  \sqrt{E_\mu } \outerproduct{\mu,+}{\mu,-}$ and
$Q^\dagger_\mu \equiv \sqrt{E_\mu } \outerproduct{\mu,-}{\mu,+}$,
and check that the supercharges satisfy the algebra of \cref{eq:SUSY_algebra}
\begin{equation}
\{Q_\alpha,Q_\beta\} = 2H \delta_{\alpha \beta},~[Q_\alpha,H] = \{Q_\alpha,P_F \} =0
\end{equation}
This means that we have an $\cN = 2$ SUSY QM system on the boundary of these fermionic SPT phases. The action of $\cT$ on the supercharges is as shown in \cref{eq:TQ2} i.e.
\begin{equation}
\cT: Q_{\pm} \mapsto \cT Q_{\pm}\cT^{-1}=  + Q_\pm.
\end{equation} 

\subsection{{$\cN=2$ SUSY on the boundary of 1+1d fermionic SPT phase with 
$\Z_4^{TF}$ symmetry}}
\label{sec:SYK-SUSY-DIII}

In \Sec{sec:SYK-SUSY-BDI}, we had shown that the 
\Sec{sec:free BDI}'s model with a bulk $\Z_2^{T} \times \Z_2^{F}$-symmetric $8N+2$ layers of 1+1d Kitaev chains,
the boundary exhibits the $\cN=2$ SUSY quantum mechanics.
In this section, we can also take the \Sec{sec:free DIII}'s  model with a bulk $\Z_4^{TF}$-symmetric $8N+2$ layers of Kitaev chains,
to analyze its boundary energy spectrum property. 
Follow the Remark \ref{BDI-remark3} in \Sec{subsec:gap-bdry-BDI},
when the number of Kitaev chains is $2 \mod 4$ and when $\cT P_F = - P_F \cT$, we have at least 
two fold generate ground states (with GSD = 2), one bosonic $| B\rangle$ and one fermionic $| F\rangle$.
In general, we have the equal number of bosonic eigenstate $\{ \ket{\mu,+}\}$ and fermionic eigenstate $\{ \ket{\mu,-}\}$ 
for  each eigenvalue $E_\mu$. We again have 
$
n_{B,\mu} = n_{F,\mu} =1 
$ as in \eq{eq:nBF}
for each $E_\mu$.
Thus we can repeat the analysis \Sec{sec:SYK-SUSY-BDI} again reaching the same conclusion that 
we have $\cN=2$ SUSY quantum mechanics, but with a different bulk symmetry $\Z_4^{TF}$ and the boundary has
$\cT^4 = -1$ and $\cT^8 = +1$. This would also result in a different action on the supercharges.

\color{black}

\subsection{Systematic framework in 1+1d}

The original fractionalized global symmetry such as $\cT$
acts on the supercharges $Q_\alpha$,
thus the $\cT$ is within a subgroup of the R symmetry
that permutes the supercharges $Q_\alpha$. Note that
the usual R symmetry is a unitary symmetry, here we need to generalize
R symmetry to include both unitary and anti-unitary symmetry (to include time-reversal).

\begin{table}[!htb]
		\begin{center}
			\hspace{-6mm}
			\begin{tabular}{|c|c|c| c| c| c|c| c| }
				\hline
				Cartan class & $k=1$   &   $k=2$  &  $k=3$ &  $k=4$ &  $k=5$ & $k=6$ &  $k=7$
			    \\
				\hhline{========}
		  	  BDI ($\nu \in \Z_8$) & - & 
		  	  	$\begin{matrix}
		  	   \mathcal{N}=2 \cr
		  	  	{ (\nu=2)} 
				\end{matrix}$ 
		  	  & - &
		  	  	$\begin{matrix}
		  	   \mathcal{N}=0 \cr
		  	  	{ (\nu=4)} 
				\end{matrix}$ 
		  	  &-&
		  	  $\begin{matrix}
		  	   \mathcal{N}=2 \cr
		  	  	{ (\nu=6)} 
				\end{matrix}$ 
		  	  &-\\
			      \hline
		  	  DIII ($\nu \in \Z_2$)
		  	  & -
		  	  & $\begin{matrix}
		  	   \mathcal{N}=2 \cr
		  	  	{ (\nu=1)} 
				\end{matrix}$
				&-&& -&&-\\
					\hline
				 AI, CI, CII ($\nu \in \Z_2$) & - & 
		  	  & - &
		  	  	$\begin{matrix}
		  	   \mathcal{N}=0 \cr
		  	  	{ (\nu=1)} 
				\end{matrix}$ 
		  	  &-&
		  	  &-\\
				\hline
				 AIII ($\nu \in \Z_4$) & - & 
		  	  	$\begin{matrix}
		  	   \mathcal{N}=2 \cr
		  	  	{(\nu=1)} 
				\end{matrix}$ 
		  	  & - &
		  	  	$\begin{matrix}
		  	   \mathcal{N}=0 \cr
		  	  	{ (\nu=2)} 
				\end{matrix}$ 
		  	  &-&
		  	  $\begin{matrix}
		  	   \mathcal{N}=2 \cr
		  	  	{ (\nu=3)} 
				\end{matrix}$ 
		  	  &-\\
			      \hline
				\hline
			\end{tabular}
		\end{center}
		\caption{We show the relations between 
		the SUSY number $\cN$, 
		the global symmetry (Cartan symmetry class 
		or Wigner-Dyson-Altland-Zirnbauer symmetry class) 
		and their SPT classes $\nu$,
		and the number $k$ of Majorana zero modes.
		For $k \in  \Z_{{\text{odd}}}^+$, the - implies that those models are not studied in our framework. 
		The blanks mean those classes are either not the generators of the SPT classification,
		or they are already	determined by the smaller $k$ of SPT class (other filled data).}
		\label{Table:SUSY}
	\end{table}

In general, we only have the $\cN=2$ SUSY quantum mechanics
when  the layer number $k = 2 \mod 4$. We can summarize our result in \Table{Table:SUSY}.

We remark again that the fact that the SYK model can be thought of as a termination of fermionic SPT phases has been observed by You, Ludwig and Xu~\cite{YouLudwigXu_SYKSPT_PhysRevB.95.115150}. Also the SUSY QM property of the SYK model was studied by Behrends and B\'{e}ri~\cite{BehrendsBeri_SUSYSYK_PhysRevLett.124.236804}. 
 However, our interpretation of SUSY is tightly connected to the generalization of the symmetry extension.

\begin{table*}[!h] 

	{

		\hspace{-8mm}
			\begin{tikzpicture}\kern-13mm[>=stealth,->,shorten >=2pt,looseness=.5,auto]
			\matrix (M)[matrix of math nodes,row sep=8mm,column sep=5mm]{
				& 
				\begin{minipage}[c]{.85in} C: \ccblue{${\SU}(2)^F$},\\ 
				No class \\
				\ccred{$\Omega^2_{{\mathrm{Spin}^h}}$}
				\end{minipage} &    &  
				\begin{minipage}[c]{.58in} A: \ccblue{$\U(1)^F$},\\ 
				No class   \\
				\ccred{$\Omega^2_{{\mathrm{Spin}^c}}$}
				\end{minipage} 
				&    &  
				{\parbox{1.6cm}{
						\begin{minipage}[c]{.8in} D: \ccblue{$\Z_2^F$}, \\ 
						$(\nu_{\rm D})$ 
						$\in$  \\
						$\Z_2$-class \\
						Arf\\
						{\tiny 1 Kitaev chain}
						\\
						\ccred{$\Omega^2_{{\mathrm{Spin}}}$}
						\end{minipage} 
				}}
				\\
				\begin{minipage}[c]{1.in}
				CI: \\
				\ccblue{$\frac{{\SU}(2) \times \Z_4^{TF}}{\Z_2^F}$}, \\
				$(\alpha)$ 
				$\in$ 
				$\Z_2$-class\\
				 $w_1^2 \sim$ 4ABK\\
				 {\tiny 4 Kitaev chains}\\
				    \ccred{$\Omega^2_{\mathrm{Pin}^{+}\times_{\Z_2^F} \SU(2)^F}$}\\
				\end{minipage}  
				&   &  
				\begin{minipage}[c]{1.in}
				AI:\\
				\ccblue{$\U(1)^F   \rtimes \Z^{{T}}_{2}$},\\
				$(\alpha)\in$
				$\Z_2$-class\\
			        $w_1^2 \sim$ 4ABK\\
			        	{\tiny 4 Kitaev chains}\\
			        	 \ccred{$\Omega^2_{\mathrm{Pin}^{\tilde{c}-}}\equiv$}\\
			         \ccred{$\Omega^2_{\mathrm{Pin}^{-}\ltimes_{\Z_2^F} \U(1)^F}$}
				\end{minipage}
				&   &  
				{\parbox{1.7cm}{
						\begin{minipage}[c]{.78in}
						BDI:\\
						\ccblue{$\Z_2^T \times \Z_2^F$},\\
						$(\nu_{\rm BDI})$ 
						$\in$  \\
						$\Z_8$-class\\
						 ABK
						\\
							{\tiny 1 Kitaev chain}\\
						 \ccred{$\Omega^2_{\mathrm{Pin}^{-}}$}
						\end{minipage}
				}}\;\;\;
				&\\
				&   & 
				{\parbox{2.5cm}{
						{\begin{minipage}[c]{2.5cm}
							AIII:\\
							\ccblue{$\frac{\U(1)^{F}_{} \times \Z^{{TF}}_{4}}{\Z_2^F}=$ ${\U(1)^{F}_{} \times \Z^{{T}}_{2}}$},\\
							$(\nu_{\text{AIII}})$\\
							$\in\Z_4$-class\\
							 {2ABK}\\
							 {\tiny 2 Kitaev chains}\\
							 \ccred{$\Omega^2_{\mathrm{Pin}^{c}}$}
							\end{minipage} }
				}}\quad\quad\quad\quad
				&   & 
				{\parbox{2.cm}{
						\begin{minipage}[c]{1.23in}
						\rm{DIII}:
						\ccblue{$\Z_4^{TF} $},\\
						$(\nu_{\rm DIII})\in $\\ 
						$\Z_{2}$-class\; \\
						$\tilde{\eta}\PD(w_1)$\\
							{\tiny 2 Kitaev chains}\\
						 \ccred{$\Omega^2_{\mathrm{Pin}^{+}}$} 
						\end{minipage}
				}}
				\\
				{\parbox{3.cm}{
						\begin{minipage}[c]{1.1in}
						CII:\\
						\ccblue{$\frac{({\U}(1) \rtimes \mathbb{Z}_4^C)}{\mathbb{Z}_2^F} \times \mathbb{Z}_2^T$},\\
						$(\nu_{\text{CII}})$ \\
						$\in$  
						$\Z_2$-class\\
						 $w_1^2 \sim$ 4ABK\\
						{\tiny 4 Kitaev chains}\\
					       \ccred{$\Omega^2_{\mathrm{Pin}^{-}\times_{\Z_2^F} \SU(2)^F}$}\\
						\end{minipage}  
				}}
				&   & 
				\begin{minipage}[c]{1.2in}
				AII: \ccblue{$\frac{\U(1)^F   \rtimes \Z^{{TF}}_{4}}{\Z_2^F}$},\\
				No class\\
				\ccred{$\Omega^2_{\mathrm{Pin}^{\tilde{c}+}}\equiv$}\\
				\ccred{$\Omega^2_{\mathrm{Pin}^{+}\ltimes_{\Z_2^F} \U(1)^F}$}
				\end{minipage}  
				\\
			};
			\foreach \a/\b in {1-2/1-4, 1-4/1-6, 2-1/2-3, 2-3/2-5, 
				3-3/3-5, 4-1/4-3, 2-1/1-2, 2-3/1-4, 
				2-5/1-6, 3-3/2-5, 3-5/1-6, 4-1/1-2, 4-1/3-3, 
				4-3/3-5, 2-1/3-3, 3-3/1-4, 4-3/1-4} {
				\draw[thick,->](M-\a)--(M-\b);
			}
			\end{tikzpicture}
	}
	\caption{
		The symmetry embedding web of 1+1d fermionic invertible topological phases relevant for 10 Cartan symmetry classes 
		(we follow the notations of Table 2 in \cite{AP_Unwinding_PRB2018} and Table 4 in \cite{1711.11587GPW} for 3+1D cases).
		The web suggests the maps between the nontrivial classes of their classifications of SPTs (namely, cobordism invariants).
		The web can also suggest a possible symmetry group extension to unwind the SPT states. We include topological invariants from
		the 2d Pin$^-$ bordism invariant Arf-Brown-Kervaire (ABK), and
		1d spin bordism invariant $\tilde{\eta}$. The PD means the Poincar\'e dual.
		The ``$n$ Kitaev chains'' mean the generator of invertible phases of given symmetries.
	\label{table:web}}
\end{table*}

\newpage
\section{Generalizations: from the lattice to continuum}
\label{sec:conclusion1}

In \Sec{sec:conclusion1},
We will make some refinement 
about the group extension (for a quick overview see Appendix \ref{sec:extension-formal}) 
and properties of
{bosonic/fermionic/time-reversal/spacetime symmetries and group extensions}
(for a quick overview see Appendix \ref{sec:continuum-symmetry})
that we used.\footnote{We relegate the two subsections \ref{sec:extension-formal} and \ref{sec:continuum-symmetry} to Appendices
due to their heavy mathematical formality. However, the readers are strongly encouraged to read 
Appendix \ref{sec:extension-formal} and \ref{sec:continuum-symmetry} first before proceed to \Sec{sec:conclusion1}.}
In \Table{table:web}, we show
a symmetry embedding web of 1+1d fermionic invertible topological phases with their symmetry groups (relevant for 10 Cartan symmetry classes), and the number of layers of Kitaev chains to construct the minimal phase.
In \Table{Table:G}, we summarize the result of fermionic symmetry group extension presented
in Sec.~\ref{sec:free fermions} and \ref{sec:interacting BDI}, also those
in \cite{AP_Unwinding_PRB2018}, 
for fermionic SPT phases in 1+1d.

\begin{table}[!htb]
		\begin{center}
			\hspace{-6mm}
			\begin{tabular}{|c|c|c| c| c|}
				\hline
				Cartan class & Symmetry $G$ & $N \to \tilde{G} \to G$   & 
				$\begin{array}{c} \text{Reduced classification}\\ 
				\text{of 1+1d $G$-iTQFT or}\\
				\text{0+1d $G$-anomaly in $\tilde{G}$} \end{array}$\\
				\hhline{=====}
				\multirow{2}{*}{BDI} & \multirow{2}{*}{$\ztwo^T \times \Z_2^F$} & 
				$\Z_2 \to \mathbb{Z}_4^T \times \Z_2^F  \to G$ 
				& $\mathbb{Z}_8 \Rightarrow \mathbb{Z}_4$ \\ \cline{3-4}
			              &  & $\Z_2 \to \mathbb{D}_8^{F,T} \to G$  & $\mathbb{Z}_8 \Rightarrow \mathbb{Z}_2$ \\ 
			              	\hline
				DIII & $\Z_4^{TF}$ & $\Z_4 \to  \bM_{16}^{F,T} \to G$  & $\mathbb{Z}_2 \Rightarrow 0$ \\
				\hline
				\multirow{2}{*}{AIII} & \multirow{3}{*}{${\U}(1)^F \times \ztwo^T$} & $\Z_2 \to {\U}(1)^{{F}} \times \mathbb{Z}_4^T \to G$  & $\mathbb{Z}_4 \Rightarrow \ztwo$ \\
				\cline{3-4}
			         &  & $\Z_2 \to  \frac{\U(1) \times \bD_8^{F,T}}{\ztwo} \to G$  & $\mathbb{Z}_4 \Rightarrow 0$ \\
				\hline
				\multirow{2}{*}{CII} & $\frac{({\U}(1)^F \rtimes {\Z}_4^C)}{{\Z}_2^F} \times {\Z}_2^T $ & 
				$\Z_2 \to\frac{({\U}(1)^F \rtimes {\Z}_4^C)}{{\Z}_2^F} \times \mathbb{Z}_4^T \to G$  & $\ztwo \Rightarrow 0$ \\
							        \cline{2-4}
					 & $\SU(2)^F \times \ztwo^T $ & 
				$\Z_2 \to \SU(2)^F  \times \mathbb{Z}_4^T \to G$  & $\ztwo \Rightarrow 0$ \\
				\hline
			\end{tabular}
		\end{center}
		\caption{Summary of unwinding 1+1d fermionic $G$-SPT of $G$-invertible topological field theory (iTQFT) phases 
		via the symmetry extension $1  \to  N \to \tilde{G} \to G  \to  1$
		and the change of classification by symmetry extension.
		It turns out that the group extension of $G$ by $\Z_2$ has the $\Z_2= \Z_2^B$ as a bosonic symmetry acting on any state by multiplying a sign $\{1,-1\}$.
		So the $\Z_4^T$ in the $\tilde{G}$ is in fact $\Z_4^T=\Z_4^{TB}$ obtained from $1 \to \Z_2^B \to \Z_4^{TB} \to \Z_2^T \to 1$, which 
		by including (Lorentz to Euclidean) spacetime symmetry
		becomes 
		$1 \to \Z_2^B \to  \rE(d) \to \O(d) \to 1$, with the $ \rE(d) =\Z_2^B \rtimes  \O(d) =\SO(d) \rtimes \Z_4^{TB}$, see details in \Sec{sec:continuum-symmetry}.}
		\label{Table:G}
	\end{table}

\begin{table}[!htb]

			\hspace{-0mm}
			\begin{tabular}{|c|c|c| c| c|}
				\hline
				Cartan class & 
				$\begin{array}{c} \text{Spacetime-Internal}\\	
				\text{Symmetry $G_{\text{Tot}}$}		
				 \end{array}$
				& 
				$\begin{array}{c} \text{Symmetry Extension}\\	
				\text{to $\tilde{G}_{\text{Tot}}$}		
				 \end{array}$
				 & 
				$\begin{array}{c} \text{Reduced classification}\\ 
				\text{of 1+1d $G$-iTQFT or}\\
				\text{0+1d $G$-anomaly in $\tilde{G}$} \end{array}$\\
				\hhline{=====}
				\multirow{3}{*}{BDI} & \multirow{2}{*}{$\Pin^-$} & 
				$\EPin $ 
				& $\mathbb{Z}_8 \Rightarrow \mathbb{Z}_4$ \\ \cline{3-4}
			              &  & $\mathbb{D}_8^{F,T}\Pin $  & $\mathbb{Z}_8 \Rightarrow \mathbb{Z}_2$  \\
			              	\hline
				DIII & $\Pin^+$ & $\bM_{16}^{F,T}\Pin$  & $\mathbb{Z}_2 \Rightarrow 0$ \\
				\hline
				\multirow{2}{*}{AIII} & \multirow{3}{*}{
				$\begin{array}{r} 
				\Pin^c=\Pin^- \times_{\Z_2^F} \U(1)^F\\
				=\Pin^+ \times_{\Z_2^F} \U(1)^F
				\end{array}
				$} & $\rE\Pin \times_{\Z_2^F} \U(1)^F $  & $\mathbb{Z}_4 \Rightarrow \ztwo$ \\
				\cline{3-4}
			         &  & 
			         $ { \bD_8^{F,T}\Pin} \times_{\Z_2^F} \U(1)^F$ 
			         & $\mathbb{Z}_4 \Rightarrow 0$ \\			       
				\hline
				\multirow{2}{*}{CII} & $\Pin^- \times_{\Z_2^F} \frac{({\U}(1)^F \rtimes {\Z}_4^C)}{{\Z}_2^F}  $ & 
				$ \rE\Pin \times_{\Z_2^F} \frac{({\U}(1)^F \rtimes {\Z}_4^C)}{{\Z}_2^F} $  & $\ztwo \Rightarrow 0$ \\
							        \cline{2-4}
					 & $\Pin^- \times_{\Z_2^F} \SU(2)^F $ & 
				$\rE\Pin \times_{\Z_2^F} \SU(2)^F $  & $\ztwo \Rightarrow 0$ \\
				\hline
			\end{tabular}
		\caption{The \emph{continuum} version description of
		\Table{Table:G}. The extended $\tilde G_{\text{Tot}}$
		includes the EPin group (firstly introduced in \Refe{WanWWZ1912.13504} and reviewed in \Sec{sec:BDI-4in8}),
		the  $\bD_8^{F,T}\Pin$ and $\bM_{16}^{F,T}\Pin$
		 (that we will introduce in \Sec{sec:D8Pin} and \Sec{sec:M16Pin}).
		}
		\label{Table:Gspacetime}
	\end{table}

We also summarize the bundle constraints 
for various symmetry groups and corresponding spacetime structures in \Table{table:SW-bundle-constraints} following \Refe{WanWWZ1912.13504}.
\begin{table}[!h]
	\centering
	\begin{tabular}{|c | c  c  c |}
		\hline
		Group & $w_1$ & $w_1^2$ & $w_2$ \\ 
		\hline
		$\SO(d)$ & 0 & 0 & free \\
		$\Spin(d)$ & 0 & 0 & 0 \\
		$\O(d)$ & free & free & free \\
		$\rE(d)$ & free & 0 & free \\
		$\Pin^+(d)$ & free & free & 0 \\
		$\Pin^-(d)$ & free & \multicolumn{2}{c|}{$w_1^2+w_2=0$} \\
		$\EPin(d)$ & free & 0 & 0 \\
		\hline
	\end{tabular}
	\caption{Follow 	\Refe{WanWWZ1912.13504}, we list down the (spacetime) bundle constraints for various spacetime structures.
	Here we denote $w_j\equiv w_j(TM)$
	are the $j$-th Stiefel-Whitney (SW) classes of the tangent bundle $TM$ of the spacetime manifold $M$.
	The product of cohomology classes
	as $w_1^2  \equiv     w_1(TM)\cup w_1(TM)$
	is the cup product. The free here means no restriction which can be a nontrivial cohomology class.
	}
	\label{table:SW-bundle-constraints}
\end{table}

In the following subsections, we show by examples, based on the knowledge of 
Appendix \ref{eq:short-exact-sequence-G} and \ref{sec:continuum-symmetry},
how to convert the discrete group version of
symmetry extensions (\Table{Table:G}) to the continuous group version of
symmetry extensions (\Table{Table:Gspacetime}), suitable for the 
continuum quantum field theory [QFT]).

\subsection{1+1d:
Lift $\Pin^-(d)$ to
$\rE\Pin(d)$ for $\nu_{\rm BDI} =4 \in  \Z_8$ class}
\label{sec:BDI-4in8}

The symmetry extension $\Z_2 \to \tilde{G}=\mathbb{Z}_4^T \times \Z_2^F  \to {G} =\mathbb{Z}_2^T \times \Z_2^F$
in \Table{Table:G},  reduces BDI class from the interacting $\Z_8$ classification to $\Z_4$ classification.
The $ \tilde{G}=\mathbb{Z}_4^T \times \Z_2^F$ 
in fact can have at least three physical interpretations of the symmetry realization, which in turn give also  three physical interpretations of $\Z_2 \to \rE\Pin(d) \to\Pin^-(d)$.
The $\rE\Pin(d)$ is firstly introduced by \Refe{WanWWZ1912.13504}.\footnote{Below we encounter and define three versions of EPin called
$\rE\Pin^{BF_-}(d)$, $\rE\Pin^{BF_+}(d)$, and $\rE\Pin^{F_+F_-}(d)$. In fact the $\rE\Pin^{F_+F_-}(d)$ is denoted as $\rE\Pin^{}(d)$ in \Refe{WanWWZ1912.13504}.
It is shown in \Refe{WanWWZ1912.13504} and in this section that the three versions of EPin are the same group.}
The three interpretations are:  
\begin{enumerate}[leftmargin=2.mm, label=\textcolor{blue}{(\arabic*)}., ref={(\arabic*)}]
\item One interpretation of \Table{Table:G}'s involves a $\cT^2=-1$ boson $B$ and a $\cT^2=1$ fermion $F_-$:
$$\Z_2^B \to \mathbb{Z}_4^{TB} \times \Z_2^{F_-} \to \mathbb{Z}_2^T \times \Z_2^{F_-},$$ 
by including SO$(d)$ which becomes \Table{Table:Gspacetime}'s
\bea \label{eq:EPinBF-}
\Z_2^B \to   \rE\Pin^{BF_-}(d)\equiv \Big( (\Z_2^{F_-} \rtimes \SO(d)) \times  \Z_2^B\Big)  \rtimes \mathbb{Z}_2^{T_{\rm E}} \to \Pin^-(d).
\eea
Note that 
 $ \rE\Pin^{BF_-}(d)\equiv\Big( (\Z_2^{F_-} \rtimes \SO(d)) \times  \Z_2^B\Big)  \rtimes \mathbb{Z}_2^{T_{\rm E}}=
 \Big( \Spin(d)\times  \Z_2^B\Big)  \rtimes \mathbb{Z}_2^{T_{\rm E}}$ has the properties \emph{specifically} here:
\begin{itemize}
\item $(\Z_2^{F_-}\rtimes \SO(d))  \rtimes \mathbb{Z}_2^{T_{\rm E}}=\Spin(d)  \rtimes \mathbb{Z}_2^{T_{\rm E}}= \Pin^-(d)$.
\item In Euclidean signature: {$\Z_2^B  \rtimes \mathbb{Z}_2^{T_{\rm E}}= \Z_4^{{T_{\rm E}}B}$}
and $\Z_2^{F_-}  \rtimes \mathbb{Z}_2^{T_{\rm E}}=\Z_4^{{T_{\rm E}} F_-}$.
\item In Lorentzian signature: $\Z_2^B  \rtimes \mathbb{Z}_2^T= \Z_4^{TB}$
and $\Z_2^{F_-}  \rtimes \mathbb{Z}_2^T=\mathbb{Z}_2^T  \times  \Z_2^{F_-} $.
\item $\big(  \SO(d) \times  \Z_2^B\big)  \rtimes \mathbb{Z}_2^{T_{\rm E}}
=  \Z_2^B \rtimes \O(d) 
= \SO(d) \rtimes  \mathbb{Z}_4^{T_{\rm E}B} =\rE(d).$
\end{itemize}

\item One interpretation of \Table{Table:G}'s  involves two types of fermions, $F_+$ with $\cT^2=-1$ and $F_-$ with $\cT^2=1$:
$$\Z_2^{F_+} \to \mathbb{Z}_4^{TF_+} \times \Z_2^{F_-}  \to \mathbb{Z}_2^T \times \Z_2^{F_-},$$ by including SO$(d)$ 
which becomes \Table{Table:Gspacetime}'s
\bea  \label{eq:EPinF}
\Z_2^{F_+} \to   \rE\Pin^{F}(d) \equiv
 \Big( \Z_2^{F_+}  \times \Z_2^{F_-} \Big) \rtimes \Big( \SO(d) \rtimes \mathbb{Z}_2^T \Big) 
  \to \Pin^-(d).
\eea
Note that 
 $ \rE\Pin^{F}(d)\equiv
  \Big( \Z_2^{F_+}  \times \Z_2^{F_-} \Big) \rtimes \Big( \SO(d) \rtimes \mathbb{Z}_2^T \Big) $ has the properties  \emph{specifically} here:
\begin{itemize}
\item 
$ \Z_2^{F_+}   \rtimes \SO(d) \cong \Z_2^{F_-}   \rtimes \SO(d) \cong \Spin(d)$.
\item In Euclidean signature: $\Z_2^{F_+}  \rtimes \mathbb{Z}_2^{T_{\rm E}}= \Z_2^{F_+}  \times \mathbb{Z}_2^{T_{\rm E}}$ and  
$\Z_2^{F_-}  \rtimes \mathbb{Z}_2^{T_{\rm E}}=\Z_4^{{T_{\rm E}}F_-} $.
\item In Lorentzian signature: $\Z_2^{F_+}  \rtimes \mathbb{Z}_2^T= \Z_4^{TF_+}$ and  $\Z_2^{F_-}  \rtimes \mathbb{Z}_2^T=\mathbb{Z}_2^T  \times  \Z_2^{F_-} $.
\item $ \Z_2^{F_{\pm}}   \rtimes \Big( \SO(d) \rtimes \mathbb{Z}_2^{T_{\rm E}} \Big)=\Pin^{\pm}(d)$.
\end{itemize}

\item Another interpretation of \Table{Table:G}'s  involves a $\cT^2=-1$ boson $B$ and a $\cT^2=-1$ fermion $F_+$:
$$\Z_2^{F_+} \to \frac{  \mathbb{Z}_4^{TB} \times \mathbb{Z}_4^{TF_+}}{\Z_2^{T}}  \to \mathbb{Z}_2^T \times \Z_2^{F_-},$$ 
by including SO$(d)$ which becomes \Table{Table:Gspacetime}'s
\bea  \label{eq:EPinBF+}
\Z_2^B \to   \rE\Pin^{BF_+}(d)\equiv \Big( (\Z_2^{F_+} \rtimes \SO(d)) \times  \Z_2^B\Big)  \rtimes \mathbb{Z}_2^{T_{\rm E}} \to \Pin^-(d).
\eea
Note that 
 $ \rE\Pin^{BF_+}(d)\equiv \Big( (\Z_2^{F_+} \rtimes \SO(d)) \times  \Z_2^B\Big)  \rtimes \mathbb{Z}_2^{T_{\rm E}} $ has the properties  \emph{specifically} here:
 \begin{itemize}
\item 
$\Pin^+(d) =
(\Z_2^{F_+}\rtimes \SO(d))  \rtimes \mathbb{Z}_2^{T_{\rm E}}=\Spin(d)  \rtimes \mathbb{Z}_2^{T_{\rm E}}=  \frac{\Spin(d)\rtimes \mathbb{Z}_{4}^{T_{\rm E}F}}{\mathbb{Z}_{2}^F} $.
\item In Euclidean signature: {$\Z_2^B  \rtimes \mathbb{Z}_2^{T_{\rm E}}= \Z_4^{{T_{\rm E}}B}$}
and $\Z_2^{F_+}  \rtimes \mathbb{Z}_2^{T_{\rm E}}=\mathbb{Z}_2^{T_{\rm E}}  \times  \Z_2^{F_+} $.
\item In Lorentzian signature: $\Z_2^B  \rtimes \mathbb{Z}_2^T= \Z_4^{TB}$
and $\Z_2^{F_+}  \rtimes \mathbb{Z}_2^T=\Z_4^{TF_+}$. 
\item $\big(  \SO(d) \times  \Z_2^B\big)  \rtimes \mathbb{Z}_2^{T_{\rm E}}
=  \Z_2^B \rtimes \O(d) 
= \SO(d) \rtimes  \mathbb{Z}_4^{T_{\rm E}B} =\rE(d).$\end{itemize}
\end{enumerate}
\Refe{WanWWZ1912.13504} finds that the three interpretations and expressions
 \eq{eq:EPinBF-}, (\ref{eq:EPinF}), and (\ref{eq:EPinBF+}) are equivalent by relabeling the group elements, so we can generally write all of them as the
same as the $\rE\Pin^{}$ extension:
\bea  \label{eq:EPin}
\Z_2 \to   \rE\Pin^{}(d) \to \Pin^-(d).
\eea
Our claim is that the 1+1d $\mathbb{Z}_2^T  \times  \Z_2^{F}$ fermionic SPTs  classified by $\Omega_2^{\Pin^-}=\Z_8$ with 
$\nu_{\rm BDI} \in \Z_8$
class 2d topological invariants over a 2-manifold ${M^2}$ with $\Pin^-$ structure
\bea \label{eq:ABK}
\exp(\ii \frac{2 \pi \nu_{\rm BDI} }{ 8} \text{ABK})\vert_{M^2}
\eea
can be trivialized at $\nu_{\rm BDI} = 4$ by an extension from $\Pin^-$  to $\rE\Pin$.
The reason is that 4ABK (4 layers of Kitaev fermionic chain) equals to the topological invariant of the first Stiefel-Whitney classes 
$w_1^2 \equiv w_1(TM)^2$ (1 layer of Haldane spin chain) 
\cite{AP_Unwinding_PRB2018}.\footnote{Here we denote 
$w_j\equiv w_j(TM)$
	are the $j$-th Stiefel-Whitney (SW) classes of the tangent bundle $TM$ of the spacetime manifold $M$. 
	 For the SW class for the vector bundle $V_G$ associated with certain group $G$ (where $G$ can be gauge group or global symmetry group), 
	 we denote it as $w_j(V_G)$ or simply $w_j(G)$. 
All the product notations between cohomology classes are cup products, such as $w_i w_j  : =   w_i(TM)w_j(TM)=  w_i(TM)\cup w_j(TM)$.
	 }
	 Namely, the two topological invariants are related
\bea \label{eq:ABK=4}
\exp(\ii \frac{2 \pi 4 }{ 8} \text{ABK})\vert_{M^2} \simeq \exp(\ii \pi (w_1)^2)\vert_{M^2}
\eea 
up to a trivially gapped fermionic sector specifying spin structures.

Let us show the \Eq{eq:EPin}'s EPin can trivialize the \Eq{eq:ABK=4}.
First, the $\rE\Pin$ structure requires both $\Pin^+$ and $\Pin^-$ to hold, while
 $\Pin^+$ requires $w_2=0$ and
  $\Pin^-$  requires $w_2+ w_1^2=0$,
  so $\rE\Pin$ requires $w_2=w_1^2=0$,
  but all can have nontrivial $w_1$ to be on unorientable manifolds.
(See \Refe{WanWWZ1912.13504} and \Table{table:SW-bundle-constraints} for a summary of the bundle constraints 
for various symmetry groups and corresponding spacetime structures.)
Second, \Eq{eq:ABK=4} also implies $ \exp(\ii \pi (w_1)^2)= \exp(\ii \pi w_2)$ in 2d $\Pin^-$,
while both expressions become trivial as 1 in $\rE\Pin$.
Thus EPin as a symmetry extension of Pin$^-$ trivializes \eq{eq:EPin}  for  $\nu_{\rm BDI} = 4$.

We can also write down how the symmetry operator S of EPin$(d)$ group acts on the state vectors of Hilbert space 
$$\cH=\cH_B \oplus \cH_{F_+}\oplus \cH_{F_-},$$
spanned by three sectors, the bosonic sector $|B\rangle $ with $T^2=+1$,
 the fermionic sector $|F_+\rangle $ with $T^2=-1$,
and the fermionic sector $|F_-\rangle $ with $T^2=+1$.
Naively one looks for the larger symmetry including 
$\O(d)$,
$\Pin^+(d)$,
and $\Pin^-(d)$ for three sectors in the continuum,
in a more general manifold on $M^{d+1}$.
But for quantum mechanics, it is more practical to define the Hilbert space on a flat space on a $\mathbb{R}^{d-1}$ with a continuous time $t$ on $\mathbb{R}^1$.
Then we find the operator S acts on state vectors living on the 
$\mathbb{R}^{d-1}$ generating the following groups for each sector:
\bea
\begin{pmatrix}
\O(d-1)\times\Z_2^T  & 0 &0\\
0& \Pin^+(d-1)\times_{\Z_2^F}\Z_4^{TF}  & 0 \\ 
0 & 0  & \Pin^-(d-1)\times \Z_2^{T} 
\end{pmatrix} \text{ acts on }
\begin{pmatrix*}[l]
|B\rangle  \\
|F_+\rangle  \\
|F_-\rangle
\end{pmatrix*}. 
\label{eq:EPin-H}
\eea
Note that $\Z_2^{F_+}$ and $\Z_2^{F_-}$ have the two generators
$$
\begin{pmatrix}
1  & 0 &0\\ 
0  & -1 &0\\ 
0  & 0 & 1
\end{pmatrix} \text{ and }
\begin{pmatrix}
1  & 0 &0\\ 
0  & 1 &0\\ 
0  & 0 & -1
\end{pmatrix}.
\text{ act on }
\begin{pmatrix*}[l]
|B\rangle  \\
|F_+\rangle  \\
|F_-\rangle
\end{pmatrix*},
$$
each sits at the $\Z_2$ center of Spin group within Pin$^+$ and Pin$^-$ groups for each. If we choose $|B\rangle $ with $T^2=-1$, we must change
\eq{eq:EPin-H}'s $\O(d-1)\times\Z_2^T$ to 
$\rE(d-1)\times_{\Z_2^B} \Z_4^{TB}$.
This \eq{eq:EPin-H} is our interpretation of $\EPin(d)$ acting on the assumed Hilbert space.

In summary,
 \Eq{eq:EPin} on $\Z_2 \to \rE\Pin(d) \to\Pin^-(d)$ 
 gives the mathematical way to understand the unwinding of 4 layers of Kitaev chains
 via the symmetry extension $\Z_2 \to \mathbb{Z}_4^T \times \Z_2^F  \to \mathbb{Z}_2^T \times \Z_2^F$
demonstrated in \Refe{AP_Unwinding_PRB2018}'s Sec.~4.

\subsection{1+1d: 
Lift $\O(d)$ to $\rE(d)$ for $\nu_{\rm CII} =1 \in  \Z_2$ or even $\nu_{\rm AIII} \in  \Z_4$ class}

\label{sec:CII-AIII-2in4}

In fact, the previous extension method in \Sec{sec:BDI-4in8} also works for 
 1+1d SPTs with the same topological invariants $ \exp(\ii \pi (w_1)^2)$
but with other symmetry groups of \Table{Table:G}, such as CII and AIII classes. In the discrete version, we have
$\Z_2 \to \mathbb{Z}_4^T  \to \mathbb{Z}_2^T$,
converting to a continuum language becomes 
$\Z_2 \to \rE(d) \to\O(d)$. 

For example, \Sec{sec:CII}'s CII class symmetry extension \Eq{eq:CII-1ext} and \Eq{eq:CII-2ext}, 
and AIII even $\nu_{\rm AIII} \in  \Z_4$ class studied in \Refe{AP_Unwinding_PRB2018}'s Sec.~4,
can be understood by including the spacetime symmetry in the continuum as 
the following extensions $N \to \tilde{G}_{\text{Tot}} \to G_{\text{Tot}}$:
\bea
\begin{array}{rcccccccc}
\text{CII: }&\Z_2 &\to& \rE\Pin \times_{\Z_2^F} \SU(2)^F &\to& \Pin^- \times_{\Z_2^F} \SU(2)^F.\\
&\Z_2 &\to& \rE\Pin \times_{\Z_2^F} \frac{({\U}(1) \rtimes {\Z}_4^C)}{{\Z}_2^F} &\to&  \Pin^- \times_{\Z_2^F} \frac{({\U}(1) \rtimes {\Z}_4^C)}{{\Z}_2^F}.\\
\text{AIII: }&\Z_2 &\to& \rE\Pin \times_{\Z_2^F} \U(1)^F &\to& \Pin^- \times_{\Z_2^F} \U(1)^F.\\
\text{In pinciple: }&\Z_2 &\to& \rE(d) &\to&  \O(d).\\
\end{array}
\eea
Here $d=2$. The above extension works because the $ \exp(\ii \pi (w_1)^2)$ is a nontrivial topological invariant in $\Omega_2^{\O}$  
but becomes trivial by pulling it back to $\Omega_2^{\rE}$ due to the constraint $(w_1)^2=0$ for ${\rE}(d)$ structure, 
see Table \ref{table:SW-bundle-constraints}. More generally, the cocycle $(w_1)^2$ becomes a coboundary, thus 
$(w_1)^2=\delta \beta_1$, which splits into one lower dimensional 1-cochain $\beta_1$ under the pullback trivialization.\footnote{More
generally in a multiplicative notation, in
terms of 2-cocycle with U(1) coefficient: 
$\omega_2=(-1)^{\int (w_1)^2}$, 
we mean the $\omega_2 = \delta \alpha_1$
can split into one lower dimensional 1-cochain $\alpha_1$ under the pullback trivialization.}

\subsection{1+1d supersymmetry extension: 
Lift $\Pin^-$ to 
$\mathbb{D}_8^{F,T} \Pin$
for even $\nu_{\rm BDI} \in  \Z_8$ class}

\label{sec:D8Pin}

We have given a discrete group interpretation of $\mathbb{D}_8^{F,T}$ of
\Table{Table:G} and the extension:
$$\Z_2^B \to \big(\mathbb{D}_8^{F,T} \equiv \mathbb{Z}_4^{T} \rtimes \Z_2^{F} \big)\to \mathbb{Z}_2^T \times \Z_2^{F}.
$$ 
Now we will like to include  the continuous $\SO(d)$ sector 
in order to
generalize our formulation to the continuum QFT.
Note that 
 $ \mathbb{D}_8^{F,T} \Pin(d)$ 
 contains the ingredients 
\bea
\left\{\begin{array}{l}
\Z_2^{F}\rtimes \SO(d) =\Spin(d),\\
\SO(d) \rtimes  \mathbb{Z}_4^{T_{}B} =\rE(d), \\
\mathbb{D}_8^{F,T} \equiv \mathbb{Z}_4^{T} \rtimes \Z_2^{F}.
\end{array}\right.
\eea
However, there is a \emph{caveat}:  
The short exact sequence 
$$1 \to \Z_2 \to \tilde{G}_{\text{Tot}}  \to\Pin^-(d) \to 1$$
is classified by $\tilde{G}_{\text{Tot}} = \Z_2  \rtimes_{\rho,\varphi} \Pin^-(d)$
controlled by no $\rho \in G_{\text{Tot}} \to \text{Aut}(\mathbb{Z}_2)=0$,
but only by a 2-cocycle 
$\varphi \in \H^2(\B\Pin^-(d) ,\Z_2)=\Z_2$:
\begin{itemize}
\item
The trivial $\varphi$ gives $\tilde{G}_{\text{Tot}} = \Z_2  \times \Pin^-(d) $. 
\item The nontrivial $\varphi =w_1^2$ gives the $\tilde{G}_{\text{Tot}} = \EPin(d)$.\\[-10mm]
\end{itemize}
So naively $\tilde{G}_{\text{Tot}}$ has only two solutions,
$\Z_2  \times \Pin^-(d) $
and $\EPin$, there is \emph{no} $\mathbb{D}_8^{F,T} \Pin$.
How can we make sense the extension
from 
$\Pin^-(d)$ to $\mathbb{D}_8^{F,T} \Pin(d)$?\footnote{We
should clarify that our $\mathbb{D}_8^{F,T} \Pin(d)$ is  
distinct from the
$\DPin(d)$ studied in \cite{Kaidi2019pzjJulioTachikawa1908.04805,Kaidi2019tyfJulioTachikawa1911.11780}.
The $\mathbb{D}_8^{F,T} \Pin(d)$ contains a $\mathbb{D}_8^{F,T}\equiv \mathbb{Z}_4^{T} \rtimes \Z_2^{F}$, where
fermion parity is not at the $\mathbb{D}_8^{F,T}$'s center. 
The $\DPin(d)$ contains a different $\mathbb{D}_8$ which is
$\mathbb{D}_8= (\Z_2^{F_+} \times \Z_2^{F_-})\rtimes_{\rho,0} \Z_2^T$
where $\rho$ is a nontrivial $\Z_2^T$ action on 
${\rm Aut}(\Z_2^{F_+} \times \Z_2^{F_-})$ with two kinds of fermion parity 
$\Z_2^{F_+} \times \Z_2^{F_-}$ at the
$\mathbb{D}_8$'s center. 
In fact, in this article, we will not pursue a Euclidean spacetime realization of
$\mathbb{D}_8^{F,T} \Pin(d)$. 
Since we do not specify the full $d$d spacetime rotational part,
we will denote it as $\mathbb{D}_8^{F,T} \Pin$ from now on.
We shall only discuss the $\mathbb{D}_8^{F,T} \Pin$ symmetry generator acts on the states in the Hilbert space.
}

The answer comes to the subtle fact that the time-reversal
$\cT$ in $\mathbb{D}_8^{F,T}$ has to switch between
bosonic and fermionic sectors, 
thus a \emph{supersymmetry extension} comes in.
We can write down how the symmetry operator S of $\mathbb{D}_8^{F,T} \Pin$ 
group acts on the state vectors of Hilbert space 
$$\cH=\cH_B \oplus \cH_{F},$$
spanned by two sectors, the bosonic sector $|B\rangle $ with $T^2=-1$,
 the fermionic sector $|F\rangle$ with $T^2=-1$.
Again it is practical to define the Hilbert space on a flat space on a $\mathbb{R}^{d-1}$ with a continuous time $t$ on $\mathbb{R}^1$.
Then we find the S acts on state vectors living on the 
$\mathbb{R}^{d-1}$ generating the following groups for each sector:\footnote{
We 
represent the complex and Majorana fermionic operators
acting on 2-dimensional Hilbert space $\begin{pmatrix*}[l]
|B\rangle  \\
|F\rangle 
\end{pmatrix*}$ as:
\bea
&&\psi\equiv
\begin{pmatrix}
0 & 1 \\
0 & 0 
\end{pmatrix}
\equiv\sigma_-
\equiv\frac{(\sigma_x+ \ii \sigma_y)}{2}
,\quad
\psi^\dagger \equiv
\begin{pmatrix}
0 & 0 \\
1 & 0 
\end{pmatrix}
\equiv\sigma_+
\equiv\frac{(\sigma_x- \ii \sigma_y)}{2},
\quad
F \equiv N_{F} \equiv 
\psi^\dagger_{} \psi_{}  
\equiv\sigma_+\sigma_-
\equiv
\begin{pmatrix}
0 & 0 \\
0 & 1 
\end{pmatrix}.\quad\quad\cr
&&{\gamma}=
\begin{pmatrix}
0 & 1 \\
1 & 0 
\end{pmatrix}
=\sigma_x,
\quad
\bar{\gamma}=
\begin{pmatrix}
0 & -\ii \\
\ii & 0 
\end{pmatrix}
=\sigma_y,
\quad
P_F = (-1)^F =
\ii
\bar{\gamma}
{\gamma}=
-\ii
{\gamma}\bar{\gamma}=
\begin{pmatrix}
1 & 0 \\
0 & -1 
\end{pmatrix}
=\sigma_z .
\eea
The following matrix representations of higher-dimensional Hilbert spaces are the direct parallel story of 2-dimensional Hilbert space.  
}
\begin{enumerate}
\item The unitary $\Z_2^F$ generator is
\bea \label{eq:Pf-matrix}
P_F = (-1)^F =\begin{pmatrix}
1 & 0\\
0& -1  
\end{pmatrix} =\sigma_z \text{, which acts on }
\begin{pmatrix*}[l]
|B\rangle  \\
|F\rangle 
\end{pmatrix*}. 
\eea
We omit a square identity matrix $\mathds{1}$ that acts on the whole Hilbert space of 
$|B\rangle$ and $|F\rangle$ (fermionic and bosonic sectors 
have equal Hilbert space dimensions),
so precisely $P_F = \sigma_z \otimes \mathds{1}$.

\item The anti-unitary $\Z_4^T$ generator, with complex conjugation $K$, is
\bea
\cT =\begin{pmatrix}
0 & - \ii \\
\ii & 0
\end{pmatrix} K 
=\sigma_y K 
\text{, which acts on }
\begin{pmatrix*}[l]
|B\rangle  \\
|F\rangle 
\end{pmatrix*}, 
\label{eq:D8Pin-T}
\eea
or precisely $\cT = (\sigma_y   \otimes \mathds{1}) K$.
The $\cT$ switches between $|B\rangle$ and $|F\rangle$. 

\item The $\mathbb{D}_8^{F,T} \equiv \mathbb{Z}_4^{TB} \rtimes \Z_2^{F_-}$
is generated by $\{\cT, P_F |\cT^4=P_F^2=1,  
P_F \cT P_F = \cT^3 = -\cT \}$. Note that
 $\cT^2=-1$ for all bosonic and fermionic states.

\item The Spin group has elements $\rm{S}_{\rm{rot}}$ acting on the Hilbert space associated with a $\mathbb{R}^{d-1}$ space:
\bea
\rm{S}_{\rm{rot}}=
\begin{pmatrix}
\SO(d-1) & 0 \\
0& \Spin(d-1)   
\end{pmatrix} \text{ acts on }
\begin{pmatrix*}[l]
|B\rangle  \\
|F\rangle 
\end{pmatrix*}. 
\label{eq:D8Pin-rot}
\eea
The $P_F$ is the $\Z_2$ center of Spin group.
\end{enumerate}
Combining the group elements $P_F$, $\cT$, and $\rm{S}_{\rm{rot}}$ together,
we generate the full structure of $\mathbb{D}_8^{F,T} \Pin$ that acts on the
assume Hilbert space 
$\cH=\cH_B \oplus \cH_{F}$ associated with the $\mathbb{R}^{d-1}$ space.
We name such an extension as a \emph{supersymmetry extension} when we require 
some of the group elements (here $\cT$) permutes 
bosonic and fermionic sectors. Formally, we propose that
by pulling back $\Pin^-$ to $\mathbb{D}_8^{F,T} \Pin$ via the  \emph{supersymmetry extension},
we can trivialize the even $\nu \in \Omega_2^{\Pin^-} =\Z_8$ whose cobordism invariant \eq{eq:ABK} can be written as:
\bea
\label{eq:ABK=2}
\exp(\ii \frac{2 \pi 2 }{ 8} \text{ABK})\vert_{M^2} \simeq \exp(\ii \frac{2 \pi  }{4} \eta'\PD(w_1(TM)))\vert_{M^2}.
\eea
The $\eta'$ is the 1d cobordism invariant of $\Omega_1^{\Spin \times_{\Z_2} \Z_4}=\Z_4$, which is a $\Z_4$ extended
version of the $\tilde \eta$ (which is the 1d invariant of $\Omega_1^{\Spin}=\Z_2$).
See more discussions about cobordism invariants in \Sec{sec:1+1d}.

\subsection{1+1d supersymmetry extension:
Lift $\Pin^+$ to
$\mathbb{M}_{16}^{F,T} \Pin$
for odd $\nu_{\rm DIII} \in  \Z_2$ class}
\label{sec:M16Pin}

We have given a discrete group interpretation of $\mathbb{M}_{16}^{F,T}$ of
\Table{Table:G}
via an extension:
$$\Z_4 \to \big(\mathbb{M}_{16}^{F,T}\equiv \mathbb{Z}_4^{TF} \rtimes \Z_4^{} \big)\to \mathbb{Z}_4^{TF}.
$$ 
Similar to \Sec{sec:D8Pin},
we  write down how the symmetry operator S of $\mathbb{M}_{16}^{F,T} \Pin(d)$ 
group acts on the state vectors of Hilbert space 
$$\cH=\cH_B \oplus \cH_{F},$$
spanned by two sectors, the bosonic sector $|B\rangle $
and the fermionic sector $|F\rangle$ both with $T^4=-1$ and $T^8=+1$.
Again it is practical to define the Hilbert space on a flat space on a $\mathbb{R}^{d-1}$ with a continuous time $t$ on $\mathbb{R}^1$.
Then we find the S acts on state vectors living on the 
$\mathbb{R}^{d-1}$ generating the following groups for each sector:
\begin{enumerate}
\item The unitary $\Z_2^F$ generator is the same as \eq{eq:Pf-matrix}.

\item The anti-unitary $\Z_8^T$ generator, with complex conjugation $K$, is
\bea
\cT =\frac{1}{\sqrt{2}}\begin{pmatrix}
0 & 1 + \ii \\
 1-\ii & 0
\end{pmatrix} K 
=(\frac{\sigma_x-\sigma_y}{\sqrt{2}}) K 
\text{, which acts on }
\begin{pmatrix*}[l]
|B\rangle  \\
|F\rangle 
\end{pmatrix*}, 
\label{eq:D8Pin-T}
\eea
or precisely $\cT = ((\frac{\sigma_x-\sigma_y}{\sqrt{2}})   \otimes \mathds{1}) K$.
The $\cT$ switches between $|B\rangle$ and $|F\rangle$. 

\item The $\mathbb{M}_{16}^{F,T}$
is generated by $\{\cT, P_F |\cT^8=P_F^2=1,  
P_F \cT P_F = \cT^5 = -\cT \}$. Note that
 $\cT^2=-\ii P_F = -\ii   (\sigma_z \otimes \mathds{1})$, while
 $\cT^4=-1$ for all bosonic and fermionic states.

\item The Spin group has elements $\rm{S}_{\rm{rot}}$ the same as
\eq{eq:D8Pin-rot}, with $P_F$ its $\Z_2$ center.
\end{enumerate}
Combining the group elements $P_F$, $\cT$, and $\rm{S}_{\rm{rot}}$ together,
we generate the full structure of $\mathbb{M}_{16}^{F,T} \Pin$ that acts on the
assume Hilbert space 
$\cH=\cH_B \oplus \cH_{F}$ associated with the $\mathbb{R}^{d-1}$ space.
Again such an extension is a \emph{supersymmetry extension} when we require 
some group element (here $\cT$) permutes 
bosonic and fermionic sectors. 
Formally, we propose that
we can trivialize the odd $\nu \in \Omega_2^{\Pin^+} =\Z_2$ class whose cobordism invariant is the same as in \eq{eq:ABK=2},
by pulling back $\Pin^+$ to $\mathbb{M}_{16}^{F,T} \Pin$ via a \emph{supersymmetry extension}.

\section{Conclusion and Higher Dimensional Generalization}
\label{sec:conclusion}

Let us discuss the higher dimensional generalization
of the \emph{fermionic symmetry extension} and \emph{supersymmetry extension}:

Earlier we have studied $\Z_2^T \times \Z_2^F$ ($\Pin^-$) and $\Z_4^{TF}$ ($\Pin^+$)
    and their \emph{fermionic symmetry extension} 
    and \emph{supersymmetry (SUSY) extension} via a
    finite group extension.
In fact, 
there are maps known as Smith homomorphisms in any dimension 
relating the following
SPT classes and symmetry groups (see 
\cite{Kapustin1406.7329, 2018arXiv180502772T, Hason2019akwRyanThorngren1910.14039} and also 
\cite{GuoJW1812.11959, WanWWZ1912.13504}):
\begin{multline} \label{eq:discrete-Smith}
(n+3){\rm{d}}\; \text{unitary }\Z_2 \times \Z_2^F\Rightarrow 
(n+2){\rm{d}}\; \text{anti-unitary }\Z_2^T \times \Z_2^F \Rightarrow\\
(n+1){\rm{d}}\;\text{unitary } \Z_4^F \Rightarrow 
n {\rm{d}}\; \text{anti-unitary }\Z_4^{TF}\Rightarrow \dots. \quad
\end{multline}
By an adding SO rotational symmetry
in terms of continuous groups, \eq{eq:discrete-Smith} becomes
\bea \label{eq:continuous-Smith}
(n+3) {\rm{d}}\; \Spin \times \Z_2 \Rightarrow 
(n+2) {\rm{d}}\; \Pin^- \Rightarrow
(n+1) {\rm{d}}\; \Spin \times_{\Z_2^F}  \Z_4 \Rightarrow 
n {\rm{d}}\; \Pin^+ \Rightarrow \dots. \quad
\eea
Smith homomorphism maps include the maps between
SPT classifications of these symmetries \cite{Kapustin1406.7329, 2018arXiv180502772T, Hason2019akwRyanThorngren1910.14039, WanWWZ1912.13504}:
\bea \label{eq:Smithmap}
\begin{array}{rrrl}
\Omega_8^{\Pin^+} =\Z_{32} \times \Z_2&\to& \Omega_7^{\Spin \times {\Z_2}}=\Z_{16},&\quad\\
\Omega_7^{\Spin \times {\Z_2}}=\Z_{16}&\to&
\Omega_6^{\Pin^-} =\Z_{16},& \quad\\
\Omega_6^{\Pin^-} =\Z_{16} &\to&
\Omega_5^{\Spin \times_{\Z_2} \Z_4}=\Z_{16},&\quad \\
\Omega_5^{\Spin \times_{\Z_2} \Z_4}=\Z_{16} &\to&
\Omega_4^{\Pin^+} =\Z_{16},& \quad\\
\Omega_4^{\Pin^+} =\Z_{16}
&\to& \Omega_3^{\Spin \times {\Z_2}}=\Z_{8},&\quad \\
\Omega_3^{\Spin \times {\Z_2}}=\Z_{8}&\to&
\Omega_2^{\Pin^-} =\Z_{8},&\quad \\
\Omega_2^{\Pin^-} =\Z_{8}&\to&
 \Omega_1^{\Spin \times_{\Z_2} \Z_4}=\Z_{4},&\quad\\
 \Omega_1^{\Spin \times_{\Z_2} \Z_4}=\Z_{4} 
 &\to&
  \Omega_0^{\Pin^+} =\Z_{2}.&\quad 
  \end{array}
\eea
In fact, as found in \Refe{WanWWZ1912.13504} and \cite{PTWtoappear},
    the $\Z_2$ extensions can trivialize many SPT classes ---
    thus this symmetry extension
    can help to construct the 
    symmetry-extended 
    gapped boundaries (lifting all boundary
    zero modes): 
\bea
&&1 \to \Z_2 \to \EPin(d) \to \Pin^{\pm} \to 1.\\
&&1 \to \Z_2 \to \Spin(d) \times\Z_4 \to 
\begin{matrix}
\Spin(d) \times\Z_2\\
\text{ or }\\
\frac{\Spin(d) \times\Z_4}{\Z_2}
\end{matrix}\to 1.
\eea    
In the following subsections, 
we make some comments about these extensions and the generalization to supersymmetry 
extensions, for various dimensions.

\subsection{1+1d}
\label{sec:1+1d}

For  $d=2$ or $1+1$d:
In the bulk part of our work, we had focused on $d=2$ or $1+1$d cases.
We have $\Z_2^T \times \Z_2^F$ SPTs classified
    by $\Omega_2^{\Pin^-}=\Z_{8}$,
    which
 involves 3d SPT cobordism invariant 
with $\nu \in \Z_8$ \cite{Kapustin1406.7329}:
 \bea
\exp( \frac{2\pi \ii \nu}{8}   \int_{M^2} \ABK). 
\eea    
$\bullet$ When $\nu=4$, we can show that the
four layers of Kitaev fermionic chain ($\nu=4$) has SPT invariants 
$4 \ABK$ relates to the bosonic Haldane spin chain 
$w_1(TM)^2$ up to a trivial gapped fermionic sector, 
$$
\exp( \frac{2\pi \ii 4}{8}   \int_{M^2} \ABK)
=\exp( {\pi \ii }  \int_{M^2} w_1(TM)^2). 
$$
such as in
$\Pin^{-}$ 
which becomes trivialized  $w_1(TM)^2=\delta \mu$ as a coboundary in EPin, 
via the group extension $1 \to \Z_2 \to \EPin(d) \to \Pin^{-} \to 1$,
see the result of \cite{WanWWZ1912.13504},
\Sec{sec:BDI-4in8}, and \Sec{sec:CII-AIII-2in4}. 
\\
$\bullet$ When $\nu=2$, we can show that
$2\ABK$ is related to $ w_1(TM) {\eta'}
\equiv {\eta'}\; \PD( w_1(TM))$ via
$$
\exp( \frac{2\pi \ii 2}{8}   \int_{M^2} \ABK)
=\exp( \frac{\pi \ii }{2}  \int_{M^2} 
 {\eta'}\;\PD( w_1(TM))). 
$$
The ${\eta'}$ is the 1d  cobordism invariant of $\Omega_1^{\Spin \times_{\Z_2} \Z_4}=\Z_4$, 
a $\Z_4$ enhancement of the $ \tilde{\eta}$. 
The $ \tilde{\eta}$ is the 1d cobordism invariant of $\Omega_1^{\Spin}=\Z_2$ which assigns
different spin structures with different values:\\
$$
\begin{cases}
\text{For the periodic (Ramond) boundary condition, $ \tilde{\eta}=1$.}\\
\text{For the anti-periodic (Neveu-Schwarz) boundary condition, $ \tilde{\eta}=0$.}
\end{cases}
$$\\
The ${\eta'}\; \PD( w_1(TM))$ is also a $\Z_4$ class like
${\eta'}$.
In \Sec{sec:free BDI}, we have shown this 2 ABK invariant 
can be trivialized by a 
\emph{SUSY extension}
involving a discrete 
$\mathbb{D}_8^{F,T}$ that we introduce (or we denote it as a $\mathbb{D}_8^{F,T}\Pin$ in the continuum).
\\
$\bullet$ When $\nu=1$, our method does not apply. But 
Dijkgraaf-Witten \cite{Dijkgraaf2018Witten1804.03275} suggests 
that 2d topological gravity can lift up a single Majorana zero mode
and thus trivialize the Kitaev chain. Other related 10 Cartan symmetry classes 
and 2d topological gravity are explored subsequently by Stanford-Witten \cite{StanfordWitten1907.03363}.

\subsection{2+1d}
    
    For $d=3$ or $2+1$d: we have $\Z_2 \times \Z_2^F$ SPTs classified
    by $\Omega_3^{\Spin \times {\Z_2}}=\Z_{8}$,
    which
 involves 3d SPT cobordism invariant 
with $\nu \in \Z_8$ \cite{Kapustin1406.7329, Putrov2016qdo1612.09298PWY}:
 \bea
\exp( \frac{2\pi \ii \nu}{8}   \int_{M^3} \ABK(\PD(A)) ), 
\eea    
where ABK is the Arf-Brown-Kervaire (ABK) invariant and $A$ is a $\Z_2$ gauge field
so it is a mod 2 cohomology class $A \in \H^1(M,\Z_2)$.\\
$\bullet$ When $\nu=4$,  
\Refe{WangWenWitten_SymmetricGapped_PRX2018, 
WanWWZ1912.13504} shows that the
$4\ABK(\PD(A))$ is related to  $A^3$. This is the same generator as a 2+1d bosonic SPTs with $\Z_2$ symmetry given by $\H^3(\B\Z_2,\U(1))=\Z_2$, namely up to a trivial gapped fermionic sector
$$
\exp( \frac{2\pi \ii 4}{8}   \int_{M^3} \ABK(\PD(A)) )
= 
\exp( {\pi \ii }   \int_{M^3} A^3). 
$$
$\bullet$ When $\nu=2$, the
$2\ABK(\PD(A))$ is related to  $A^2 {\eta}'\equiv {\eta}'(\PD(A^2))$.
Namely,
$$
\exp( \frac{2\pi \ii 2}{8}   \int_{M^3} \ABK(\PD(A)) )
= 
\exp( \frac{\pi \ii }{2}   \int_{M^3} {\eta}'(\PD(A^2))).
$$
{Again, 
the ${\eta'}$ is the 1d
cobordism invariant of $\Omega_1^{\Spin \times_{\Z_2} \Z_4}=\Z_4$.
By pulling back $\Spin \times {\Z_2}$
to $\Spin \times {\Z_4}$, we can trivialize $A^2$ by a $\Z_2$ extension, thus we also trivialize all the even classes $\nu=4,2$ at once  
\cite{WanWWZ1912.13504, PTWtoappear}.}\\
$\bullet$ When $\nu=1$ or odd, it will be interesting to check whether our \emph{SUSY extension} method applies.

\subsection{3+1d}
\label{sec:3+1d}
%
For $d=4$ or $3+1$d: We have $\Z_4^{TF}$ SPTs 
 classified
    by $\Omega_4^{\Pin^+ }=\Z_{16}$,
    which
 involves 4d SPT cobordism invariant 
with $\nu \in \Z_{16}$ \cite{Kitaev2015, Kapustin1406.7329}:
 \bea
\exp( \frac{2\pi \ii \nu}{16}   \int_{M^4} \eta), 
\eea    
where $\eta$ is the famous eta invariant.\\
$\bullet$ When $\nu=8$, the $8 \eta$ is related to  $w_1(TM)^4$ up to trivial gapped fermions, namely 
$$
\exp( \frac{2\pi \ii 8}{16}   \int_{M^4} \eta)
=
\exp(\pi \ii \int_{M^4} w_1(TM)^4). 
$$
$\bullet$ When $\nu=4$, the $4 \eta$  is related to  $w_1(TM)^3 {\eta}' 
\equiv 
{\eta}' (\PD(w_1(TM)^3)$, namely
$$
\exp( \frac{2\pi \ii 4}{16}   \int_{M^4} \eta)
=
\exp( \frac{\pi \ii }{2}  \int_{M^4} {\eta}' (\PD(w_1(TM)^3)). 
$$
$\bullet$ When $\nu=2$, the $2 \eta$  is related to  $w_1(TM)^2 \ABK 
\equiv 
\ABK (\PD(w_1(TM)^2)$, namely
$$
\exp( \frac{2\pi \ii 2}{16}   \int_{M^4} \eta)
=
\exp( \frac{\pi \ii }{4}  \int_{M^4} {\ABK} (\PD(w_1(TM)^2)). 
$$
By pulling back $\Pin^+$
to $\EPin$, we can trivialize $w_1(TM)^2$ by a $\Z_2$ extension, thus we also trivialize all the even classes $\nu=8,4,2$ at once \cite{WanWWZ1912.13504, PTWtoappear}.\\
$\bullet$
When $\nu=1$ or odd, it will be interesting to check whether our \emph{SUSY extension} method applies.

\subsection{4+1d}

For $d=5$ or $4+1$d:
We have $\Z_4^{F}$ SPTs 
 classified
    by $\Omega_5^{\Spin \times_{\Z_2} \Z_4}=\Z_{16}$,
    which
 involves 5d SPT cobordism invariant 
with $\nu \in \Z_{16}$ 
\cite{2018arXiv180502772T, GuoJW1812.11959, WanWWZ1912.13504}:
 \bea
\exp( \frac{2\pi \ii \nu}{16}   \int_{M^5} \eta(\PD(A))). 
\eea    
$\bullet$
When $\nu=8$, 
the $8 \eta(\PD(A)))$ 
is related to  $A^5$ up to trivial gapped fermions, namely 
$$
\exp( \frac{2\pi \ii 8}{16}   \int_{M^5} \eta(\PD(A))))
=
\exp(\pi \ii \int_{M^5} A^5). 
$$
$\bullet$
When $\nu=4$, the $4 \eta(\PD(A)))$ is related to $A^4 {\eta}' 
\equiv 
{\eta}' (\PD(A^4))$, namely 
$$
\exp( \frac{2\pi \ii 4}{16}   \int_{M^5} \eta(\PD(A))))
=
\exp(\frac{\pi \ii}{2} \int_{M^5}{\eta}' (\PD(A^4))). 
$$
$\bullet$ When $\nu=2$, the $2 \eta(\PD(A)))$ 
is related to $A^3 \Arf 
\equiv 
\ABK (\PD(A^3))$, namely 
$$
\exp( \frac{2\pi \ii 2}{16}   \int_{M^5} \eta(\PD(A))))
=
\exp( \frac{\pi \ii }{4}  \int_{M^5} \ABK (\PD(A^3))). 
$$
By pulling back ${\Spin \times_{\Z_2} \Z_4}$
to ${\Spin \times \Z_4}$, we can trivialize $A^3$ by a $\Z_2$ extension, thus we also trivialize
all the even classes $\nu=8,4,2$ at once \cite{WanWWZ1912.13504, PTWtoappear}.
\\
$\bullet$
When $\nu=1$ or odd, it will be interesting to check whether our SUSY extension method applies.

These results of 4d
$\Z_{16}$ 't Hooft anomalies 
have applications to the 3+1d Beyond the Standard Model physics and non-perturbative interactions 
\cite{Garcia-EtxebarriaMontero1808.00009, JW2006.16996, JW2008.06499, RazamatTong2009.05037}.
Our work suggests that the $\Z_2$ symmetry extension \cite{WanWWZ1912.13504, PTWtoappear}
and possible \emph{SUSY extensions}
can help to compensate the $\nu \in \Z_{16}$ 't Hooft anomaly via: 
$$
\begin{cases}
\text{a 3+1d symmetry-extended TQFT by trivializing the $\nu \in \Z_{16}$ anomaly via the symmetry-extension \cite{WangWenWitten_SymmetricGapped_PRX2018}.}\\
\text{a 3+1d symmetric anomalous gapped TQFT with the $\nu \in \Z_{16}$ 't Hooft anomaly.}
\end{cases}
$$
The 3+1d symmetric anomalous gapped TQFT with the $\nu \in \Z_{16}$ 't Hooft anomaly may only require a $\Z_2$ gauge group
if the symmetry extension only requires a $\Z_2$-extension.

If there is no 't Hooft anomaly, namely $\nu = 0 \mod {16}$, then
we can in principle have a 3+1d symmetric gap boundary preserving the ${\Spin \times_{\Z_2} \Z_4}$ symmetry
with a trivial order (no topological order, no TQFT, no symmetry-breaking order, etc.). See the related work on 
the 3+1d physics along this research direction 
\cite{Wen2013ppa1305.1045, You2014oaaYouBenTovXu1402.4151, YX14124784,
Kikukawa1710.11618, WangWen1809.11171, Catterall2010.02290}.

\subsection{5+1d}
For $d=6$ or $5+1$d:
We have $\Z_2^T \times \Z_2^F$ SPTs classified
    by $\Omega_6^{\Pin^-}=\Z_{16}$,
    which
 involves 6d SPT invariant 
with $\nu \in \Z_{16}$ \cite{PTWtoappear}:
 \bea
\exp( \frac{2\pi \ii \nu}{16}   \int_{M^6} \eta(\PD(w_1(TM)^2))). 
\eea   
$\bullet$
When $\nu=8$, the $8 \eta(\PD(w_1(TM)^2)))$ is related to $w_1(TM)^6$
up to trivial gapped fermions, namely 
$$
\exp( \frac{2\pi \ii 8}{16}   \int_{M^6} \eta(\PD(w_1(TM)^2))))
=
\exp(\pi \ii \int_{M^6} w_1(TM)^6). 
$$
$\bullet$
When $\nu=4$, the $4 \eta(\PD(w_1(TM)^2)))$ 
 is related to $w_1(TM)^5 {\eta}' 
\equiv 
{\eta}' (\PD( w_1(TM)^6))$, namely 
$$
\exp( \frac{2\pi \ii 4}{16}   \int_{M^6} \eta(\PD(w_1(TM)^2))))
=
\exp(\frac{\pi \ii}{2} \int_{M^6} {\eta}' (\PD( w_1(TM)^5))). 
$$
$\bullet$
When $\nu=2$, the $2 \eta(\PD(w_1(TM)^2)))$ is related to $w_1(TM)^4 \ABK 
\equiv 
\ABK (\PD( w_1(TM)^4))$, namely 
$$
\exp( \frac{2\pi \ii 2}{16}   \int_{M^6} \eta(\PD(w_1(TM)^2))))
=
\exp( \frac{\pi \ii }{4}  \int_{M^6} \ABK (\PD(w_1(TM)^4))). 
$$
$\bullet$
When $\nu=1$, we have the $\eta(\PD(w_1(TM)^2)))$.
However, by pulling back $\Pin^-$
to $\EPin$, we can trivialize $ w_1(TM)^2$ by a $\Z_2$ extension, thus we also trivialize
all these classes $\nu=8,4,2,1$ and any $\nu \in \Z_{16}$ at once \cite{WanWWZ1912.13504, PTWtoappear}.

\subsection{6+1d}
\label{sec:6+1d}

For $d=7$ or $6+1$d:
We have $\Z_2 \times \Z_2^F$ SPTs classified
    by $\Omega_7^{\Spin \times \Z_2}=\Z_{16}$,
    which
 involves 7d SPT invariant 
with $\nu \in \Z_{16}$ \cite{PTWtoappear}:
 \bea
\exp( \frac{2\pi \ii \nu}{16}   \int_{M^7} \eta(\PD(A^3))). 
\eea   
$\bullet$
When $\nu=8$, the $8 \eta(\PD(A^3)))$ 
 is related to $ A^7$
up to trivial gapped fermions, namely 
$$
\exp( \frac{2\pi \ii 8}{16}   \int_{M^7} \eta(\PD(A^3))))
=
\exp(\pi \ii \int_{M^7}  A^7). 
$$
$\bullet$
When $\nu=4$, the $4 \eta(\PD(A^3)))$  is related to  
$A^6 \tilde{\eta} 
\equiv 
\tilde{\eta} (\PD( A^6))$, namely 
$$
\exp( \frac{2\pi \ii 4}{16}   \int_{M^7} \eta(\PD(A^3))))
=
\exp(\frac{\pi \ii}{2} \int_{M^7} {\eta}' (\PD( A^6))). 
$$
$\bullet$
When $\nu=2$, the $2 \eta(\PD(A^3)))$  is related to  $A^5 \ABK 
\equiv 
\ABK (\PD( A^5))$, namely 
$$
\exp( \frac{2\pi \ii 2}{16}   \int_{M^7} \eta(\PD(A^3))))
=
\exp( \frac{\pi \ii }{4}  \int_{M^7} \ABK (\PD(A^5))). 
$$
$\bullet$
When $\nu=1$, 
 we have the $\eta(\PD(A^3))$.
However, by pulling back ${\Spin \times \Z_2}$
to ${\Spin \times \Z_4}$, we can trivialize $A^3$ by a $\Z_2$ extension, thus we also trivialize
all these classes $\nu=8,4,2, 1$ and any $\nu \in \Z_{16}$
at once \cite{WanWWZ1912.13504, PTWtoappear}.


\subsection{Smooth (DIFF) manifolds to topological (TOP) manifolds}
\label{sec:DIFF-TOP} 

\Sec{sec:3+1d} shows that the following extension can trivialize the cobordism invariant of $\Omega_4^{\Pin^+ }=\Z_{16}$
when $\nu$ is even ($2,4,6,8,10,12,14,16$) $\in \Z_{16}$: 
\bea
\Z_2 \to \EPin \to {\Pin^+ },
\eea
because we can trivialize the $w_1^2(TM)$ in $\EPin$.
The EPin is introduced in  \Refe{WanWWZ1912.13504} and reviewed in \Sec{sec:BDI-4in8}.
Is there any way we can trivialize when the $\nu$ is odd ($1,3,5,7,9,11,13,15$) $\in \Z_{16}$?

\noindent
$\bullet$ One known way is by the construction of 
\Refe{Fidkowski1305.5851}, which shows that the 3d $\SO(3)_{3,\pm}$ Chern-Simons (CS) theory can saturate the 'Hooft anomaly of 
$\nu = \pm 3 \in \Z_{16}$. The $\pm$ in 3d $\SO(3)_{3,\pm}$ CS means two version of time-reversal symmetry assignment. 

\noindent
$\bullet$ 
Here we provide another possibility by looking into more general categories of manifolds.
The bordism group $\Omega_4^{\Pin^+ }=\Z_{16}$ classifies the 4d smooth and differentiable (DIFF) manifolds with ${\Pin^+}$ structures.
However, \Refe{kirby1990pin} looks at the bordism group
\bea
\Omega_4^{{\rm Top} \Pin^+ }=\Z_{8} \times \Z_2
\eea
which classifies the 4d non-triangulable topological (Top) manifolds also with ${\Pin^+}$ structures.
It turns out that what survives as $\Z_{8}$ in  $\Omega_4^{{\rm Top} \Pin^+ }$ is the even $\nu$ class in $\Omega_4^{\Pin^+ }=\Z_{16}$.
Which means all the odd $\nu$ class in $\Omega_4^{\Pin^+ }=\Z_{16}$ is ``trivialized'' in the more general category of topological manifolds in 
$\Omega_4^{{\rm Top} \Pin^+ }$.\footnote{The remained $\Z_2$ in $\Omega_4^{{\rm Top} \Pin^+ }=\Z_{8} \times \Z_2$
is given by the nontrivial Kirby-Siebenmann class $\kappa \in \H^4(M,\Z_2)$, whose manifold generator is the Freedman's E8 manifold.
The E8 manifold is a unique compact, simply connected 
4d topological manifold whose intersection form is the 
positive-definite even unimodular rank-8 
matrix of the 
E8 lattice.  
Th E8 manifold does not have a smooth DIFF or piecewise linear PL structure.}

Since EPin trivializes the even $\nu \in \Z_{16}$ of 
$\Omega_4^{\Pin^+ }=\Z_{16}$, 
while the ${{\rm Top} \Pin^+ }$ trivializes 
odd $\nu \in \Z_{16}$ of $\Omega_4^{\Pin^+ }=\Z_{16}$,
therefore we expect that the combined categories of topological manifolds with the extended EPin structure,
which we name ${{\rm Top} \EPin }$, 
can trivialize all $\nu \in \Z_{16}$ of $\Omega_4^{\Pin^+ }=\Z_{16}$.
For future work, it will be illuminating to study the bordism group
$\Omega_4^{{\rm Top} \EPin }$ further. 

\subsection{Intrinsic fermionic {gapless} topological phase protected by supersymmetry}
\label{sec:gaplessSPT}

Recent work \cite{ThorngrenVerresen2008.06638} points out the relation between 
(1) the boundary of $D+1$ dimensional SPTs whose 't Hooft anomaly can be trivialized by symmetry extension,
and 
(2) the intrinsic {gapless} $D$ dimensional  SPTs living in one lower dimension.

\Refe{ThorngrenVerresen2008.06638} focus on the bosonic SPT example, here we can propose a 
novel intrinsically fermionic {``gapless''} SPTs protected by supersymmetry. 

Take the 0+1d boundary of the 1+1d 2-layer of Kitaev chains, we had found that 
$$
\begin{cases}
\text{the bulk $\Z_2^T \times \Z_2^F$ symmetry can be extended to
the boundary $\mathbb{D}_8^{F,T}$ with 't Hooft anomaly trivialized.}\\
\text{the bulk $\Z_4^{TF}$ symmetry can be extended to
the boundary $\mathbb{M}_{16}^{F,T}$ with 't Hooft anomaly trivialized.}
\end{cases}
$$
So there should be a 0+1d intrinsic fermionic {``gapless''} SPTs protected by supersymmetry
$\mathbb{D}_8^{F,T}$ or $\mathbb{M}_{16}^{F,T}$, living in 0+1d (without the requirement of an extra dimension 1+1d).
``Gapless'' in 0+1d only means degenerate zero energy modes.
How to find such a 0+1d intrinsic fermionic {``gapless''} SPTs with degenerate zero energy modes?
We can consider $N = 0 \mod 8$-layer  of Kitaev chains, such as the 8-layer of 1+1d Kitaev chains.
The 8-layer of 1+1d Kitaev chains can be gapped in the bulk with a trivial order without any SPTs.
Thus, the 0+1d boundary of 1+1d 8-layer Kitaev chains must be intrinsically 0+1d.

Say we consider the Hilbert space of 8 Majorana fermion operators: 
$\gamma_1,\gamma_{1'}, \gamma_2, \gamma_{2'},
\gamma_3, \gamma_{3'}, \gamma_4, \gamma_{4'}$
(on the boundary of 8-layer of Kitaev chains). 
The 8 Majorana fermions have at most $\GSD = 2^4=16$ ground states.
But we can turn on the 0+1d time-reversal symmetric Hamiltonian: \\[-8mm]
\begin{enumerate}[leftmargin=2.mm, label=\textcolor{blue}{(\arabic*)}., ref={(\arabic*)}]
\item Consider the 0+1d Hamiltonian 
\bea 
H'_{BB} &= &(\gamma_1 \gamma_{1'} \gamma_2 \gamma_{2'})
+   ( \gamma_3 \gamma_{3'} \gamma_4 \gamma_{4'} ) 
+ (\vec{S}^\al_B \cdot \vec{S}^\bt_B)\cr
&=&
(- P_{F,1} \; P_{F,2} )
+   (- P_{F,3} \; P_{F,4} ) 
+ (\vec{S}^\al_B \cdot \vec{S}^\bt_B) \equiv
H'_{B,11'22'}+H'_{B,33'44'}
+(\vec{S}^\al_B \cdot \vec{S}^\bt_B). \quad \label{eq:H'BB}
\eea
The $P_{F,a}=-\ii \gamma_a \gamma_{a'}$ is the fermion parity from the $a$-th Majorana pairs, for $a=1,2,3,4$.
The $H'_{BB}$  leaves us a single bosonic ground state ($\GSD=1$) with a total fermion parity 
$P_F=P_{F,1}P_{F,2}P_{F,3}P_{F,4}=+1$.
The reasoning is the following.\\[-8mm]
\begin{itemize}[leftmargin=-1.mm]
\item
The first interaction term $H'_{B,11'22'}\equiv (\gamma_1 \gamma_{1'} \gamma_2 \gamma_{2'})$ in $H'_{BB}$ energetically selects the bosonic ground state 
sector in $P_{F,1} \; P_{F,2}=+1$ out of
the $\GSD = 2^2=4$ from the first four Majorana $\gamma_1,\gamma_{1'}, \gamma_2, \gamma_{2'}$. 
There are two bosonic sectors: $P_{F,1}=P_{F,2}=+1$ and $P_{F,1}=P_{F,2}=-1$. In short, 
we gap the other two fermionic sectors out of $\GSD =4$ to leave with only bosonic $\GSD = 2$ remained for the first four Majorana modes. 
\item
The second interaction term $H'_{B,33'44'}\equiv ( \gamma_3 \gamma_{3'} \gamma_4 \gamma_{4'} )$ in $H'_{BB}$ 
also selects the bosonic ground state sector in $P_{F,3} \; P_{F,4}=+1$
out of the $\GSD = 2^2=4$ from the second four Majorana $\gamma_3,\gamma_{3'}, \gamma_4, \gamma_{4'}$. 
We gap the other two fermionic sectors out of $\GSD =4$ to leave with only bosonic $\GSD = 2$ remained for the last four Majorana modes. 
\item So far the first two interaction terms $H'_{B,11'22'}+H'_{B,33'44'}$
in $H'_{BB}$ energetically select 4 bosonic ground states out of the original 16 states.
Out of the bosonic $\GSD = 4$, the first four Majorana modes form a bosonic doublet, while the last Majorana modes form another bosonic doublet.
Each bosonic doublet is the same degree of freedom as a $J=\frac{1}{2}$ spin (as a representation of SU(2)) at the one end of Haldane spin-1 chain.
We can write these two SU(2) bosonic spin-$\frac{1}{2}$ operators as, $\vec{S}^\al_B$ and $\vec{S}^\bt_B$, respectively. 
Since the two bosonic doublets can form a spin-0 singlet and a spin-1 triplet (via $\frac{1}{2}_B \otimes \frac{1}{2}_B = 0_B \oplus 1_B$),
we can gap the spin-1 triplet via the third interaction term, the Heisenberg-type interaction $(\vec{S}^\al_B \cdot \vec{S}^\bt_B)$.
Thus turning on the full $H'_{BB}$, we end up with only a bosonic sector of spin-0 singlet with $\GSD = 1$.
Let us call this single bosonic ground state sector $\ket{B_0}$.
\item
In fact the $H'_{B}$ of \eq{eq:H'BB} can be rewritten as 14 terms of 
four-Majorana interactions similar to \Refe{FidkowskiKitaev_PhysRevB.83.075103_2011},
firstly known by Fidkowski-Kitaev.
\end{itemize}
\item We can also tune the  0+1d Hamiltonian via different interactions: 
\bea 
H'_{FF} &= &  (-\gamma_1 \gamma_{1'} \gamma_2 \gamma_{2'})
+  (- \gamma_3 \gamma_{3'} \gamma_4 \gamma_{4'} ) 
+ (\vec{S}^\al_F \cdot \vec{S}^\bt_F)\cr
&=&
( P_{F,1} \; P_{F,2} )
+   ( P_{F,3} \; P_{F,4} ) 
+ (\vec{S}^\al_F \cdot \vec{S}^\bt_F)
 \equiv
H'_{F,11'22'}+H'_{F,33'44'}
+(\vec{S}^\al_F \cdot \vec{S}^\bt_F). \label{eq:H'FF}
\eea
The $H'_{F,11'22'}$ gives two fermionic ground states 
with $( P_{F,1} \; P_{F,2} )=-1$, and the 
$H'_{F,33'44'}$ gives another two fermionic ground states with $( P_{F,3} \; P_{F,4} )=-1$.
Here we write these two SU(2) fermionic spin-$\frac{1}{2}$ operators as, $\vec{S}^\al_F$ and $\vec{S}^\bt_F$, respectively. 
However, the full fermion parity of each of these $\GSD = 4$ is still 
$P_F=P_{F,1}P_{F,2}P_{F,3}P_{F,4}=+1$ thus bosonic.
We can further gap the spin-1 triplet (3 ground states) via the spin-1 interaction in $\frac{1}{2}_F \otimes \frac{1}{2}_F = 0_B \oplus 1_B$,
which can be constructed out of two fermionic spin-$\frac{1}{2}$ doublets as $(\vec{S}^\al_F \cdot \vec{S}^\bt_F)$.
Thus turning on the full $H'_{FF}$, we still end up with only a bosonic sector of spin-0 singlet with $\GSD = 1$.
\item An interesting question is whether we can tune Hamiltonian ending up with only a fermionic sector with $\GSD = 1$?
Consider the 0+1d Hamiltonian 
\bea 
H'_{BF} &= &(\gamma_1 \gamma_{1'} \gamma_2 \gamma_{2'})
+   ( -\gamma_3 \gamma_{3'} \gamma_4 \gamma_{4'} ) 
+ (\vec{S}^\al_B \cdot \vec{S}^\bt_B)\cr
&=&
(- P_{F,1} \; P_{F,2} )
+   (+ P_{F,3} \; P_{F,4} ) 
+ (\vec{S}^\al_B \cdot \vec{S}^\bt_F) \equiv
H'_{B,11'22'}+H'_{F,33'44'}
+(\vec{S}^\al_B \cdot \vec{S}^\bt_F). \quad \label{eq:H'BF}
\eea
The $H'_{B,11'22'}$ gives two bosonic ground states 
with $( P_{F,1} \; P_{F,2} )=+1$, and the 
$H'_{F,33'44'}$ gives another two fermionic ground states with $( P_{F,3} \; P_{F,4} )=-1$.
However, the full fermion parity of each of these $\GSD = 4$ is  
$P_F=P_{F,1}P_{F,2}P_{F,3}P_{F,4}=-1$ fermionic.
We further gap the spin-1 triplet (3 ground states) via the spin-1 interaction $\frac{1}{2}_B \otimes \frac{1}{2}_F = 0_F \oplus 1_F$
constructed out of one bosonic and one fermionic spin-$\frac{1}{2}$ doublet as $(\vec{S}^\al_B \cdot \vec{S}^\bt_F)$.
Thus turning on the full $H'_{BF}$, we end up with only a fermionic sector of spin-0 singlet with $\GSD = 1$.
Let us call this single fermionic ground state sector $\ket{F_0}$.

Similarly, we can also define $H'_{FB}$ analogous to \eq{eq:H'BF},
which also ends up with only a fermionic sector of spin-0 singlet with $\GSD = 1$.

\end{enumerate}

The combined Hamiltonian above, by tuning the coupling appropriately  
\bea
H' = g_{BB} \; H'_{BB}
+
g_{FF}  \; H'_{FF}
+
g_{BF}  \; H'_{BF} + g_{FB}  \; H'_{FB} +\dots
\eea
can energetically select ground states out of the original 16 ground states.
With certain appropriate couplings, 
when the bosonic and fermionic states, $\ket{B_0}$ and $\ket{F_0}$,  are both favored as ground states, we have $\GSD =2$.
We can imagine a process when the $\ket{B_0}$ is favored 
on one side of coupling choice, 
while the $\ket{F_0}$ is favored 
on the other side of coupling choice (both have  $\GSD =1$);
while at some critical coupling, we have the $\GSD =2$ formed by a 2-dimensional Hilbert space 
$\cH=\cH_B \oplus \cH_F = \{\ket{B_0}, \ket{F_0} \}$.
Although we are studying the boundary of 8-layer Kitaev chains,
we find with appropriate interactions,
we land on the same 2-dimensional Hilbert space 
$\cH= \{\ket{B_0}, \ket{F_0} \}$
for 2-layer of Kitaev chain studied in \Sec{sec:free fermions} and \Sec{sec:SUSYQM}!

As mentioned, the boundary of 8-layer Kitaev chains is intrinsically 0+1d.
Thus, this theory at the critical coupling with a Hilbert space $\cH= \{\ket{B_0}, \ket{F_0} \}$, 
can be regarded as \emph{a critical ``gapless'' intrinsically 0+1d fermionic SPTs
protected by supersymmetry} (here gapless only means the degenerate ground states in 0+1d)! 
Depends on the assignment of time-reversal $\cT$ symmetry shown in \Sec{sec:free fermions} ,
the supersymmetry here can be $\bD^{F,T}_8$ or $\bM^{F,T}_8$.

We leave the generalization of this understanding on 
\emph{a critical ``gapless'' intrinsically fermionic SPTs
protected by supersymmetry}
in higher-dimensions for future work.

\subsection{Final comments and future directions}
\label{sec:Finalcomments}

{We conclude by 
enlisting some additional comments and future directions:}

\begin{enumerate}

\item \emph{All intrinsically fermionic 1+1d SPT phases have supersymmetric 0+1d boundaries}:\\
$\bullet$ In the present work, especially in \Sec{sec:free fermions} and \Sec{sec:SUSYQM}, we have shown by examples that
when the number of layers of 1+1d Kitaev chains is $8N+2$ or $8N+6$ (namely the layer number is $2 \mod 4$),
then the 0+1d boundary theory exhibits at least an $\cN=2$ supersymmetric quantum mechanics.
In the sense that the boundary energy spectrum always have bosonic energy eigenstate $\ket{\mu,+}$
and fermionic energy eigenstate $\ket{\mu,-}$ 
paired together for every eigenenergy $E_\mu$. 

$\bullet$ In a companion work \cite{APJW-PRL}, we prove that {``all intrinsically fermionic 1+1d SPT phases have supersymmetric 0+1d boundaries.''}
Another way to rephrase this result in a mathematically precise statement is the following: 

Consider a 2d fermionic bordism group $\Omega_2^{G_{f,\text{total}}}$ with 
a total group ${G_{f,\text{total}}}$
obtained from the extension
of a normal subgroup of fermionic internal $G_f$ global symmetry, 
and the $d$-dimensional spacetime rotational symmetry 
(say $\O(d)/\SO(d)$, etc.)\footnote{The Spin/Pin group can be obtained via the 
$\Z_2^F$-extension of O and SO groups (see Appendix \ref{sec:extension-formal}).}: 
\bea
&&1\to G_f \to  G_{f,\text{total}} \to \O(d) \to 1, \quad \text{ with time-reversal symmetry}. \cr
\text{or }&&1\to G_f \to  G_{f,\text{total}} \to \SO(d) \to 1, \quad \text{ without time-reversal symmetry}.
\eea
The fermionic internal $G_f$ contains the fermion parity $\Z_2^F$ and the bosonic internal $G_b$, 
via a central extension:
\bea
1\to \Z_2^F \to G_f \to G_b \to 1.
\eea
The intrinsically fermionic SPTs, 
corresponds to the cobordism invariants (representing some invertible TQFTs in field theory) 
of $\Omega_2^{G_{f,\text{total}}}$, but those do \emph{not overlap} 
with cobordism invariants of fermionic invertible topological order's bordism group $\Omega_2^{\Spin}(pt)$,
\emph{nor} overlap with
those of bosonic SPT's bordism group $\Omega_2^{G_{b,\text{total}}}$. Here the ${G_{b,\text{total}}}$ is obtained from:
\bea
&&1\to G_b \to  G_{b,\text{total}} \to \O(d) \to 1, \quad \text{ with time-reversal symmetry}. \cr
\text{or }&&1\to G_b \to  G_{b,\text{total}} \to \SO(d) \to 1, \quad \text{ without time-reversal symmetry}.
\eea
We can prove that the 0+1d boundary of these 1+1d cobordism invariants (namely, the invertible TQFTs) of intrinsically fermionic SPTs 
must be supersymmetric \cite{APJW-PRL}. 

    \item \emph{Boundary supersymmetry and the symmetry extension}:\\
    Various previous works pointed out the fractionalized boundary symmetry of 
   fermionic SPTs may be related to supersymmetry \cite{Grover1206.1332, Ponte1206.2340, Grover1301.7449, Gu1308.2488, ZerfMaciejko1605.09423, YuJiabin1902.07407}.\footnote{Naively \Refe{Gu1308.2488} results look similar to ours, such as Gu's superalgebra mentioned in his equation (39). However,
   we can tell the \emph{differences} of our result and Gu's \Refe{Gu1308.2488} by comparing the commutativity of time reversal $\cT$ and fermion parity $P_f$.
    \Refe{Gu1308.2488} shows the commutative $\cT P_f = +P_f \cT$ (which has no SUSY, in our definition).
   Our work shows the anti-commutative $\cT P_f = - P_f \cT$ which implies the SUSY quantum mechanics.}
    
    In the present work, we point out that on the boundary of intrinsic fermionic SPTs,
    the non-commutative properties between fermion parity $P_f$ and other symmetry generator 
    (such as time reversal $\cT$)  implies the boundary supersymmetry.
    (In a companion work \cite{APJW-PRL}, we present a precise proof for this statement.)
    We also provide a framework based on a generalization of symmetry extension in terms of group extension \cite{WangWenWitten_SymmetricGapped_PRX2018}, 
    to the case of an extended total group involving supersymmetry --- which we name it a supersymmetry extension.

 \item \emph{Quantum criticality and emergent supersymmetry}:
 In \Sec{sec:gaplessSPT}, we point out a potential relation between the 0+1d boundary supersymmetric degenerate ground states 
 can be regarded as a 0+1d intrinsically gapless fermionic SPTs (here gapless means degenerate in 0+1d),
indeed also as a 0+1d critical theory on the boundary of the 1+1d trivial class of SPTs (by tuning the 0+1d boundary Hamiltonian). 
 This result may be generalizable to other higher dimensions, thus connecting to the literature
 \cite{Grover1206.1332, Ponte1206.2340, Grover1301.7449, ZerfMaciejko1605.09423, YuJiabin1902.07407} 
of {quantum criticality and supersymmetry} on the boundary of fermionic SPTs. 
Hopefully our approach can provide a more systematic and mathematically precise framework. 

\item \emph{'t Hooft anomalies on the boundary of worldsheet or worldvolume in string theory}:
In a different language setting. we can also apply our results to
the fermionic SPTs living on the 1+1d worldsheet of string theory.
The 0+1d boundary of worldsheet may have various 't Hooft anomalies related to the reflection, time-reversal $\cT$, and other symmetries.
These 't Hooft anomalies impose the constraints on the consistency of worldsheet and string theories \cite{Witten1605.02391, Kaidi2019pzjJulioTachikawa1908.04805, Kaidi2019tyfJulioTachikawa1911.11780, Montero2008.11729, Kaidi2010.10521}.
The constraints are tightly connected to the string landscape vs swampland program, where the \emph{cobordism conjecture} 
plays an important role to restrict the consistency of quantum gravity \cite{McNamara1909.10355}.

\item \emph{Higher-symmetry extension and fermionic symmetry extension construction of symmetric gapped boundaries of SPTs}:
The present work focus on the ordinary 0-form global symmetry acting on bosonic and fermionic Hilbert spaces.
We can also attempt to include higher-form generalized global symmetries \cite{Gaiotto2014kfa1412.5148}.
In fact the symmetry extension method had be generalized to a higher-symmetry extension 
\cite{Tachikawa1712.09542 ,Wan1812.11955, Wan1812.11968, Wan2019oyr1904.00994},
also include  fermionic symmetry extension \cite{Wang1801.05416, GuoJW1812.11959, Kobayashi2019lep1905.05391} --- 
these methods are helpful to construct symmetric gapped boundaries of SPTs, as explained in \cite{WangWenWitten_SymmetricGapped_PRX2018}.
By including the supersymmetry extension, we hope to also 
construct more general boundary theories of fermionic SPTs in higher dimensions.

\end{enumerate}

\section{Acknowledgments}
AP acknowledges useful discussions with Subhro Bhattacharjee, Ashvin Vishwanath,  Nat Tantivasadakarn, J.P. Ang and Tarun Grover. 
JW thanks conversations with Jun Hou Fung, Enno Ke{\ss}ler, Jacob McNamara, Miguel Montero, Pavel Putrov, Ryan Thorngren, Cumrun Vafa, and Yizhuang You,
and especially thanks Zheyan Wan and Yunqin Zheng for the previous collaboration and the enlightening discussions,
in particularly on \Refe{WanWWZ1912.13504}'s Section 6.
AP receives funding from the Simon's foundation through the ICTS-Simons postdoctoral fellowship. 
JW was initially supported by NSF Grant PHY-1606531 and Institute for
Advanced Study when this work started in 2018.
JW is also later supported by NSF Grant DMS-1607871 ``Analysis, Geometry and Mathematical Physics'' and Center for Mathematical Sciences and Applications at Harvard University.

\appendix

\section{Tantivasadakarn-Vishwanath model and the basis change}
\label{app:Nat_basis_change}

Here, we give details of how the model in \cref{sec:interacting BDI} was arrived at. We start by writing down the model constructed by Tantivasadakarn and Vishwanath in\cite{NatAshvin_PhysRevB.98.165104}. They start with the Hilbert space of one fermion on odd sites ($\gamma, \bar{\gamma}$), one fermion on odd sites ($\eta, \bar{\eta}$) and one qubit on even sites and the following trivial Hamiltonian
\begin{equation}
H_0 = -\sum_{\text{even}~j} \sigma^x_j - \ii \sum_{\text{even}~j} \bar{\eta}_j  \eta_j -\ii \sum_{\text{odd}~j}\bar{\gamma}_j \gamma_j.
\end{equation}
The SPT Hamiltonian is obtained from $H_0$ by a finite depth quantum circuit $W$. Up to an inconsequential difference in minus signs, we can write $W$ considered by the authors of \cite{NatAshvin_PhysRevB.98.165104} as a two-layer FDUC:
\begin{align}
W &= W_2 W_1. \\
W_1 &= \prod_{\text{even}~j} \left[ \outerproduct{0}{0}_j + \outerproduct{1}{1}_j \ii \bar{\eta}_j \xi_{j j+1} \gamma_{j+1}\right]. \\
W_2 &= \prod_{\text{even}~j} \left[ \outerproduct{0}{0}_j + \outerproduct{1}{1}_j \ii  \gamma_{j-1} \bar{\eta}_j\right].
\end{align}
 Where $\ket{0/1}$ are the eigenstates of $\sigma^z$ with eigenvalues $(-1)^{0/1}$. The $\xi_{j j+1} = \pm 1$ are $\ztwo$ valued background fields that denote the spin structure to be Ramond (periodic) or Neveu-Schwarz (anti-periodic) depending on whether $\prod_j \xi_{j j+1} = +1$ or $\prod_j \xi_{j j+1} = -1$ respectively. The time reversal symmetry $\cT$ is 
 \begin{equation}
 \cT = \prod_{\text{even}~j} \sigma^x_j ~\cK.
 \end{equation}
 It can easily be checked that $\cT$ does not commute with $W_1$ or $W_2$. However, depending on the spin structure, $\cT$ commutes (Neveu-Schwarz) or anti-commutes (Ramond) with $W$
 \begin{equation}
 \cT W \cT^{-1} = - \left(\prod_j \xi_{j j+1}\right) W.
 \end{equation}
 Regardless of spin structure, the SPT Hamiltonian $H= W H_0 W^\dagger$ shown below commutes with $\cT$. 
 \begin{equation}
 H = -\ii \sum_{\text{odd}~j} \sigma^z_{j-1} \bar{\gamma}_j \gamma_j \sigma^z_{j+1} + \sum_{\text{odd}~j} \xi_{j+1 j+2} \gamma_j (\sigma^z_{j+1} \sigma^x_{j+1}) \gamma_{j+2} \sigma^z_{j+3} -\ii \sum_{\text{even}~j} \bar{\eta}_j \eta_j.
 \end{equation}
 Notice that the term $\sum_{\text{even}~j} \bar{\eta}_j \eta_j$ remains unchanged from its trivial form and only plays an intermediary role in defining the FDUC, $W$. It is ignored in the final form by the authors of \cite{NatAshvin_PhysRevB.98.165104}. However, we shall retain it. To get the form of the Hamiltonian and symmetry presented in the main text, we redefine the Hilbert space by considering both the fermions to live on the sites and qubits to live on the links with the following relabeling
 \begin{equation}
 (\bar{\gamma}_{2k+1}, \gamma_{2k+1}) \mapsto (\bar{\gamma}_{\downarrow,k},\gamma_{\downarrow,k}),~~~ (\bar{\eta}_{2k+2}, \eta_{2k+2}) \mapsto(\bar{\gamma}_{\uparrow,k},\gamma_{\uparrow,k}), ~~\vec{\sigma}_{2k+2} \mapsto \vec{\sigma}_{k, k+1}.
 \end{equation}
 Finally, by performing a change-of-basis using $W_2$, we get the form of the Hamiltonian and symmetry operators used in the main text. 
 \begin{align}
 W_2 H W_2^\dagger &= -\ii \sum_j \left[\sigma^z_{j,j+1} \left( \bar{\gamma}_{\uparrow,j}\gamma_{\uparrow,j} +  \bar{\gamma}_{\downarrow,j+1}\gamma_{\downarrow,j+1}\right) +   
 {\sigma^x_{j,j+1}  \bar{\gamma}_{\uparrow,j}\gamma_{\downarrow,j+1}} \right]. \\
 W_2 \cT W_2^\dagger &= \prod_j \left(\ii \gamma_{\downarrow,j} \bar{\gamma}_{\uparrow,j}\right)  \sigma^x_{j,j+1} ~\cK.
 \end{align}

\section{Appendices for Section 6}
\subsection{{Group extension: Formal setup}}
\label{sec:extension-formal}

Below we illuminate further the group extension used throughout our work:
\bea \label{eq:short-exact-sequence-G}
1 \to N \to \tilde{G} \to G \to 1.
\eea
We will later include also the continuous spacetime symmetry $\SO(d)$ in the continuum limit\footnote{{By the 
continuous spacetime symmetry, we mean the Lorentz symmetry $\SO(d-1,1)$ or Euclidean rotational symmetry $\SO(d)$ in the continuum limit.
In the condensed matter system with a discretized lattice, the spacetime symmetry, such as continuous rotations, is not usually manifested. We may regard this 
continuous spacetime symmetry as an emergent symmetry at an intermediate energy scale of some effective field theories, 
\emph{below} the ultraviolet (UV) high-energy lattice cutoff, 
but \emph{above} the infrared (IR) low-energy field theory.
In this work, we mainly do $d=2$ or 1+1d, however we will also discuss general $d$  in \Sec{sec:conclusion}.
}}
in terms of
\bea  \label{eq:short-exact-sequence-Gtot}
1 \to N \to \tilde{G}_{\text{Tot}} \to G_{\text{Tot}} \to 1,
\eea
where 
$\tilde{G}_{\text{Tot}}$ and $G_{\text{Tot}}$ can be written in terms of the spacetime-internal symmetry group 
form of $\frac{G_{\text{spacetime}} \ltimes {G}_{\text{internal}}}{N_{\text{shared}}}$. In 
\Table{Table:G}, \Table{Table:Gspacetime}, and discussions below,
we shorthand the short exact sequences in \Eq{eq:short-exact-sequence-G}
and \Eq{eq:short-exact-sequence-Gtot} as $N \to \tilde{G} \to G$ and $N \to \tilde{G}_{\text{Tot}} \to G_{\text{Tot}}$ for simplicity.

{To precisely specify an extended group 
\bea 
\tilde{G} =  N \rtimes_{\rho, \varphi} G
\eea 
in  of \eq{eq:short-exact-sequence-G}, we need the following data 
\begin{itemize}
\item The normal subgroup $N$ and the quotient group $Q$.
\item $\rho: G \to {\rm Aut}( N)$. If and only if a nontrivial $\rho$ implies a non-central extension \eq{eq:short-exact-sequence-G}. 
\item A 2-cocycle $\varphi \in \H^2(\B G, N)$ in the cohomology group with the classifying space $\B G$
mapping to values in $N$ (usually defined for an Abelian group) 
is solved from the cocycle condition, in the additive form, 
$$ 
0=\delta\varphi(g_1,g_2,g_3)=
\big(\rho(g_1).\varphi(g_2,g_3)\big)-\varphi(g_1g_2,g_3)+\varphi(g_1,g_2g_3)-\varphi(g_1,g_2)
$$
or in the multiplicative form $\delta\varphi(g_1,g_2,g_3)=\frac{\big(\rho(g_1).\varphi(g_2,g_3)\big) \varphi(g_1,g_2g_3) }{\varphi(g_1g_2,g_3)\varphi(g_1,g_2)} 
$.
\end{itemize}
With two group elements in $ \tilde{G}$ specified by a doublet  $(n_1,g_1)$ and $(n_2,g_2) \in (N,G)$, 
we have the group law
$$(n_1,g_1)(n_2,g_2)=(n_1+\rho(g_1) . n_2+\varphi(g_1,g_2), \;g_1\cdot g_2).$$
Here we use the addition $+$ for the group operation in $N$, and the product $\cdot$ for the group operation in $G$.
Here the 2-cocycle $\varphi(g_1,g_2)$ specifies how much the group extension is twisted.
We illustrate the precise $\tilde{G} =  N \rtimes_{\rho, \varphi} G$ 
for the examples of \Table{Table:G} in a footnote.}\footnote{Here let us illustrate 
the mathematically precise $\tilde{G} =  N \rtimes_{\rho, \varphi} G$ for the examples in \Table{Table:G}:
\begin{enumerate}
\item  $\mathbb{Z}_4^T \times \Z_2^F =  \Z_2 \rtimes_{0,\varphi } ( \Z_2^T \times \Z_2^F)$ has no $\rho$ but a nontrivial $\varphi \in \H^2(\B \mathbb{Z}_2^T, \mathbb{Z}_2)=\mathbb{Z}_2$. We may denote this generator 
$\varphi =w_T^2$ with $w_T$ can be regarded as a time-reversal background field as 
a generator from $\H^1(\B \mathbb{Z}_2^T, \mathbb{Z}_2)=\Z_2$.
\item  $\mathbb{D}_8^{F,T} =  \Z_2 \rtimes_{0,\varphi } ( \Z_2^T \times \Z_2^F)$ has no $\rho$ but a nontrivial $\varphi = w_T A_F
\in \H^2(\B ( \Z_2^T \times \Z_2^F), \mathbb{Z}_2)=\mathbb{Z}_2^3$
with $A_F$ 
can be regarded as a fermion-parity background field as 
a generator from
 $\H^1(\B  \Z_2^F, \mathbb{Z}_2)=\Z_2$. 
But for  $\mathbb{D}_8^{F,T} = \Z_4^T  \rtimes_{\rho,0} \Z_2^F$, it has a nontrivial $\rho: \Z_2 \to \text{Aut}(\Z_4)=\Z_2$
thus not a central extension, but no $\phi$.
\item  $\bM_{16}^{F,T} = \Z_4 \rtimes_{0,\varphi} \Z_4^{TF}$ has 
no $\rho: \Z_4 \to \text{Aut}(\Z_4)=\Z_2$
and a nontrivial odd class $\varphi 
\in \H^2(\B  \Z_4^{TF}, \Z_4)=\mathbb{Z}_4$.
\item Other groups in \Table{Table:G}  obtained by adding extra U(1) and SU(2) 
in group extension twisted by ${(\rho, \varphi)}$ 
are similar in the spirit as the previous examples. 
\end{enumerate}
We can also use the information below to specify the group laws for $\tilde g_1 =(n_1, g_1) \in \tilde{G}$
and $\tilde g_2 =(n_2, g_2) \in \tilde{G}$, with $n_i \in N$ and $g_i \in G$ as
$\tilde  g_1 \tilde  g_2 =(n_1, g_1)(n_2, g_2)= 
(n_1+ \rho(g_1) . n_2 + \varphi(g_1, g_2),g_1 + g_2)$.
}

\subsection{{Bosonic, Fermionic, Time Reversal, Spacetime Symmetries and Group Extensions}}
\label{sec:continuum-symmetry}

We start from elaborating the familiar groups: 
the continuous rotation $\SO(d)$,\footnote{The
major part of our paper focus on $d=2$ or 1+1d, but now we keep
the general $d$ notation in order to generalize our approach to
higher dimensions in \Sec{sec:conclusion}.} 
the reflection $\Z_2^{T_{\rm E}}$ or time-reversal $\Z_2^T$,  and the fermion parity $\Z_2^F$ 
(See useful discussions in \cite{Cordova2017vab1711.10008, 1711.11587GPW}).
Then we can write down the short exact sequences relating the familiar groups to 
$\O(d)$, $\Spin(d)$, and $\Pin^{\pm}(d)$, by symmetry group extension along the horizontal and vertical directions
of:
\begin{equation}
\xymatrix{
&1 \ar[d] &1 \ar[d] & 1 \ar[d]\\
& \mathbb{Z}_{2}^{F} \ar[d] & \mathbb{Z}_{2}^{F} \ar[d]  & 1 \ar[d]\\
1\ar[r]& \Spin(d)\ar[r] \ar[d]&\Pin^{\pm}(d) \ar[d] \ar[r] & \mathbb{Z}_{2}^{T_{\rm E}}  \ar[r] \ar[d] &1\\
1\ar[r]& \SO(d) \ar[r] \ar[d]&\O(d)\ar[r]^{{\det}=\pm 1}  \ar[d]& \mathbb{Z}_{2}^{T_{\rm E}} \ar[r] \ar[d] &1\\
& 1 &1 & 1
}
\end{equation}

Some remarks about these symmetry groups:
\begin{enumerate}[leftmargin=2.mm, label=\textcolor{blue}{\arabic*}., ref={\arabic*}]
\item  
The $\SO(d)$ is a spatial rotation group.  The general linear group $\GL(d,\mathbb{R})$
is the group of the rank-$d$ invertible matrix with all the entry with $\mathbb{R}$ coefficients as group elements,
together with the operation of ordinary matrix multiplication.
The special linear group $\SL(d,\mathbb{R})$
is the subgroup of $\GL(d,\mathbb{R})$, then
\bea
\SO(d)=\{g_{\SO(d)} \in {\SL(d,\mathbb{R})} \;\; \vert \;\; g_{\SO(d)}  g_{\SO(d)}^{\rm T}=g_{\SO(d)}^{\rm T} g_{\SO(d)} =1 \},
\eea
with the transpose operation ${\rm T}$ and its determinant $\det(g_{\SO(d)})= 1$.

\item  
The $\Spin(d)=  \Z_2^F \rtimes  \SO(d)$ is from the spatial rotation $\SO(d)$ extended/graded by the fermion parity $\Z_2^F$.
For $d>2$, $\Spin(d)$ is double cover of $\SO(d)$ and simply-connected, thus the (unique) universal cover of $\SO(d)$.
So its first homotopy group is $\pi_1(\Spin(d))=0$. 
The centers of $\Spin(d)$ and $\SO(d)$ for $d\geq 2$ are:
\bea
Z(\Spin(d))= 
\begin{cases}
\Z_2^2,  & d= 0 \mod 4 \\
\Z_2,  & d=1 \mod 4 \\
\Z_4, & d=2 \mod 4 \\
\Z_2,  & d=3 \mod 4 
 \end{cases}.
\quad
Z(\SO(d))= 
\begin{cases}
\Z_2 ,  & d= 0 \mod 2 \\
1,  & d=1 \mod 2 
 \end{cases}.
\eea

\item The $\O(d)=  \SO(d) \rtimes \Z_2^{T_{\rm E}} $ is 
\bea
\O(d)=\{g_{\O(d)} \in {\GL(d,\mathbb{R})} \;\; \vert \;\; g_{\O(d)}  g_{\O(d)}^{\rm T}=g_{\O(d)}^{\rm T} g_{\O(d)} =1 \}
\eea
from the time reversal $\Z_2^{T_{\rm E}}$ extended by the spatial rotation $\SO(d)$:
\bea
 \O(d) \cong
 \begin{cases}
  \SO(d) \rtimes \Z_2^{T_{\rm E}}, & d \in ~\text{\text{even}}\\
   \SO(d) \times \Z_2^{T_{\rm E}}, & d \in ~\text{\text{odd}}
 \end{cases}.
\eea
When $d$ is odd, the $\Z_2^{T_{\rm E}} =\{+1, - 1\}$ along the diagonal element, exactly matches the determinant map
$\det(g_{\O(d)})=\pm 1$. 
The center of $\O(d)$ for $d\geq 2$ is:
$
Z(\O(d))=\Z_2^{T_{\rm E}}.
$

\item 
The $\Pin^{\pm}(d)=  \Z_2^F \rtimes  \O(d)$ are double covers of $\O(d)$.
For physics purpose, we need to interpret the spacetime symmetry in the Lorentzian signature (i.e., Minkowski signature).
Let us clarify the notations of groups in Euclidean signature (time-reversal $\cT_{\rm E}$)
and Lorentzian signature (time-reversal $\cT$):\footnote{Follow \eq{eq:Z4TF}, for $1 \to \Z_2^F \to \tilde{G} \to \Z_2^T \to 1$,
we have two possible extended $\tilde{G}=\Z_4^{TF}$ or $\Z_2^{T} \times \Z_2^{F}$. We can also replace $T$ to $T_{\rm E}$ so
$\tilde{G}=\Z_4^{T_{\rm E} F}$ or $\Z_2^{T_{\rm E}} \times \Z_2^{F}$. }
\bea \label{eq:Pin+}
\Pin^{+}(d)&\cong& \Spin(d)\rtimes \mathbb{Z}_{2}^{T_{\rm E}}. 
~~ \quad\quad \cT_{\rm E}^{2}=+1 \text{ or }  \cT_{ }^{2}=(-1)^F. ~~  \\
&& \Z_2^{T_{\rm E}} \times \Z_2^F  \quad\quad \quad\text{ or } \quad \Z_4^{TF}  \text{ for discrete subgroups}. \nn\\[2mm]  
\label{eq:Pin-}
\Pin^{-}(d)&\cong& \frac{\Spin(d)\rtimes \mathbb{Z}_{4}^{T_{\rm E}F}}{\mathbb{Z}_{2}^F}. 
~~ \quad\quad  \cT_{\rm E}^{2}=(-1)^F \text{ or }  \cT_{}^{2}=+1. ~~  \\
&& \Z_4^{T_{\rm E} F}  \quad\quad\quad\quad\quad \text{ or } \quad  \Z_2^{T} \times \Z_2^F  \text{ for discrete subgroups}. \nn
\eea
In Euclidean signature, the $\Z_2^{{T_{\rm E}}}$ is generated by a reflection $\cT_{\rm E}$.
The reflection $\cT_{\rm E} \in \O(d)$ has $\cT_{\rm E}^2=+1$.
The pullback from $\cT_{\rm E} \in \O(d)$ to
$\cT_{\rm E} \in \Pin^{\pm}(d)$ via the $\Z_2^F$ extension
shows that
$\cT_{\rm E}^2=+1$ for $\Pin^{+}(d)$ 
and $\cT_{\rm E}^2=(-1)^F$ for $\Pin^{-}(d)$.

In Lorentzian signature, 
the reflection $\cT_{\rm E}$ becomes the time-reversal $\cT$ by a Wick rotation \cite{Kapustin1406.7329}.
The Wick rotation introduces the imaginary $\ii$ sign to the Dirac gamma matrix $\gamma_{\rm E} \to \gamma_0 =\ii \gamma_{\rm E}$ of the temporal component in order to fit the 
Lorentzian version of Clifford algebra $(\gamma_0)^2=-(\gamma_{\rm E})^2$,
so $\cT_{}^2=  (\ii\cT_{\rm E})^2=- \cT_{\rm E}^2$,
which suggests that
$\cT_{}^2=(-1)^F$ for $\Pin^{+}(d)$ 
and $\cT_{}^2=1$ for $\Pin^{-}(d)$.

The centers of $\Pin^{+}(d)$ and $\Pin^{-}(d)$ for $d\geq 2$ are:
\bea \label{eq:Pin-center}
Z(\Pin^+(d))= 
\begin{cases}
\Z_2 ,  & d= 0 \mod 4 \\
\Z_2^2,  & d=1 \mod 4 \\
\Z_2, & d=2 \mod 4 \\
\Z_4,  & d=3 \mod 4 
 \end{cases}.
\quad
Z(\Pin^-(d))= 
\begin{cases}
\Z_2 ,  & d= 0 \mod 4 \\
\Z_4,  & d=1 \mod 4 \\
\Z_2, & d=2 \mod 4 \\
\Z_2^2 ,  & d=3 \mod 4 
 \end{cases}.
\eea
\item When $d$ is an odd integer, because of $\O(d) =\SO(d) \times \Z_2^{T_{\rm E}}$,
we can simplify \eq{eq:Pin+} and \eq{eq:Pin-}, in the Euclidean or Lorentzian signatures 
(with time coordinates ${T_{\rm E}}$ and ${T}$ respectively), 
 to:
\bea
\Pin^{+}(d)\cong \begin{cases}  \Spin(d)\times \mathbb{Z}_{2}^{T_{\rm E}}, 
& d=+1 ~(\text{mod}~4) \\ 
(\Spin(d)\times \mathbb{Z}_{4}^{T_{\rm E} F})/\mathbb{Z}_{2}^F,   
& d=+3 ~(\text{mod}~4)\end{cases}~.\\
\Pin^{-}(d)\cong \begin{cases}  
(\Spin(d)\times \mathbb{Z}_{4}^{T_{\rm E}F})/\mathbb{Z}_{2}^F, 
& d=+ 1 ~(\text{mod}~4)\\ 
\Spin(d)\times \mathbb{Z}_{2}^{T_{\rm E}}, 
&  d=+3 ~(\text{mod}~4)
\end{cases}~.
\eea
These forms match the center of $\Pin^{\pm}$ groups in \eq{eq:Pin-center}.
Furthermore, we can still find in these expressions, 
$\Pin^{+}(d)$ and $\Pin^{-}(d)$
contain the discrete subgroups
$ \Z_2^{T_{\rm E}} \times \Z_2^F$
and $ \Z_4^{T_{\rm E}F}$ respectively. 

\item The $\rE(d)$ is defined in \cite{FreedHopkins1604.06527,WanWWZ1912.13504, Wan2019oyr1904.00994}
which is a subgroup of $\O(d) \times \Z_4$,
described by
\bea
\rE(d)=\{ (M, j) \in (\O(d), \Z_4) \; \vert \; \det M = j^2\}
\eea
in our interpretations, we take $\Z_4= \Z_4^{T B}$ which is from
$1 \to   \Z_2^{B} \to   \Z_4^{T B}  \to   \Z_2^{T}  \to  1$.\footnote{{For bosonic case, 
we can identify the Lorentz and Euclidean time reversal symmetries directly (without extra $-1$ sign),
thus $\Z_4^{T B}= \Z_4^{T_{\rE} B}$ and 
$\Z_2^{T} = \Z_2^{T_{\rE} }$.}}
The $\rE(d)$ sits at the following extensions:
$1 \to \Z_2^{B} \to \rE(d) \to \O(d) \to 1$
and $1 \to \SO(d)  \to \rE(d) \to \Z_4^{TB} \to 1$  \cite{Wan2019oyr1904.00994}.
\end{enumerate}

For the later convenience, we will specify two species of fermions $F_-$, $F_+$, and their boson bound states $B=F_+F_-$;
their time-reversal $\cT$ and spacetime symmetries in the continuum correspond to:
\bea\hspace{-5mm}
\text{Fermion } F_-&:& \cT^2 = +1,\;  \cT_{\rm E}^2 = -1, \quad (\Z_2^{F_-}\rtimes \SO(d))  \rtimes \mathbb{Z}_2^{T_{\rm E}}=
\Spin(d)  \rtimes \mathbb{Z}_2^{T_{\rm E}}
= \Pin^-(d).\quad\quad\quad \\
\text{Fermion }  F_+&:& \cT^2 = -1, \;  \cT_{\rm E}^2 = +1,  \quad (\Z_2^{F_+}\rtimes \SO(d))  \rtimes \mathbb{Z}_2^{T_{\rm E}}= 
{\frac{\Spin(d)\rtimes \mathbb{Z}_{4}^{T_{\rm E}F}}{\mathbb{Z}_{2}^F}}
= \Pin^+(d).\quad\quad\quad\\
\text{Boson }  B&:& {\cT^2 = -1,\;  \cT_{\rm E}^2 = -1,  \quad
\big(  \SO(d) \times  \Z_2^B\big)  \rtimes \mathbb{Z}_2^{T_{\rm E}}=
{  \SO(d)   \rtimes \mathbb{Z}_4^{T_{\rm E}B}}= \rE(d).}
\eea
%

%

\bibliography{references}

\providecommand{\href}[2]{#2}\begingroup\raggedright\begin{thebibliography}{10}

\bibitem{ChenGuLiuWen_GroupChomology_PhysRevB.87.155114}
X.~Chen, Z.-C. Gu, Z.-X. Liu and X.-G. Wen, \emph{Symmetry protected
  topological orders and the group cohomology of their symmetry group},
  \href{http://dx.doi.org/10.1103/PhysRevB.87.155114}{\emph{Phys. Rev. B} {\bf
  87} 155114 (2013 Apr)}.

\bibitem{Senthil1405.4015}
T.~{Senthil}, \emph{{Symmetry-Protected Topological Phases of Quantum Matter}},
  \href{http://dx.doi.org/10.1146/annurev-conmatphys-031214-014740}{\emph{Annual
  Review of Condensed Matter Physics} {\bf 6} 299--324 (2015 Mar.)},
  [\href{https://arxiv.org/abs/1405.4015}{{\tt arXiv:1405.4015}}].

\bibitem{AP_Unwinding_PRB2018}
A.~Prakash, J.~Wang and T.-C. Wei, \emph{{Unwinding Short-Range Entanglement}},
  \href{http://dx.doi.org/10.1103/PhysRevB.98.125108}{\emph{Phys. Rev. B} {\bf
  98} 125108 (2018)}, [\href{https://arxiv.org/abs/1804.11236}{{\tt
  arXiv:1804.11236}}].

\bibitem{WangWenWitten_SymmetricGapped_PRX2018}
J.~Wang, X.-G. Wen and E.~Witten, \emph{{Symmetric Gapped Interfaces of SPT and
  SET States: Systematic Constructions}},
  \href{http://dx.doi.org/10.1103/PhysRevX.8.031048}{\emph{Phys. Rev. X} {\bf
  8} 031048 (2018)}, [\href{https://arxiv.org/abs/1705.06728}{{\tt
  arXiv:1705.06728}}].

\bibitem{Witten1605.02391}
E.~Witten, \emph{{The ''Parity'' Anomaly On An Unorientable Manifold}},
  \href{http://dx.doi.org/10.1103/PhysRevB.94.195150}{\emph{Phys. Rev. B} {\bf
  94} 195150 (2016)}, [\href{https://arxiv.org/abs/1605.02391}{{\tt
  arXiv:1605.02391}}].

\bibitem{Tachikawa1712.09542}
Y.~Tachikawa, \emph{{On gauging finite subgroups}},
  \href{http://dx.doi.org/10.21468/SciPostPhys.8.1.015}{\emph{SciPost Phys.}
  {\bf 8} 015 (2020)}, [\href{https://arxiv.org/abs/1712.09542}{{\tt
  arXiv:1712.09542}}].

\bibitem{Komargodski_SciPostPhys.6.1.003_2019}
Z.~Komargodski, A.~Sharon, R.~Thorngren and X.~Zhou, \emph{{Comments on Abelian
  Higgs Models and Persistent Order}},
  \href{http://dx.doi.org/10.21468/SciPostPhys.6.1.003}{\emph{SciPost Phys.}
  {\bf 6} 3 (2019)}.

\bibitem{VishwanathSentil_surfaceTI_PhysRevX.3.011016}
A.~Vishwanath and T.~Senthil, \emph{Physics of three-dimensional bosonic
  topological insulators: Surface-deconfined criticality and quantized
  magnetoelectric effect},
  \href{http://dx.doi.org/10.1103/PhysRevX.3.011016}{\emph{Phys. Rev. X} {\bf
  3} 011016 (2013 Feb)}.

\bibitem{LukaszAshvin_TSC_PRX2013}
L.~Fidkowski, X.~Chen and A.~Vishwanath, \emph{Non-abelian topological order on
  the surface of a 3d topological superconductor from an exactly solved model},
  \href{http://dx.doi.org/10.1103/PhysRevX.3.041016}{\emph{Phys. Rev. X} {\bf
  3} 041016 (2013 Nov)}.

\bibitem{AKLT_PhysRevLett.59.799}
I.~Affleck, T.~Kennedy, E.~H. Lieb and H.~Tasaki, \emph{Rigorous results on
  valence-bond ground states in antiferromagnets},
  \href{http://dx.doi.org/10.1103/PhysRevLett.59.799}{\emph{Phys. Rev. Lett.}
  {\bf 59} 799--802 (1987 Aug)}.

\bibitem{LiebSchultzMattis1961fr}
E.~H. Lieb, T.~Schultz and D.~Mattis, \emph{{Two soluble models of an
  antiferromagnetic chain}},
  \href{http://dx.doi.org/10.1016/0003-4916(61)90115-4}{\emph{Annals Phys.}
  {\bf 16} 407--466 (1961)}.

\bibitem{AP_LSM}
A.~Prakash, \emph{An elementary proof of 1d lsm theorems},
  \href{https://arxiv.org/abs/2002.11176}{{\tt arXiv:2002.11176}}.

\bibitem{Sachdev1992fkSYK9212030}
S.~Sachdev and J.~Ye, \emph{{Gapless spin fluid ground state in a random,
  quantum Heisenberg magnet}},
  \href{http://dx.doi.org/10.1103/PhysRevLett.70.3339}{\emph{Phys. Rev. Lett.}
  {\bf 70} 3339 (1993)}, [\href{https://arxiv.org/abs/cond-mat/9212030}{{\tt
  arXiv:cond-mat/9212030}}].

\bibitem{WanWang1812.11967}
Z.~Wan and J.~Wang, \emph{{Higher anomalies, higher symmetries, and cobordisms
  I: classification of higher-symmetry-protected topological states and their
  boundary fermionic/bosonic anomalies via a generalized cobordism theory}},
  \href{http://dx.doi.org/10.4310/AMSA.2019.v4.n2.a2}{\emph{Ann. Math. Sci.
  Appl.} {\bf 4} 107--311 (2019)},
  [\href{https://arxiv.org/abs/1812.11967}{{\tt arXiv:1812.11967}}].

\bibitem{SchnyderRyu_classification_doi:10.1063/1.3149481}
A.~P. Schnyder, S.~Ryu, A.~Furusaki and A.~W.~W. Ludwig, \emph{Classification
  of topological insulators and superconductors},
  \href{http://dx.doi.org/10.1063/1.3149481}{\emph{AIP Conference Proceedings}
  {\bf 1134} 10--21 (2009)},
  [\href{https://arxiv.org/abs/https://aip.scitation.org/doi/pdf/10.1063/1.3149481}{{\tt
  arXiv:https://aip.scitation.org/doi/pdf/10.1063/1.3149481}}].

\bibitem{Kitaev_periodic_doi:10.1063/1.3149495}
A.~Kitaev, \emph{Periodic table for topological insulators and
  superconductors}, \href{http://dx.doi.org/10.1063/1.3149495}{\emph{AIP
  Conference Proceedings} {\bf 1134} 22--30 (2009)},
  [\href{https://arxiv.org/abs/https://aip.scitation.org/doi/pdf/10.1063/1.3149495}{{\tt
  arXiv:https://aip.scitation.org/doi/pdf/10.1063/1.3149495}}].

\bibitem{Gu1201.2648}
Z.-C. Gu and X.-G. Wen, \emph{{Symmetry-protected topological orders for
  interacting fermions: Fermionic topological nonlinear \ensuremath{\sigma}
  models and a special group supercohomology theory}},
  \href{http://dx.doi.org/10.1103/PhysRevB.90.115141}{\emph{Phys. Rev. B} {\bf
  90} 115141 (2014)}, [\href{https://arxiv.org/abs/1201.2648}{{\tt
  arXiv:1201.2648}}].

\bibitem{Kitaev2015}
A.~Kitaev, \emph{{ Homotopy-theoretic approach to SPT phases in action: Z(16)
  classification of three-dimensional superconductors, in Symmetry and Topology
  in Quantum Matter Workshop, Institute for Pure and Applied Mathematics,
  University of California, Los Angeles, California.
  \href{http://www.ipam.ucla.edu/abstract/?tid=12389}{http://www.ipam.ucla.edu/abstract/?tid=12389}}},
  .

\bibitem{Kapustin1701.08264}
A.~Kapustin and R.~Thorngren, \emph{{Fermionic SPT phases in higher dimensions
  and bosonization}},
  \href{http://dx.doi.org/10.1007/JHEP10(2017)080}{\emph{JHEP} {\bf 10} 080
  (2017)}, [\href{https://arxiv.org/abs/1701.08264}{{\tt arXiv:1701.08264}}].

\bibitem{WangGu1703.10937}
Q.-R. Wang and Z.-C. Gu, \emph{{Towards a Complete Classification of
  Symmetry-Protected Topological Phases for Interacting Fermions in Three
  Dimensions and a General Group Supercohomology Theory}},
  \href{http://dx.doi.org/10.1103/PhysRevX.8.011055}{\emph{Phys. Rev. X} {\bf
  8} 011055 (2018)}, [\href{https://arxiv.org/abs/1703.10937}{{\tt
  arXiv:1703.10937}}].

\bibitem{GaiottoJohnson-Freyd1712.07950}
D.~Gaiotto and T.~Johnson-Freyd, \emph{{Symmetry Protected Topological phases
  and Generalized Cohomology}},
  \href{http://dx.doi.org/10.1007/JHEP05(2019)007}{\emph{JHEP} {\bf 05} 007
  (2019)}, [\href{https://arxiv.org/abs/1712.07950}{{\tt arXiv:1712.07950}}].

\bibitem{WangGu1811.00536}
Q.-R. Wang and Z.-C. Gu, \emph{{Construction and classification of symmetry
  protected topological phases in interacting fermion systems}},
  \href{https://arxiv.org/abs/1811.00536}{{\tt arXiv:1811.00536}}.

\bibitem{Kapustin1406.7329}
A.~Kapustin, R.~Thorngren, A.~Turzillo and Z.~Wang, \emph{{Fermionic Symmetry
  Protected Topological Phases and Cobordisms}},
  \href{http://dx.doi.org/10.1007/JHEP12(2015)052}{\emph{JHEP} {\bf 12} 052
  (2015)}, [\href{https://arxiv.org/abs/1406.7329}{{\tt arXiv:1406.7329}}].

\bibitem{FreedHopkins1604.06527}
D.~S. {Freed} and M.~J. {Hopkins}, \emph{{Reflection positivity and invertible
  topological phases}}, {\emph{ArXiv e-prints} (2016 Apr.)},
  [\href{https://arxiv.org/abs/1604.06527}{{\tt arXiv:1604.06527}}].

\bibitem{1711.11587GPW}
M.~{Guo}, P.~{Putrov} and J.~{Wang}, \emph{{Time reversal, SU(N) Yang-Mills and
  cobordisms: Interacting topological superconductors/insulators and quantum
  spin liquids in 3 + 1 D}},
  \href{http://dx.doi.org/10.1016/j.aop.2018.04.025}{\emph{Annals of Physics}
  {\bf 394} 244--293 (2018 Jul)}, [\href{https://arxiv.org/abs/1711.11587}{{\tt
  arXiv:1711.11587}}].

\bibitem{GuoJW1812.11959}
M.~Guo, K.~Ohmori, P.~Putrov, Z.~Wan and J.~Wang, \emph{{Fermionic Finite-Group
  Gauge Theories and Interacting Symmetric/Crystalline Orders via Cobordisms}},
  \href{http://dx.doi.org/10.1007/s00220-019-03671-6}{\emph{Communications in
  Mathematical Physics} 1073?1154 (2018 Jan)},
  [\href{https://arxiv.org/abs/1812.11959}{{\tt arXiv:1812.11959}}].

\bibitem{ChiuRyuSchnyderRyu_RevModPhys.88.035005}
C.-K. Chiu, J.~C.~Y. Teo, A.~P. Schnyder and S.~Ryu, \emph{Classification of
  topological quantum matter with symmetries},
  \href{http://dx.doi.org/10.1103/RevModPhys.88.035005}{\emph{Rev. Mod. Phys.}
  {\bf 88} 035005 (2016 Aug)}.

\bibitem{1405.7689}
J.~C. Wang, Z.-C. Gu and X.-G. Wen, \emph{{Field theory representation of
  gauge-gravity symmetry-protected topological invariants, group cohomology and
  beyond}}, \href{http://dx.doi.org/10.1103/PhysRevLett.114.031601}{\emph{Phys.
  Rev. Lett.} {\bf 114} 031601 (2015)},
  [\href{https://arxiv.org/abs/1405.7689}{{\tt arXiv:1405.7689}}].

\bibitem{Wen2016ddy1610.03911}
X.-G. Wen, \emph{{Zoo of quantum-topological phases of matter}},
  \href{http://dx.doi.org/10.1103/RevModPhys.89.041004}{\emph{Rev. Mod. Phys.}
  {\bf 89} 041004 (2017)}, [\href{https://arxiv.org/abs/1610.03911}{{\tt
  arXiv:1610.03911}}].

\bibitem{Kitaev_majorana_2001}
A.~Y. Kitaev, \emph{Unpaired majorana fermions in quantum wires},
  \href{http://dx.doi.org/10.1070/1063-7869/44/10s/s29}{\emph{Physics-Uspekhi}
  {\bf 44} 131--136 (2001 oct)}.

\bibitem{FidkowskiKitaev_PhysRevB.81.134509_2010}
L.~Fidkowski and A.~Kitaev, \emph{Effects of interactions on the topological
  classification of free fermion systems},
  \href{http://dx.doi.org/10.1103/PhysRevB.81.134509}{\emph{Phys. Rev. B} {\bf
  81} 134509 (2010 Apr)}.

\bibitem{FidkowskiKitaev_PhysRevB.83.075103_2011}
L.~Fidkowski and A.~Kitaev, \emph{Topological phases of fermions in one
  dimension}, \href{http://dx.doi.org/10.1103/PhysRevB.83.075103}{\emph{Phys.
  Rev. B} {\bf 83} 075103 (2011 Feb)}.

\bibitem{M16}
``M16.'' \url{https://groupprops.subwiki.org/wiki/M16}.

\bibitem{Gu1308.2488}
Z.-C. Gu, \emph{{Fractionalized time reversal, parity and charge conjugation
  symmetry in topological superconductor: a possible origin of three
  generations of neutrinos and mass mixing}},
  \href{http://dx.doi.org/10.1103/PhysRevResearch.2.033290}{\emph{Phys. Rev.
  Res.} {\bf 2} 033290 (2020)}, [\href{https://arxiv.org/abs/1308.2488}{{\tt
  arXiv:1308.2488}}].

\bibitem{NatAshvin_PhysRevB.98.165104}
N.~Tantivasadakarn and A.~Vishwanath, \emph{Full commuting projector
  hamiltonians of interacting symmetry-protected topological phases of
  fermions}, \href{http://dx.doi.org/10.1103/PhysRevB.98.165104}{\emph{Phys.
  Rev. B} {\bf 98} 165104 (2018 Oct)}.

\bibitem{APJW-PRL}
A.~Prakash and J.~Wang, \emph{Boundary supersymmetry of 1+1 d fermionic spt
  phases},  \href{https://arxiv.org/abs/2011.12320}{{\tt arXiv:2011.12320}}.

\bibitem{Witten_IntroSUSY_1983}
E.~Witten, \emph{Introduction to Supersymmetry}.
\newblock Springer US, Boston, MA, 1983,
  \href{http://dx.doi.org/10.1007/978-1-4613-3655-6\_7}{10.1007/978-1-4613-3655-6\_7}.

\bibitem{CooperKhareSukhatme_SUSYQM_1995}
F.~Cooper, A.~Khare and U.~Sukhatme, \emph{Supersymmetry and quantum
  mechanics},
  \href{http://dx.doi.org/10.1016/0370-1573(94)00080-m}{\emph{Physics Reports}
  {\bf 251} 267–385 (1995 Jan)}.

\bibitem{WITTEN_SUSYQM_1981513}
E.~Witten, \emph{Dynamical breaking of supersymmetry},
  \href{http://dx.doi.org/https://doi.org/10.1016/0550-3213(81)90006-7}{\emph{Nuclear
  Physics B} {\bf 188} 513 -- 554 (1981)}.

\bibitem{WITTEN_SUSYQM_1982253}
E.~Witten, \emph{Constraints on supersymmetry breaking},
  \href{http://dx.doi.org/https://doi.org/10.1016/0550-3213(82)90071-2}{\emph{Nuclear
  Physics B} {\bf 202} 253 -- 316 (1982)}.

\bibitem{WittenSUSY1982}
E.~Witten, \emph{{Supersymmetry and Morse theory}}, {\emph{J. Diff. Geom.} {\bf
  17} 661--692 (1982)}.

\bibitem{BehrendsBeri_SUSYSYK_PhysRevLett.124.236804}
J.~Behrends and B.~B\'eri, \emph{{Supersymmetry in the Standard
  Sachdev-Ye-Kitaev Model}},
  \href{http://dx.doi.org/10.1103/PhysRevLett.124.236804}{\emph{Phys. Rev.
  Lett.} {\bf 124} 236804 (2020)},
  [\href{https://arxiv.org/abs/1908.00995}{{\tt arXiv:1908.00995}}].

\bibitem{KapustinTurzilloYou_PhysRevB.98.125101_2018}
A.~Kapustin, A.~Turzillo and M.~You, \emph{Spin topological field theory and
  fermionic matrix product states},
  \href{http://dx.doi.org/10.1103/PhysRevB.98.125101}{\emph{Phys. Rev. B} {\bf
  98} 125101 (2018 Sep)}.

\bibitem{YouLudwigXu_SYKSPT_PhysRevB.95.115150}
Y.-Z. You, A.~W.~W. Ludwig and C.~Xu, \emph{Sachdev-ye-kitaev model and
  thermalization on the boundary of many-body localized fermionic
  symmetry-protected topological states},
  \href{http://dx.doi.org/10.1103/PhysRevB.95.115150}{\emph{Phys. Rev. B} {\bf
  95} 115150 (2017 Mar)}.

\bibitem{Kitaevtalk}
A.~Kitaev, \emph{{A simple model of quantum holography (part 1 and 2)}},
  {\emph{online.kitp.ucsb.edu
  \href{https://online.kitp.ucsb.edu/online/entangled15/kitaev/}{https://online.kitp.ucsb.edu/online/entangled15/kitaev/}
  \href{https://online.kitp.ucsb.edu/online/entangled15/kitaev2/}{https://online.kitp.ucsb.edu/online/entangled15/kitaev2/}}
  (2015)}.

\bibitem{MaldacenaStanford1604.07818}
J.~Maldacena and D.~Stanford, \emph{{Remarks on the Sachdev-Ye-Kitaev model}},
  \href{http://dx.doi.org/10.1103/PhysRevD.94.106002}{\emph{Phys. Rev. D} {\bf
  94} 106002 (2016)}, [\href{https://arxiv.org/abs/1604.07818}{{\tt
  arXiv:1604.07818}}].

\bibitem{KitaevSuh1711.08467}
A.~Kitaev and S.~J. Suh, \emph{{The soft mode in the Sachdev-Ye-Kitaev model
  and its gravity dual}},
  \href{http://dx.doi.org/10.1007/JHEP05(2018)183}{\emph{JHEP} {\bf 05} 183
  (2018)}, [\href{https://arxiv.org/abs/1711.08467}{{\tt arXiv:1711.08467}}].

\bibitem{WanWWZ1912.13504}
Z.~Wan, J.~Wang and Y.~Zheng, \emph{{Higher Anomalies, Higher Symmetries, and
  Cobordisms II: Lorentz Symmetry Extension and Enriched Bosonic/Fermionic
  Quantum Gauge Theory}}, {\emph{Ann. Math. Sci. Appl.} {\bf 5} (2020 (to
  appear))}, [\href{https://arxiv.org/abs/1912.13504}{{\tt arXiv:1912.13504}}].

\bibitem{Kaidi2019pzjJulioTachikawa1908.04805}
J.~Kaidi, J.~Parra-Martinez and Y.~Tachikawa, \emph{{GSO projections via SPT
  phases}},  \href{https://arxiv.org/abs/1908.04805}{{\tt arXiv:1908.04805}}.

\bibitem{Kaidi2019tyfJulioTachikawa1911.11780}
J.~Kaidi, J.~Parra-Martinez and Y.~Tachikawa, \emph{{Topological
  Superconductors on Superstring Worldsheets}},
  \href{https://arxiv.org/abs/1911.11780}{{\tt arXiv:1911.11780}}.

\bibitem{2018arXiv180502772T}
Y.~{Tachikawa} and K.~{Yonekura}, \emph{{Why are fractional charges of
  orientifolds compatible with Dirac quantization?}}, {\emph{arXiv e-prints}
  arXiv:1805.02772 (2018 May)}, [\href{https://arxiv.org/abs/1805.02772}{{\tt
  arXiv:1805.02772}}].

\bibitem{Hason2019akwRyanThorngren1910.14039}
I.~Hason, Z.~Komargodski and R.~Thorngren, \emph{{Anomaly Matching in the
  Symmetry Broken Phase: Domain Walls, CPT, and the Smith Isomorphism}},
  \href{http://dx.doi.org/10.21468/SciPostPhys.8.4.062}{\emph{SciPost Phys.}
  {\bf 8} 062 (2020)}, [\href{https://arxiv.org/abs/1910.14039}{{\tt
  arXiv:1910.14039}}].

\bibitem{PTWtoappear}
P.~Putrov, R.~Thorngren, Z.~Wan and J.~Wang, \emph{{(in preparation)}},
  {\emph{to appear}}.

\bibitem{Dijkgraaf2018Witten1804.03275}
R.~Dijkgraaf and E.~Witten, \emph{{Developments in Topological Gravity}},
  \href{http://dx.doi.org/10.1142/S0217751X18300296}{\emph{Int. J. Mod. Phys.
  A} {\bf 33} 1830029 (2018)}, [\href{https://arxiv.org/abs/1804.03275}{{\tt
  arXiv:1804.03275}}].

\bibitem{StanfordWitten1907.03363}
D.~Stanford and E.~Witten, \emph{{JT Gravity and the Ensembles of Random Matrix
  Theory}},  \href{https://arxiv.org/abs/1907.03363}{{\tt arXiv:1907.03363}}.

\bibitem{Putrov2016qdo1612.09298PWY}
P.~Putrov, J.~Wang and S.-T. Yau, \emph{{Braiding Statistics and Link
  Invariants of Bosonic/Fermionic Topological Quantum Matter in 2+1 and 3+1
  dimensions}}, \href{http://dx.doi.org/10.1016/j.aop.2017.06.019}{\emph{Annals
  Phys.} {\bf 384} 254--287 (2017)},
  [\href{https://arxiv.org/abs/1612.09298}{{\tt arXiv:1612.09298}}].

\bibitem{Garcia-EtxebarriaMontero1808.00009}
I.~Garcia-Etxebarria and M.~Montero, \emph{{Dai-Freed anomalies in particle
  physics}}, \href{http://dx.doi.org/10.1007/JHEP08(2019)003}{\emph{JHEP} {\bf
  08} 003 (2019)}, [\href{https://arxiv.org/abs/1808.00009}{{\tt
  arXiv:1808.00009}}].

\bibitem{JW2006.16996}
J.~Wang, \emph{{Anomaly and Cobordism Constraints Beyond Standard Model:
  Topological Force}},  \href{https://arxiv.org/abs/2006.16996}{{\tt
  arXiv:2006.16996}}.

\bibitem{JW2008.06499}
J.~Wang, \emph{{Anomaly and Cobordism Constraints Beyond Grand Unification:
  Energy Hierarchy}},  \href{https://arxiv.org/abs/2008.06499}{{\tt
  arXiv:2008.06499}}.

\bibitem{RazamatTong2009.05037}
S.~S. Razamat and D.~Tong, \emph{{Gapped Chiral Fermions}},
  \href{https://arxiv.org/abs/2009.05037}{{\tt arXiv:2009.05037}}.

\bibitem{Wen2013ppa1305.1045}
X.-G. Wen, \emph{{A lattice non-perturbative definition of an SO(10) chiral
  gauge theory and its induced standard model}},
  \href{http://dx.doi.org/10.1088/0256-307X/30/11/111101}{\emph{Chin. Phys.
  Lett.} {\bf 30} 111101 (2013)}, [\href{https://arxiv.org/abs/1305.1045}{{\tt
  arXiv:1305.1045}}].

\bibitem{You2014oaaYouBenTovXu1402.4151}
Y.~You, Y.~BenTov and C.~Xu, \emph{{Interacting Topological Superconductors and
  possible Origin of $16n$ Chiral Fermions in the Standard Model}},
  \href{https://arxiv.org/abs/1402.4151}{{\tt arXiv:1402.4151}}.

\bibitem{YX14124784}
Y.-Z. You and C.~Xu, \emph{Interacting topological insulator and emergent grand
  unified theory},
  \href{http://dx.doi.org/10.1103/physrevb.91.125147}{\emph{Phys. Rev. B} {\bf
  91} 125147 (2015 Mar.)}, [\href{https://arxiv.org/abs/1412.4784}{{\tt
  arXiv:1412.4784}}].

\bibitem{Kikukawa1710.11618}
Y.~Kikukawa, \emph{{On the gauge invariant path-integral measure for the
  overlap Weyl fermions in $\underline{16}$ of SO(10)}},
  \href{http://dx.doi.org/10.1093/ptep/ptz115}{\emph{PTEP} {\bf 2019} 113B03
  (2019)}, [\href{https://arxiv.org/abs/1710.11618}{{\tt arXiv:1710.11618}}].

\bibitem{WangWen1809.11171}
J.~Wang and X.-G. Wen, \emph{{Nonperturbative definition of the standard
  models}},
  \href{http://dx.doi.org/10.1103/PhysRevResearch.2.023356}{\emph{Phys. Rev.
  Res.} {\bf 2} 023356 (2020)}, [\href{https://arxiv.org/abs/1809.11171}{{\tt
  arXiv:1809.11171}}].

\bibitem{Catterall2010.02290}
S.~Catterall, \emph{{Chiral Lattice Theories From Staggered Fermions}},
  \href{https://arxiv.org/abs/2010.02290}{{\tt arXiv:2010.02290}}.

\bibitem{Fidkowski1305.5851}
L.~Fidkowski, X.~Chen and A.~Vishwanath, \emph{{Non-Abelian Topological Order
  on the Surface of a 3D Topological Superconductor from an Exactly Solved
  Model}}, \href{http://dx.doi.org/10.1103/PhysRevX.3.041016}{\emph{Phys. Rev.
  X} {\bf 3} 041016 (2013)}, [\href{https://arxiv.org/abs/1305.5851}{{\tt
  arXiv:1305.5851}}].

\bibitem{kirby1990pin}
R.~C. Kirby and L.~R. Taylor, \emph{Pin structures on low-dimensional
  manifolds}, {\emph{London Mathematical Society Lecture Note Series} {\bf 2}
  177--242 (1990)}.

\bibitem{ThorngrenVerresen2008.06638}
R.~Thorngren, A.~Vishwanath and R.~Verresen, \emph{{Intrinsically Gapless
  Topological Phases}},  \href{https://arxiv.org/abs/2008.06638}{{\tt
  arXiv:2008.06638}}.

\bibitem{Grover1206.1332}
T.~Grover and A.~Vishwanath, \emph{{Quantum Criticality in Topological
  Insulators and Superconductors: Emergence of Strongly Coupled Majoranas and
  Supersymmetry}},  \href{https://arxiv.org/abs/1206.1332}{{\tt
  arXiv:1206.1332}}.

\bibitem{Ponte1206.2340}
P.~Ponte and S.-S. Lee, \emph{{Emergence of supersymmetry on the surface of
  three dimensional topological insulators}},
  \href{http://dx.doi.org/10.1088/1367-2630/16/1/013044}{\emph{New J. Phys.}
  {\bf 16} 013044 (2014)}, [\href{https://arxiv.org/abs/1206.2340}{{\tt
  arXiv:1206.2340}}].

\bibitem{Grover1301.7449}
T.~Grover, D.~Sheng and A.~Vishwanath, \emph{{Emergent Space-Time Supersymmetry
  at the Boundary of a Topological Phase}},
  \href{http://dx.doi.org/10.1126/science.1248253}{\emph{Science} {\bf 344}
  280--283 (2014)}, [\href{https://arxiv.org/abs/1301.7449}{{\tt
  arXiv:1301.7449}}].

\bibitem{ZerfMaciejko1605.09423}
N.~Zerf, C.-H. Lin and J.~Maciejko, \emph{{Superconducting quantum criticality
  of topological surface states at three loops}},
  \href{http://dx.doi.org/10.1103/PhysRevB.94.205106}{\emph{Phys. Rev. B} {\bf
  94} 205106 (2016)}, [\href{https://arxiv.org/abs/1605.09423}{{\tt
  arXiv:1605.09423}}].

\bibitem{YuJiabin1902.07407}
J.~Yu, R.~Roiban, S.-K. Jian and C.-X. Liu, \emph{{Finite-Scale Emergence of
  2+1D Supersymmetry at First-Order Quantum Phase Transition}},
  \href{http://dx.doi.org/10.1103/PhysRevB.100.075153}{\emph{Phys. Rev. B} {\bf
  100} 075153 (2019)}, [\href{https://arxiv.org/abs/1902.07407}{{\tt
  arXiv:1902.07407}}].

\bibitem{Montero2008.11729}
M.~Montero and C.~Vafa, \emph{{Cobordism Conjecture, Anomalies, and the String
  Lamppost Principle}},  \href{https://arxiv.org/abs/2008.11729}{{\tt
  arXiv:2008.11729}}.

\bibitem{Kaidi2010.10521}
J.~Kaidi, \emph{{Stable Vacua for Tachyonic Strings}},
  \href{https://arxiv.org/abs/2010.10521}{{\tt arXiv:2010.10521}}.

\bibitem{McNamara1909.10355}
J.~McNamara and C.~Vafa, \emph{{Cobordism Classes and the Swampland}},
  \href{https://arxiv.org/abs/1909.10355}{{\tt arXiv:1909.10355}}.

\bibitem{Gaiotto2014kfa1412.5148}
D.~Gaiotto, A.~Kapustin, N.~Seiberg and B.~Willett, \emph{{Generalized Global
  Symmetries}}, \href{http://dx.doi.org/10.1007/JHEP02(2015)172}{\emph{JHEP}
  {\bf 02} 172 (2015)}, [\href{https://arxiv.org/abs/1412.5148}{{\tt
  arXiv:1412.5148}}].

\bibitem{Wan1812.11955}
Z.~Wan and J.~Wang, \emph{{Adjoint QCD$_4$, Deconfined Critical Phenomena,
  Symmetry-Enriched Topological Quantum Field Theory, and Higher
  Symmetry-Extension}},
  \href{http://dx.doi.org/10.1103/PhysRevD.99.065013}{\emph{Phys. Rev.} {\bf
  D99} 065013 (2019)}, [\href{https://arxiv.org/abs/1812.11955}{{\tt
  arXiv:1812.11955}}].

\bibitem{Wan1812.11968}
Z.~Wan, J.~Wang and Y.~Zheng, \emph{{New higher anomalies, SU(N)
  Yang\textendash{}Mills gauge theory and $\mathbb{CP}^{\mathrm{N}-1}$ sigma
  model}}, \href{http://dx.doi.org/10.1016/j.aop.2020.168074}{\emph{Annals
  Phys.} {\bf 414} 168074 (2020)},
  [\href{https://arxiv.org/abs/1812.11968}{{\tt arXiv:1812.11968}}].

\bibitem{Wan2019oyr1904.00994}
Z.~Wan, J.~Wang and Y.~Zheng, \emph{{Quantum 4d Yang-Mills Theory and
  Time-Reversal Symmetric 5d Higher-Gauge Topological Field Theory}},
  \href{http://dx.doi.org/10.1103/PhysRevD.100.085012}{\emph{Phys. Rev.} {\bf
  D100} 085012 (2019)}, [\href{https://arxiv.org/abs/1904.00994}{{\tt
  arXiv:1904.00994}}].

\bibitem{Wang1801.05416}
J.~Wang, K.~Ohmori, P.~Putrov, Y.~Zheng, Z.~Wan, M.~Guo et~al.,
  \emph{{Tunneling Topological Vacua via Extended Operators: (Spin-)TQFT
  Spectra and Boundary Deconfinement in Various Dimensions}},
  \href{http://dx.doi.org/10.1093/ptep/pty051}{\emph{PTEP} {\bf 2018} 053A01
  (2018)}, [\href{https://arxiv.org/abs/1801.05416}{{\tt arXiv:1801.05416}}].

\bibitem{Kobayashi2019lep1905.05391}
R.~Kobayashi, K.~Ohmori and Y.~Tachikawa, \emph{{On gapped boundaries for SPT
  phases beyond group cohomology}},
  \href{http://dx.doi.org/10.1007/JHEP11(2019)131}{\emph{JHEP} {\bf 11} 131
  (2019)}, [\href{https://arxiv.org/abs/1905.05391}{{\tt arXiv:1905.05391}}].

\bibitem{Cordova2017vab1711.10008}
C.~Cordova, P.-S. Hsin and N.~Seiberg, \emph{{Global Symmetries, Counterterms,
  and Duality in Chern-Simons Matter Theories with Orthogonal Gauge Groups}},
  \href{http://dx.doi.org/10.21468/SciPostPhys.4.4.021}{\emph{SciPost Phys.}
  {\bf 4} 021 (2018)}, [\href{https://arxiv.org/abs/1711.10008}{{\tt
  arXiv:1711.10008}}].

\end{thebibliography}\endgroup
\bibliographystyle{arXiv-new}

\end{document}